\documentclass[hyper,12pt,letterpaper]{JHEP3}
\pdfoutput=1

\usepackage{graphicx}
\usepackage{latexsym,amsmath,amsfonts,amssymb}
\usepackage{mathrsfs}
\usepackage[makeroom]{cancel}
\usepackage{bbm}
\usepackage{bm}
\usepackage{subfigure}
\usepackage{paralist}
\usepackage{url}

\newcommand{\eg}{\textit{e.g.}}

\newcommand{\ie}{\textit{i.e.}}

\newcommand{\MOD}[1]{{\;(\text{mod} \;#1)}}

\numberwithin{equation}{section}

\newcommand{\be}{\begin{equation}} 
\newcommand{\ee}{\end{equation}}
\newcommand{\bea}{\begin{equation} \begin{aligned}} \newcommand{\eea}{\end{aligned} \end{equation}}

\newcommand{\bit}{\begin{itemize}} 
\newcommand{\eit}{\end{itemize}}

\newcommand{\cC}{\mathcal{C}}

\newcommand{\cF}{\mathcal{F}}

\newcommand{\cH}{\mathcal{H}}
\newcommand{\cI}{\mathcal{I}}

\newcommand{\cM}{\mathcal{M}}
\newcommand{\cN}{\mathcal{N}}

\newcommand{\cS}{\mathcal{S}}

\newcommand{\cV}{\mathcal{V}}
\newcommand{\cW}{\mathcal{W}}

\newcommand{\Z}{\mathbb{Z}}
\newcommand{\C}{\mathbb{C}}
\newcommand{\R}{\mathbb{R}}

\newcommand{\bb}{{\bb b}}

\newcommand{\dilog}{{\text{Li}_2}}

\title{Partition functions on 3d circle bundles and their gravity duals}

\author{Chiara Toldo$^{\,  \sharp, \natural}$ and Brian Willett$^{\natural}$\\
{}$^{\sharp}$ Columbia University in the City of New York\\
Pupin Hall, 538 West 120th Street, New York City, NY 10027\\
{}$^{\natural}$ Kavli Institute for Theoretical Physics\\
 University of California, Santa Barbara, CA 93106
}

\preprint{}

\abstract{The partition function of a three-dimensional $\mathcal{N} =2$ theory on the manifold $\mathcal{M}_{g,p}$, an $S^1$ bundle of degree $p$ over a closed Riemann surface $\Sigma_g$, was recently computed via supersymmetric localization.  In this paper, we compute these partition functions at large $N$ in a class of quiver gauge theories with holographic M-theory duals.  We provide the supergravity bulk dual having as conformal boundary such three-dimensional circle bundles. These configurations are solutions to  $\mathcal{N}=2$ minimal gauged supergravity and pertain to the class of Taub-NUT-AdS and Taub-Bolt-AdS preserving $1/4$ of the supersymmetries. We discuss the conditions for the uplift of these solutions to M-theory, and compute the on-shell action via holographic renormalization.  We show that the uplift condition and on-shell action for the Bolt solutions are correctly reproduced by the large $N$ limit of the partition function of the dual superconformal field theory.  In particular, the $\Sigma_g \times S^1 \cong \mathcal{M}_{g,0}$ partition function, which was recently shown to match the entropy of $AdS_4$ black holes, and the $S^3 \cong \mathcal{M}_{0,1}$ free energy, occur as special cases of our formalism, and we comment on relations between them.}

\begin{document}

\tableofcontents

\section{Introduction}

Recently, there has been much progress in performing exact, nonperturbative computations for superconformal field theories (SCFTs) on curved manifolds via the technique of supersymmetric localization (see the review \cite{Pestun:2016zxk} and references therein).  Such methods have greatly developed in the past several years, providing a tool to study a wide variety of SCFTs in various dimensions and backgrounds, leading to non-trivial tests of holography and other known dualities. 

In particular, recently these techniques have been successfully  applied to the computation of the partition function of three-dimensional superconformal Chern-Simons-matter theories on $\Sigma_g \times S^1$ in presence of background magnetic flux for the R- and flavor symmetries through the Riemann surface, $\Sigma_g$ \cite{Nekrasov:2014xaa,Benini:2015noa,Benini:2016hjo,Closset:2016arn}. By performing a partial topological twist \cite{Witten:1988ze} on $\Sigma_g$, one obtains the so-called ``topologically twisted Witten index'' \cite{Benini:2015noa}.  This was shown in  \cite{Benini:2015eyy} to reproduce in the large $N$ limit\footnote{Upon a suitable extremization of the index with respect to the fugacities, which corresponds to the attractor mechanism in the gravity side.} the macroscopic entropy of supersymmetric magnetic AdS$_4$ black holes in theories of 4d FI-gauged supergravity. These black hole configurations, first found in \cite{Cacciatori:2009iz}, consist of M2-branes wrapped around $\Sigma_g$, and thus they implement the partial topological twist for the QFT describing the low-energy dynamics of the M2-branes.

This recent success led to several extensions and developments. First of all, the entropy matching was performed on more general supergravity backgrounds, including dyonic black holes \cite{Benini:2016rke}, black hole configurations arising from massive IIA supergravity truncations \cite{Hosseini:2017fjo,Benini:2017oxt} and solutions with hyperbolic horizon \cite{Cabo-Bizet:2017jsl}.  Moreover, unexpected relations were discovered between the topologically twisted index on $S^2 \times S^1$ and the corresponding partition function on $S^3$  in the large $N$ limit \cite{Hosseini:2016tor}. 
Specifically, the computation of the twisted index involves, as an intermediate step, the computation of the twisted  superpotential, or Bethe potential, as a function of the flavor fugacities \cite{Benini:2015noa}. Then this quantity was shown to coincide, for a suitable mapping of parameters, with the large $N$ limit of the $S^3$ partition function of the same $\mathcal{N}=2$ theory \cite{Hosseini:2016tor}.
Given that the partition function on $S^3$ of three-dimensional superconformal theories, such as the ABJM theory \cite{Aharony:2008ug}, has been extensively studied, and has connections to entanglement entropy and the $F$-theorem \cite{Jafferis:2011zi,Freedman:2013ryh,Casini:2012ei}, it is natural to ask if such a correspondence has a deeper meaning.

In parallel with these developments, a new class of partition functions for general $3d$ $\cN=2$ gauge theories was computed in \cite{Closset:2017zgf} utilizing a three-dimensional uplift of the $2d$ $A$-model \cite{witten1988}.  These partition functions are defined on the manifold $\mathcal{M}_{g,p}$, a $U(1)$ bundle of Chern degree $p \in \mathbb{Z}$ over a Riemann surface $\Sigma_g$,
\be
S^1 \overset{p}{\rightarrow} \mathcal{M}_{g,p}  \rightarrow \Sigma_g \;,
\ee
where $g \in \mathbb{Z}_{\geq 0}$ denotes the genus of the Riemann surface. This set of manifolds includes in particular the three-sphere $S^3$ and the product spaces $\Sigma_g \times S^1$
\be
\mathcal{M}_{0,1} \simeq S^3 \,, \qquad  \mathcal{M}_{g,0} \simeq \Sigma_g \times S^1\,.
\ee
Thus, these partition functions include both the topologically twisted index of \cite{Nekrasov:2014xaa,Benini:2015noa,Benini:2016hjo,Closset:2016arn} and the round $S^3$ partition function of \cite{Kapustin:2009kz,Jafferis:2010un,Hama:2010av} as special cases.  This then provides a natural framework to address the relation between the topologically twisted index and $S^3$ partition function.  At the same time, constructing explicit supergravity backgrounds whose boundary is such a circle bundle and computing their renormalized on-shell action provides a viable holographic check for these field theory computations.

In more detail, the partition function on $\cM_{g,p}$ can be computed by a sum over supersymmetric ``Bethe vacua,'' \cite{Nekrasov:2009uh},
\be \label{Z_intro_fn} Z_{\cM_{g,p}}(m_i,s_i) = \sum_{I \in \cS_{BE}} {\cF^I}(m_i)^p {\cH^I}(m_i)^{g-1} {\Pi_i^I}(m_i)^{s_i}  \;,\ee
where the index $I$ runs over the set $\cS_{BE}$ of vacua of the theory.  Here $m_i$ and $s_i$ are, respectively, real masses and fluxes for background flavor symmetry gauge fields, and $\cF^I$, $\cH^I$, and $\Pi_i^I$ are certain functions appearing in the $3d$ uplift of the $A$-model, described in Section \ref{sec:field theory} below.  We will argue that, for a class of quiver gauge theories with holographically dual M-theory descriptions, in the large $N$ limit this sum can be approximated by a single dominant term, $I_{dom}$, and we find the result:

\be \label{Z_intro} \log Z_{\cM_{g,p}}(m_i) \approx p \log \cF^{I_{dom}}(m_i) + (g-1) \log  {\cH^{I_{dom}}}(m_i) + s_i \log {\Pi_i^{I_{dom}}}(m_i)  \;, \ee
leading to a very simple dependence on the geometric and flux parameters.  We find the partition function exhibits the expected $N^{3/2}$ scaling, and reproduce and generalize the results of \cite{Benini:2016hjo,Benini:2015eyy} in the case $p=0$.  However, we find that a large $N$ solution exists only under certain conditions on the mass and flux parameters.  In the case of $\cM_{0,1}=S^3$, these conditions differ from those under which previous large $N$ computations of the $S^3$ partition function were carried out, \eg, in  \cite{Jafferis:2011zi}, and we comment on this discrepancy in Section \ref{sec:holo_checks} below.

We reproduce this result holographically, by providing supergravity backgrounds having boundary $\mathcal{M}_{g,p}$ in the framework of minimal $\mathcal{N}=2$ $U(1)$ gauged supergravity. Such solutions can be embedded locally in 11d on 7-dimensional Sasaki-Einstein manifolds. We construct Euclidean regular solutions which preserve 1/4 of the supersymmetries and have appropriately quantized magnetic flux. Starting from the analysis of \cite{Martelli:2011fw,Martelli:2012sz,Martelli:2013aqa}, we find that the boundary can be filled with multiple gravity configurations, with different topology.  In particular, for the boundary $S^3$ case we can have regular ``NUT'' solutions, with topology $\R^4$, and for $S^3/\Z_p$ one finds mildly singular NUT$/\Z_p$ solutions.  On the other hand, for general $\mathcal{M}_{g,p}$ we find regular ``Bolt'' solutions, with topology $\mathcal{O}(-p) \rightarrow \Sigma_g$. The different topology has non-trivial consequences for the uplift of these solutions. Indeed, while there are no requirements for the NUT solution to lift to M-theory, the Bolt uplifts to eleven dimensions only for certain values of $g$ and $p$, depending on the geometrical properties of the internal Sasaki-Einstein  7-manifold.  Interestingly, the same constraints are recovered in the field theory computation by setting all fluxes equal, thus reproducing the universal twist which corresponds to minimal gauged supergravity\footnote{This holds provided that the reduction on $Y_7$ does not contain Betti vector multiplets in its spectrum. There are additional subtleties in the quantization condition of the fluxes of the Betti vectors that we do not treat here.}.

The computation of the on-shell action of these two distinct bulk solutions is obtained via standard techniques of holographic renormalization. The resulting on-shell action for the NUT configuration
coincides with the free energy of the corresponding theory on $S^3$. The renormalized on-shell action for Bolt solutions is instead of the form
 \be  \label{bolt_intro}
 I_{Bolt\pm} =  \frac{\sqrt2 \pi N^{3/2} }{12} \sqrt{ \frac{\text{Vol}(S^7)}{ \, \text{Vol}(Y_7) }  } \left( 4(1-g) \mp p \right) \,. 
 \ee
with the additional constraint
\be
\pm p + 2(g-1) =0 \quad \MOD{I(Y_7)} \;,
\ee
where $I(Y_7)$ is the Fano index of the internal 7-manifold. In particular, for $g=0$ we retrieve the results of \cite{Martelli:2012sz}.

 We are able to show that, for this solution in minimal gauged supergravity, the on-shell action in the gravity side matches with the partition function of the corresponding field theory \eqref{Z_intro}
\be
- \log Z_{\cM_{g,p}} = I_{Bolt} \;,
\ee
as expected. In case of trivial fibration, $p=0$, our formulas \eqref{Z_intro}   and  \eqref{bolt_intro} find agreement with those of  \cite{Azzurli:2017kxo}. In this particular case the on-shell action of the Euclidean solution coincides with the entropy of supersymmetric 1/4 BPS black holes with constant scalars and higher genus horizon\footnote{See also the recent analysis of \cite{Halmagyi:2017hmw,Cabo-Bizet:2017xdr} for further computations regarding the equivalence between renormalized on-shell action and BPS black hole entropy}.

Along with the matching with the Bolt solutions, we study the relation between the $S^3$ partition function as computed by \cite{Jafferis:2011zi} and the result we obtain for the $\mathcal{M}_{g,p}$ partition function, in light of the result of \cite{Hosseini:2016tor}. In particular, we elaborate on how the interesting relation  between the extremal value of the twisted superpotential and the large $N$ partition function on $S^3$, discovered in \cite{Hosseini:2016tor}, fits in our framework by relating these both to the partition function on the lens space $S^3/\mathbb{Z}_2$.

The main text of the paper is organized as follows: in Section \ref{sec:field theory} we provide the details of the computation of the large $N$ partition function of a class of $\mathcal{N}=2$ $3d$ quiver gauge theories on $\mathcal{M}_{g,p}$, focusing on the example of the ABJM theory. In Section \ref{sec:sugra} we describe Euclidean minimal gauged supergravity solution whose boundary is $\mathcal{M}_{g,p}$, and examine their supersymmetry properties, along with their moduli space for regularity. Moreover, we compute the on-shell action via holographic renormalization. In Section \ref{sec:holo_checks} we show the matching between the renormalized on-shell action of the Bolt solutions and the partition function of the dual field theory for the ABJM theory.  In Section \ref{sec:gq}, we consider more general quiver gauge theories, including the $V^{5,2}$ theory, and describe the truncation to minimal supergravity for these theories, obtaining a generalization of the universal twist of \cite{Azzurli:2017kxo}.  In Section \ref{sec:s3comp}, we discuss the relation between the twisted superpotential and $S^3$ partition function observed by \cite{Hosseini:2016tor}, and relate these to the lens space partition function.  Finally in Section \ref{sec:disc}, we discuss some open issues and future directions.  Several appendices complete this paper, and they are devoted to the construction of the explicit Killing spinor for the supergravity solutions, to the description of their moduli space, and to the explicit details for the computation of the partition function $Z_{\mathcal{M}_{g,p}}$.


\section{$\cM_{g,p}$ partition function at large $N$}
\label{sec:field theory}

We start in this section by discussing the computation of the $\cM_{g,p}$ partition function for $3d$ $\cN=2$ field theories.  We first describe the supersymmetric background on $\cM_{g,p}$ and review the computation for general, finite $N$ theories, and then turn to the large $N$ computation for a class of $U(N)$ quiver gauge theories with M-theory duals.

\subsection{Supersymmetric background on $\cM_{g,p}$}
\label{sec:mgpbg}

Following \cite{Closset:2017zgf}, we consider manifolds which are $U(1)$ bundles over a Riemann surface, $\Sigma_g$, with $p \in \Z$ the Chern-degree of the bundle, which we take to be non-zero in this subsection.  On this space we take the following metric:\footnote{Here we have redefined $\cC \rightarrow  - \cC$ relative to the background in \cite{Closset:2017zgf} to facilitate comparison to the asymptotic supergravity solution in the next section.}

\be ds^2 = \beta^2 (d\psi - \cC(z,\bar{z}) )^2 + 2 g_{z \bar{z}} dz d\bar{z}  \;,\ee
where $z,\bar{z}$ are local coordinates on $\Sigma_g$, with $g_{z \bar{z}}$ the metric on $\Sigma_g$, $\psi \sim \psi+2 \pi$ is a coordinate along the $U(1)$ fiber, which has length $2 \pi \beta$, and $\cC$ is a locally defined $1$-form on $\Sigma_g$, satisfying:

\be \label{cq} \frac{1}{2\pi} \int_{\Sigma_g} d\cC = p  \;. \ee
We may define a Killing vector $K=\frac{1}{\beta} \partial_\psi$ pointing along the $U(1)$ fiber, or equivalently, a $1$-form:

\be \label{etadef} \eta = K_\mu dx^\mu = \beta(d\psi - \cC(z,\bar{z}) ) \;.\ee

To preserve supersymmetry on this space, we must turn on additional fields in the background supergravity multiplet \cite{Klare:2012gn,Closset:2012ru}.  These include the R-symmetry gauge field, $A_\mu^R$, a scalar, $H$, and a vector, $V_\mu$.  These lead to the Killing spinor equation:

\be \label{ks1} (\nabla_\mu - i A_\mu^R) \zeta = -\frac{1}{2} H \gamma_\mu \zeta + \frac{i}{2} V_\mu \zeta - \frac{1}{2} \epsilon_{\mu \nu \rho} V^\nu \gamma^\rho \zeta  \;.\ee
We find a solution on the above geometry, in the local coordinates above, when we take:
\be H = i (p \beta +  \kappa), \;\;\; V_\mu = -(2p \beta + \kappa) \eta_\mu, \;\;\; A^R_\mu = \frac{1}{8} \epsilon_\mu^{\nu \rho} \eta_\rho \partial_\nu (\log \det g)+ p \beta \eta_\mu +\partial_\mu s  \;.\ee
In \cite{Closset:2017zgf} the scalar parameter $\kappa$ was set to zero, but for comparison to the supergravity background below we will take $\kappa=-2 p \beta$; these choices will lead to the same Killing spinor $\zeta$.  Here the last term in $A^R_\mu$ corresponds to a contribution from a flat connection, and we will describe this in more detail below.

Although in principle we may take an arbitrary smooth metric on $\Sigma_g$, and an arbitrary connection $\cC$ subject to $\eqref{cq}$, for concreteness, and to compare to the bulk supergravity solution, we will consider constant curvature metrics and connections:
\be \label{sgds} ds^2 = \beta^2 (d \psi - p \; a(\theta,\phi))^2 +  \left\{ \begin{array}{ccc} \frac{1}{4}(d\theta^2 + \sin^2\theta d\phi^2) \,\,\, & \text{for} \,\,\,& g=0, \\ \frac{1}{2}(d\theta^2 + d\phi^2) \,\,\, &\text{for} \,\,\,&  g=1  \\  \frac{1}{4(g-1)}(d\theta^2 + \sinh^2 \theta d\phi^2) \,\,\, &\text{for} \,\,\, &  g>1 \end{array} \right.
\ee
where the connection $a(\theta,\phi)$ is given by:
\be
a(\theta,\phi) =  \left\{ \begin{array}{ccc}  - \frac{1}{2} \cos \theta d\phi \,\,\, & \text{for} \,\,\,& g=0, \\\theta d \phi \,\,\, &\text{for} \,\,\,&  g=1  \\ \frac{1}{2(g-1)}\cosh \theta d\phi \,\,\, &\text{for} \,\,\, &  g>1 \end{array} \right.
\ee
In all cases, $\phi \sim \phi+2 \pi$ is an angular coordinate.  For $g=0$, $\theta \in [0,\pi]$ and this is the usual round metric on $S^2$.  For $g=1$, we identify $\theta \sim \theta+1$, obtaining the flat, rectangular metric on the torus.  For $g>1$, the second and third terms in \eqref{sgds} describe the metric on the hyperbolic plane, $H^2$, and we form $\Sigma_g$ by taking an appropriate quotient of the hyperbolic plane by a Fuchsian subgroup \cite{1986PhR...143..109B}, with a fundamental domain $D_g$.  In all cases we have normalized the metric on $\Sigma_g$ so that $\text{vol}(\Sigma_g)=\pi$.  The connection $a$ has curvature $da$ proportional to the volume form on $\Sigma_g$, and satisfies:

\be \label{adef} \frac{1}{2 \pi} \int_{\Sigma_g} da = 1 \;. \ee

In this constant curvature background, the background supergravity fields are $H = - i p \beta$ and $V_\mu = 0$, and the R-symmetry gauge field is:

\be \label{rgf} A^R =(-p \beta - \frac{g-1}{p \beta} ) \eta + (g-1) \gamma \;. \ee
Here $\gamma$ is a flat connection with Chern class generating $\Z_p \subset H_2(\cM_{g,p})$, so that this gauge field has torsion flux $g-1\MOD{p}$.  Then the Killing spinor equation, \eqref{ks1}, becomes:

\be  \label{ks2} (\nabla_\mu - i A_\mu^R) \zeta = \frac{i p \beta}{2} \gamma_\mu \zeta  \;.\ee

\subsection*{Background gauge fields}

In addition to the fields in the background supergravity multiplet, we may include background gauge multiplets coupled to the global symmetries of the theory.  If $i=1,\cdots,r_H$ runs over a basis of the Cartan of the flavor symmetry group, $H$, then we may turn on background gauge multiplets $\cV_i$ in configurations labeled by: 

\be m_i = i \beta (\sigma_i + i {\alpha}_i) \in \C, \;\;\;\; s_i \in \Z  \;,\ee
where $\sigma_i$ is the real scalar in the background gauge multiplet, and the gauge field $A_i$ is given by:
\be \label{pba} A_i = \alpha_i \eta + \pi^*(s_i a)  \;,\ee
where $\eta$ and $a$ are as in \eqref{etadef} and \eqref{adef}, and $\pi^*$ is the pullback along the projection map $\pi:\cM_{g,p} \rightarrow \Sigma_g$.  Explicitly, we can write, for $p \neq 0$:
\be \label{bgfgf} A_i = ( \alpha_i - \frac{s_i}{\beta p})\eta +s_i \gamma  \;,\ee
with $\gamma$ as in \eqref{rgf}, so that the gauge field has torsion flux $s_i  \MOD{p}$.

The $\cM_{g,p}$ partition function we compute below will be functions of the parameters $m_i$ and $s_i$, depending holomorphically on the former.  Note that shifting:
\be \label{flxshift}  m_i \rightarrow m_i+1, \;\;\; s_i \rightarrow s_i+p \ee
does not change the connection $A_i$ (modulo gauge transformations), and we will see below this is an invariance of the partition function.

\subsection{Computation of $\cM_{g,p}$ partition function}
\label{sec:mgpcomp}

In this subsection we review the computation of the $\cM_{g,p}$ partition function for a general, finite $N$, $3d$ gauge theory, in preparation for computing the partition function at large $N$ in the next subsection.  We refer to \cite{Closset:2017zgf} and references therein for more details.

There are two equivalent methods to compute the $\cM_{g,p}$ partition function.  First, one may couple the UV action of the $3d$ gauge theory to the supersymmetric background on $\cM_{g,p}$ discussed above, along the lines of \cite{Closset:2012ru}, and use localization to reduce the path integral to a finite dimensional integral.  As shown in \cite{Closset:2017zgf}, one arrives at the following integral formula for a theory with gauge group $G$ with rank $r_G$ and Weyl group $W$:

\bea \label{mgpint} &Z_{\cM_{g,p}}(m_i,s_i) \\&= \frac{1}{|W|} \sum_{{\frak m}_a \in \Z^{r_G}} \int_{C_{JK}} d^{r_G} u \cF({u}_a,m_i)^p \cH({u}_a,m_i)^{g-1} {\Pi_i}({u}_a,m_i)^{s_i}  {\Pi_a}(u_a,m_i)^{\frak m_a} {\text H}(u_a,m_i)  \;.\eea
Here $u_a,{\frak m}_a$, $a=1,\cdots,r_G$ are holonomies and fluxes, respectively, for the gauge field, and similarly for $m_i,s_i$, $i=1,\cdots,r_H$ for background gauge fields coupled to the flavor symmetry group, $H$.  The functions $\cF$, $\cH$, $\Pi$, and ${\text H}$ depend on the field content of the theory, and are defined in \eqref{opdef} below.  The integral is taken over a compact contour $C_{JK}$,  the so-called ``Jeffrey-Kirwan'' contour \cite{JK1995,Benini:2013xpa}; we refer to \cite{Closset:2017zgf} for the precise definition.

In the case $p \neq 0$, this contour integral can be deformed into one over a non-compact ``Coulomb'' contour, $C_{Coul}$, and one obtains the equivalent formula:

\bea \label{mgpcint} &Z_{\cM_{g,p \neq 0}}(m_i,s_i) \\&= \frac{1}{|W|} \sum_{{\frak m}_a \in \Z_p^{r_G}} \int_{C_{Coul}} d^{r_G} u \cF({u}_a,m_i)^p \cH({u}_a,m_i)^{g-1} {\Pi_i}({u}_a,m_i)^{s_i}  {\Pi_a}(u_a,m_i)^{\frak m_a} {\text H}(u_a,m_i)  \;,\eea
where the fluxes ${\frak m}_a$ now takes values in $\Z_p$ rather than $\Z$.  Roughly speaking\footnote{More precisely, this statement is true only when suitable conditions on the R-charges of the chiral multiplets are satisfied; more generally the contour may be deformed to pass around certain poles coming from the contributions of the chirals.  See \cite{Closset:2017zgf} for more details.},  $C_{coul} = i \R^{r_G}$ is the imaginary slice in the complex $u$ plane, and upon making the identification $u \rightarrow i \sigma$, one recovers the usual integral formula for the $S^3$ partition function  \cite{Kapustin:2009kz,Jafferis:2010un,Hama:2010av} in the case $g=0,p=1$.  We return to the connection to previous computations of the $S^3$ partition function in Sections \ref{sec:holo_checks} and \ref{sec:s3comp} below.

The second method starts from the observation that the $\cM_{g,p}$ partition function is computed by a certain $2d$ topological quantum field theory (TQFT) on the base space, $\Sigma_g$, of the fiber bundle.  Specifically, this TQFT is the ``$A$-twist'' of the $2d$ $\cN=(2,2)$ theory obtained by compactifying the $3d$ gauge theory on a circle and studying the low energy effective action.  Then, on general grounds, we expect the partition function to be given by a sum over the supersymmetric vacua of the $3d$ theory  on a circle.  Explicitly, one finds:

\be \label{mgpsv} Z_{\cM_{g,p}}(m_i,s_i) = \sum_{\hat{u}_a \in \cS_{BE}} \cF(\hat{u}_a,m_i)^p \cH(\hat{u}_a,m_i)^{g-1} {\Pi_i}(\hat{u}_a,m_i)^{s_i}\;, \ee
involving the same functions as in  \eqref{mgpint}.  We will describe this formula in more detail below.

These two methods can be shown to be equivalent for an arbitrary $3d$ gauge theory \cite{Closset:2017zgf}.  For the purpose of taking the large $N$ limit, we will utilize the sum-over-vacua formula, \eqref{mgpsv}, for the remainder of this section.  However, despite these formulas being equivalent for finite $N$, there are some subtleties in relating the large $N$ limit obtained by the two methods.  We will return to this issue in Section \ref{sec:holo_checks}.

Let us now describe the formula \eqref{mgpsv} in more detail, starting by reviewing the compactification on $S^1$ and the vacuum structure of the resulting system.

\subsection*{Twisted superpotential and Bethe vacua on $\R^2 \times S^1_\beta$}

Given a $3d$ $\cN=2$ gauge theory, we may place it on $\R^2 \times S^1_\beta$, obtaining at low energies an effective $2d$ $\cN=(2,2)$ description.  Then the vacuum structure of the theory is determined by the ``Bethe equations'' \cite{Nekrasov:2009uh}:

\be \label{BE}  \exp\bigg( 2 \pi i \frac{\partial \cW(u_a,m_i)}{\partial u_a} \bigg) = 1 , \;\;\;\;\; a=1,\cdots,r_G \;,\ee
where $\cW(u_a,m_i)$ is the effective twisted superpotential of this effective $2d$ $\cN=(2,2)$ system, defined in 
\eqref{wfn} below.  Here we define:

\be u_a = i \beta ( \sigma_a +  i({A_t})_a) , \;\;\; a=1,\cdots,r_G , \;\;\;\;\;\;\; m_i = i \beta ( \sigma^{BG}_i +  i({A_t}^{BG})_i) , \;\;\; i=1,\cdots,r_H ,\ee
where $\sigma$ is the real scalar in the dynamical gauge multiplet, and $A_t$ the component of the gauge field along $S^1_\beta$, and we expand them in a basis of the Cartan subalgebra of $G$, and similarly for the parameters, $m_i$, $i=1,\cdots,r_H$, for background gauge multiplets coupled to the flavor symmetry, $H$.  Note that $u_a \sim u_a+1$ and $m_i \sim m_i + 1$ due to large gauge transformations around $S^1_\beta$.

The twisted superpotential, $\cW(u_a,m_i)$, depends on the matter content and UV Lagrangian of the $3d$ theory.  We consider a general theory with gauge group $G$ of rank $r_G$, and chiral multiplets in some representation of $G$.  We expand the chiral multiplets in weights of $G$, such that they have charges $Q_\alpha^a$, $a=1,\cdots,r_G$, $\alpha=1,\cdots,M$, where $M$ is the dimension of the space of chiral multiplets.  The chiral multiplets also have charges, $S_\alpha^i$, $i=1,\cdots,r_H$,under the  global symmetry group, $H$, which may be restricted by superpotential terms in the Lagrangian.  Finally, we allow Chern-Simons terms $k^{ab}$, $k^{ij}$, and $k^{ai}$, for the gauge, flavor, and mixed CS terms, respectively.   Then the twisted superpotential is given by the following function of $u_a$ and $m_i$:

\bea  \label{wfn} \cW(u_a,&m_i) = \sum_{\alpha=1}^M \cW_\chi(Q_\alpha^a u_a + S_\alpha^i m_i)  \\
& +\frac{1}{2}\sum_{a,b=1}^{r_G} k^{ab} u_a (u_b + \delta_{ab}) + \frac{1}{2}\sum_{i,j=1}^{r_H} k^{ij} m_i (m_j + \delta_{ij}) + \sum_{a=1}^{r_G} \sum_{i=1}^{r_H} k^{ai} u_a m_i \;. \eea
Here $\cW_\chi(u)$ is the contribution of a chiral multiplet, regulated with a level $-\frac{1}{2}$ CS term to ensure gauge invariance, given by:

\be \cW_\chi(u) = \frac{1}{(2 \pi i)^2} \dilog(e^{2 \pi i u})\;. \ee
With this definition of the chiral multiplet contribution, the bare CS levels, $k^{ab}$, $k^{ij}$, and $k^{ai}$, must all be integers to ensure gauge invariance.  

We note that the twisted superpotential in general has branch cuts, and is only defined modulo shifts of the form:

\be \label{branches} \cW \rightarrow \cW + n^a u_a + n^i m_i + n \;, \ee
for $n^a,n^i,n \in \Z$.  However, one can check that \eqref{BE} is invariant under these shifts, and is a polynomial equation in the variables $x_a=e^{2 \pi i u_a}$ and $\mu_i = e^{2 \pi i m_i}$.  

For non-abelian gauge theories, one must discard solutions which are not acted on freely by the Weyl group, $W$, as supersymmetry is broken at these putative vacua, and we consider the remaining solutions up to Weyl symmetry.  Then we define the set of Bethe vacua as:

\be \label{sbe} \cS_{BE} = \{ \hat{u}_a \; \big| \;  \exp\bigg( 2 \pi i \frac{\partial \cW}{\partial u_a} \bigg) = 1 , \;\;\;\;\; a=1,\cdots,r_G, \;\;\; w \cdot \hat{u}_a \neq \hat{u}_a, \forall w \in W\}/W \;.\ee

\subsection*{Ingredients in the $\cM_{g,p}$ partition function}

With this background, let us now return to the computation of the $\cM_{g,p}$ partition function.  The coupling of this theory to the curved background of $\cM_{g,p}$ depends on the choice of a $U(1)_R$ symmetry, which is used to perform a partial topological twist along the $\Sigma_g$ directions.  Since we will introduce a non-trivial flux for this R-symmetry, we must pick the R-charges, $r_\alpha$, of the chiral multiplets to be integers, so that they live in a well-defined vector bundles over $\cM_{g,p}$.  Given such a choice of R-symmetry, we define the ``effective dilaton,'' $\Omega(u_a,m_i)$:
$$ \Omega(u_a,m_i) = \frac{1}{2 \pi i} \sum_{\alpha=1}^M (r_\alpha-1) \log(1-e^{2 \pi i (Q_\alpha^a u_a + S_\alpha^i m_i) }) + \sum_{a=1}^{r_G} k^R_a u_a + \sum_{i=1}^{r_H} k^R_i m_i + \frac{1}{2} k^{RR} $$

\be + \frac{1}{2 \pi i} \sum_{\alpha \in Ad(G)}\log(1-e^{2 \pi i \alpha(u)}) \;,\ee
where $k^R_a,k^R_i,k^{RR} \in \Z$ correspond to contact terms involving the R-symmetry.  Here the second line is the contributions from the $W$-bosons of the gauge group, and the sum is over the weights of the adjoint representation of $G$.  Note the $W$-bosons do not contribute to $\cW$.

Then, as argued in \cite{Closset:2017zgf}, the $\cM_{g,p}$ partition function is given by the following sum over Bethe vacua:

\be \label{wforma} Z_{\cM_{g,p}}(m_i,s_i) = \sum_{\hat{u}_a \in \cS_{BE}} \cF(\hat{u}_a,m_i)^p \cH(\hat{u}_a,m_i)^{g-1} {\Pi_i}(\hat{u}_a,m_i)^{s_i} \;,\ee
where the ``fibering operator,'' $\cF$, ``handle-gluing operator,'' $\cH$, and ``flux operators,'' $\Pi_i$, are defined in terms of the twisted superpotential, $\cW$, and effective dilaton, $\Omega$, defined above:

\be \label{opdef} \cF(u_a,m_i) = \exp \bigg( 2 \pi i \bigg( \cW(u_a,m_i) - \sum_a u_a \frac{\partial \cW}{\partial u_a} - \sum_a m_i \frac{\partial \cW}{\partial m_i} \bigg) \bigg) \ee

$$ \cH(u_a,m_i) = e^{2 \pi i \Omega(u_a,m_i)}  {\text H} , \;\;\;\; {\text H} = \det_{a,b} \frac{\partial^2 \cW}{\partial u_a \partial u_b}, \;\;\;\;\;\; \Pi_i(u_a,m_i) = \exp \bigg( 2 \pi i \frac{\partial \cW}{\partial m_i} \bigg) \;.$$
Such a formula arises due the topological invariance along $\Sigma_g$, which implies that the operations of gluing a handle to $\Sigma_g$, adding a unit of flux for the $S^1$ fibration, or adding a unit of flavor symmetry flux, are all implemented by {\it local} operators, $\cH$, $\cF$, and $\Pi_i$, respectively, giving rise to the simple formula \eqref{wforma}.

To take a simple example, the $\cM_{g,p}$ partition function of a single chiral multiplet is:
\be \label{zchi} Z_{\chi,\cM_{g,p}}(m,s,r)= \cF_\chi(m)^p \Pi_\chi(m)^{s + (r-1)(g-1)} \;,\ee
where $m,s$, and $r$ are the its mass, flavor symmetry flux, and R-charge, respectively, and:
\be \cF_\chi(m) = \exp\bigg(\frac{1}{2 \pi i} \dilog(e^{2 \pi i m}) + m \log(1-e^{2 \pi i m})\bigg) ,  \;\;\; \Pi_\chi(m) = \frac{1}{1-e^{2 \pi i m}} \;.\ee
Note this depends only on the combination:
\be \label{elldef} {\ell} \equiv s + (r-1) (g-1) \;.\ee
In other words, a shift of the R-charge, $r \rightarrow r+c$, is equivalent to a shift of the flavor symmetry flux, $s \rightarrow s+c(g-1)$, and reflects a mixing of the R-symmetry with this flavor symmetry.  If we take:
\be \label{fluxshift} (m,\ell) \rightarrow (m+1,\ell + p)\;, \ee
and use the difference equation:
\be \label{fde1} \cF_\chi(m+1) =\cF_\chi(m) \Pi_\chi(m)^{-1} \;,\ee
we see the partition function, \eqref{zchi}, is invariant.  This is consistent with the invariance of the background gauge field, \eqref{bgfgf}, under this shift of parameters, and reflects the fact that, for $p \neq 0$, the fluxes are torsion, and take values in $\Z_p$.   

For a general gauge theory, we may explicitly write the summand in \eqref{wforma} as:\footnote{Here for simplicity we work in a basis of the flavor symmetry group where $m_i$ corresponds to the mass of the $i$th chiral multiplet, and we take the flavor and R-symmetry CS terms to vanish.  We may also treat the contribution of the vector multiplets as that of an $R$-charge $2$ chiral multiplet in the adjoint representation of $G$.}
\bea \label{mig} \cF({u}_a,m_i)^p &\cH({u}_a,m_i)^{g-1} {\Pi_i}({u}_a,m_i)^{s_i}  \\
&= e^{\pi i k^{ab} p u_a u_b} {\text H}^{g-1}  \prod_i \cF_\chi(Q_i^a u_a + m_i)^p \Pi_\chi(Q_i^a u_a + m_i)^{s_i + (r_i-1)(g-1)} \;.\eea

\subsection*{On-shell twisted superpotential}

We may conveniently construct the terms in the sum above using the ``on-shell'' twisted superpotential and effective dilaton, defined by:
\be \cW^I(m_i) = \cW(\hat{u}^I_a,m_i), \;\;\;\ \Omega^I(m_I) = \Omega(\hat{u}_a^I,m_i) + \frac{1}{2 \pi i} \log \det \frac{\partial^2 \cW}{\partial u_a \partial u_b}\bigg|_{u_a \rightarrow 
\hat{u}_a^I} \;, \ee
where the index $I$ runs over $\cS_{BE}$.  Here there is a branch cut ambiguity in defining $\cW^I$, but this is partially fixed by imposing:

\be \frac{\partial \cW}{\partial u_a}(\hat{u}_a) = 0 \;, \ee
which is a stronger condition than \eqref{BE}, and fixes the freedom to shift $\cW$ by $n^a u_a$.  Then one has:

\be \label{ossp} \cF^I(m_i) \equiv \cF(\hat{u}^I_a,m_i) = \exp\bigg( 2 \pi i \bigg( \cW^I - \sum_i m_i \frac{\partial \cW^I}{\partial m_i} \bigg) \bigg),\ee
$$ \Pi_i^I(m_i) \equiv  \Pi_i(\hat{u}^I_a,m_i) = \exp\bigg( 2 \pi i \frac{\partial \cW^I}{\partial m_i} \bigg), \;\;\;\;\; \cH^I(m_i) = \cH(\hat{u}^I_a,m_i) = \exp \big(2 \pi i \Omega^I \big) \;. $$
One can check that the remaining branch cut ambiguities in $\cW^I$ and $\Omega^I$ drop out of these expressions, and they are well-defined.  Then we may construct the partition function as:
\be \label{wformb} Z_{\cM_{g,p}}(m_i,s_i) = \sum_{I \in \cS_{BE}} {\cF^I(m_i)}^p {\cH^I(m_i)}^{g-1} {\Pi^I_i(m_i)}^{s_i} \; .\ee

\subsection{Large $N$ computation}
\label{sec:largensol}

We will be interested in computing this partition function for a large $N$ gauge theory.  In the case $p=0$, this problem was studied in \cite{Benini:2015eyy}. There they found that, although the number of Bethe vacua, $|\cS_{BE}|$, grows with $N$, in many cases there is a single vacuum, with index $I_{dom}$, whose contribution is dominant compared to all other terms in \eqref{wformb}.\footnote{More precisely, there need not be a single, strictly dominant vacuum, as other vacua which contribute at the same order will introduce extra logarithmic corrections to $\log Z$, which will be suppressed relative to the leading $N^{3/2}$ behavior we find below.}  When this occurs, we expect that \eqref{wformb} may be approximated as:

$$ Z_{\cM_{g,p}} \approx {\cF^{I_{dom}}}^p  {\Pi_i^{I_{dom}}}^{s_i} {\cH^{I_{dom}}}^{g-1} $$

\be  \label{mgpdom}  \Rightarrow \frac{1}{2 \pi i} \log Z_{\cM_{g,p}} \approx p \big( \cW^{I_{dom}}  - \sum_i m_i \partial_i \cW^{I_{dom}} \big)+ \sum_i s_i \partial_i \cW^{I_{dom}} + (g-1) \Omega^{I_{dom}} \;.\ee
Note, in particular, that the partition function has a very simple dependence on the geometric parameters, $g$ and $p$, and the fluxes, $s_i$.  In cases where the theory has a holographic dual, this suggests the holographic free energy has a similar simple dependence on these parameters, which is rather non-trivial.  Below we will verify this relation holds quite generally.

\subsubsection{$U(N)$ quiver gauge theories}
\label{sec:qd}

In this section we will focus on the ABJM model \cite{Aharony:2008ug}, which we describe in more detail below.  This is a special case of a more general class of $U(N)$ quiver gauge theories.  The ingredients in the computation of the twisted superpotential and $\cM_{g,p}$ partition function for these quivers is very similar, so we describe these general ingredients in the next few subsections, returning to a more detailed analysis of these theories in Section \ref{sec:gq}.

Specifically, the class of theories we will discuss, following \cite{Jafferis:2011zi,Hosseini:2016tor}, consists of $\cN=2$ quiver gauge theories with several $U(N)$ gauge factors, labeled by an index $\alpha=1,\cdots,n$.  We allow bifundamental chiral multiplets connecting two  gauge groups, (anti-)fundamental chiral multiplets in a single gauge group, and Chern-Simons levels,  $k_\alpha$, for the $\alpha$th gauge group.  However, we impose the following restrictions:

\begin{itemize}
\item The sum of all Chern-Simons levels is zero:
\be \sum_{\alpha=1}^n k_\alpha = 0 \;.\ee
\item For each gauge node, $\alpha$, there is a superpotential constraint which imposes that, for all bifundamental chiral mutiplets with a leg in this node:
\be \label{bfconst} \sum_{I \in \alpha} Q_I = 0 , \;\;\;\;\;\; \sum_{I \in \alpha} (r_I-1) + 2 = 0 \;, \ee
where $Q_I$ is the charge of the $I$th such bifundamental chiral multiplet under any flavor symmetry, and $r_I$ is its R-charge.  Here adjoint chirals are counted twice in the sum.
\item The number of bifundamental chiral multiplets entering a node is the same as the number exiting the node.
\item The total number of fundamental and anti-fundamental chiral multiplets in the quiver are equal.
\end{itemize}
These restrictions are to ensure the theory has a well behaved M-theory dual description at large $N$, with a characteristic $N^{3/2}$ scaling of the number of degrees of freedom.

For such quivers, following \cite{Benini:2015eyy}, we will take the following large $N$ ansatz for the eigenvalues $u^\alpha_a$:

\be \label{lnansatz} u^\alpha_a = v^\alpha_a + i N^{1/2} t_a , \;\;\;\; a=1,\cdots,N ,\;\; \alpha=1,\cdots,n \;. \ee
In the large $N$ limit the eigenvalues become dense, and we may parameterize them by the continuous variable $t$, defining:
\be \rho(t) = \frac{1}{N} \sum_{a=1}^N \delta(t-t_a) \;, \ee
and corresponding functions, $v^\alpha(t)$.  

Our strategy in the rest of this section is as follows.  First, we compute the twisted superpotential at large $N$, using the above ansatz, and find the eigenvalue distribution which extremizes it.  Then, as in \eqref{mgpdom}, we may assume that the dominant contribution to the Bethe sum computing the $\cM_{g,p}$ partition function is determined by this extremal distribution.  Thus, we evaluate the summand in \eqref{mgpdom} at this extremal distribution to compute the leading behavior of the $\cM_{g,p}$ partition function.  

\subsubsection{The twisted superpotential at large $N$}

We start by reviewing the computation of the twisted superpotential at large $N$, as first computed in \cite{Benini:2015eyy} for the ABJM theory, and studied for more general quivers of the above type in \cite{Hosseini:2016tor}.\footnote{Let us state the relation between the notations used here and those used in \cite{Benini:2015eyy,Hosseini:2016tor}.  We have:
\be u_a^{them} = 2 \pi u_a^{us} , \;\;\;\; \Delta_i^{them}=2 \pi m_i^{us}\;\;\;\; \text{and}\;\;\;\; \cV_{them} = -(2 \pi)^2 \cW_{us} \;.\ee
These changes propagate into the large $N$ ansatz, \eg, $t_{them} = 2\pi t_{us}$, $\rho_{them}=\frac{1}{2 \pi }\rho_{us}$, etc..}

For a given set of eigenvalues, $u^\alpha_a$, approximated by the distributions $\rho(t)$ and $v^{\alpha}(t)$ above, we may compute the value of the effective twisted superpotential at these eigenvalues as a functional:

\be \cW[\rho(t),v^\alpha(t),m_i]. \ee

Let us briefly summarize the various ingredients in the functional $\cW[\rho,v^\alpha]$, as computed in \cite{Benini:2015eyy,Hosseini:2016tor}.  We review the derivation of these ingredients in Appendix \ref{sec:pfdetails}.  First, the CS terms, which satisfy $\sum_\alpha k_\alpha=0$, contribute:

\be \cW_{CS} = i N^{3/2} \int dt \rho(t) \sum_\alpha k_\alpha t v^\alpha \;.\ee

Next, a bifundamental chiral multiplet connecting the $\alpha$th and $\beta$th groups contributes:
\be \cW_{bif}  = i N^{3/2} \int dt \rho(t)^2 g([ \delta v + m]) \;, \ee
where we defined $\delta v = v^\alpha - v^\beta$, $m$ is the mass of the bifundamental, and we have introduced the following notation for the ``fractional part,'' $[u]$, of a complex number $u$:
\be [u]=u-n, \;\; n \in \Z, \;\; \text{such that} \; 0<\text{Re}([u]) \leq 1 \;. \ee
The function $g(u)$ is given by:
\be g(u) = - \frac{1}{12} u (u-1) (2u-1) \;.\ee
Here we have imposed the constraint \eqref{bfconst}, which implies:
\be \sum_{I \in \alpha} m_I  = 0 , \;\;\; \alpha=1,\cdots,n\;.\ee 
We also impose that the total number of incoming and outgoing edges at each node in the quiver are equal.  Note then that the $(v^\alpha)^3$ terms contributed by $\cW_{bif}$ will cancel, so that the functional is in general quadratic in the $v^\alpha$.

In addition, there are contributions to $\cW$ from a bifundamental chiral that are subleading in $N$, but whose derivatives with respect to the $v^\alpha$ get large near special points in parameter space, and so they affect the extremization of $\cW$.  Specifically, these contributions become important when $\delta v_I +m_I  = \hat{n}_I\in \Z$ for some bifundamental chiral multiplet, with index $I$.  Then if we write:
\be \label{Yform} \delta v_I(t) +m_I  = \hat{n}_I + C e^{-2 \pi N^{1/2}Y_I(t)} \;,\ee
for some positive function $Y_I(t)$, one finds an additional ``tail contribution:''
\be \frac{\delta \cW_{tail}}{\delta (\delta v_I)} = \cdots - i N^{3/2} \int \rho(t) Y_I(t) \;. \ee
Finally, an (anti-)fundamental chiral multiplet contributes:

\be \cW_{fun} = i N^{3/2} \int dt \rho(t) \frac{1}{2} |t| ([\pm v + m] - \frac{1}{2})  \;,\ee
with the $+$ $(-)$ sign for a fundamental (anti-fundamental) chiral.

\subsubsection{Extremal value and the ABJM theory}
\label{sec:extval}

As described above, we will need to find the eigenvalue distribution which extremizes $\cW$.  To do this we vary the functional $\cW[\rho,v^\alpha]$ with respect to $\rho(t)$ and the $v^\alpha(t)$.  We also include a Lagrange multiplier term, $i N^{3/2} \mu \big(  \big(\int dt \rho\big) -1 \big)$, to impose correct normalization of $\rho$.  The solution is in general defined piecewise, bounded by points where $\delta v_I+ m_I$ becomes an integer, after which $\delta v_I$ becomes locked to this value to leading order, varying at subleading order as in (\ref{Yform}). 

Let us consider as our main example the ABJM theory \cite{Aharony:2008ug}.  This has $U(N)_{k} \times U(N)_{-k}$ gauge group, with two bifundamentals in the $(N,\bar{N})$, with masses $m_{1,2}$, and two in the $(\bar{N},N)$ representation, with masses $m_{3,4}$.  We assume $k>0$; the case with $k<0$ can be obtained by exchanging the two gauge groups.  This theory includes a quartic superpotential which imposes the following constraints on the masses:
\be \label{abjmsp} \sum_{i=1}^4 m_i = 0 \;. \ee
The twisted superpotential is periodic under $m_i \rightarrow m_i+1$ (up to branch jumps), and so depends only on the fractional part of the masses, $[m_i]$.  The functional we obtain for ABJM turns out to depend only on $\delta v = v_1 -v_2$.  We look for solutions with:\footnote{The only other essentially different possibility is to have a solution with, \eg, $-[m_1]<\delta v<-[m_2]$, but one may check that there are no solutions of this form.}
\be -[m_{1,2}] < \delta v < [m_{3,4}] \;. \ee
Then we simply have $[ \pm \delta v + m_i]= \pm \delta v+[m_i]$ (where here and below, for ``$\pm$'' we take $+$ for $i=1,2$ and $-$ for $i=3,4$), and then the functional becomes:

\be \label{wabjm} \cW_{ABJM} = i N^{3/2} \int dt \bigg( k \rho \delta v t + \rho^2 \underset{\begin{subarray}{c}
 i=1 \\
-:i=1,2 \\ +:i=3,4 
 \end{subarray}}{\sum^4} g( \pm \delta v + [m_i]) \bigg) \ee
$$ = i N^{3/2} \int dt \bigg( k \rho \delta v t + \rho^2 \bigg( (1-\frac{1}{2} \sum_i [m_i] ) {\delta v}^2 - \frac{1}{2}\sum_i(\pm [m_i]([m_i]-1) ) \delta v + \sum_i g([m_i]) \bigg) \bigg) \;.$$

The extremal distribution was first derived in \cite{Benini:2015eyy}.  Note that \eqref{abjmsp} imposes that $\sum_i [m_i]=1,2$ or $3$.  Then one finds the following solution when $\sum_i [m_i] = 1$ (here we also assume $[m_1]<[m_2]$ and $[m_3]<[m_4]$):

\be \label{abjmev} \rho(t) = \left\{ \begin{array}{cc} 
\displaystyle\frac{\mu + [m_3] k t}{([m_4]-[m_3])([m_1]+[m_3])([m_2]+[m_3])} & t_{--} < t < t_{-} \\
\displaystyle \frac{\mu - ([m_1] [m_2] - [m_3] [m_4]) k t }{([m_1]+[m_3])([m_1]+[m_4])([m_2]+[m_3])([m_2]+[m_4])} & t_{-} < t < t_{+} \\
\displaystyle  \frac{\mu-[m_1] k t }{([m_2]-[m_1])([m_1]+[m_3])([m_1]+[m_4])} & t_{+} < t < t_{++} \end{array} \right. \ee

\be \delta v(t) = \left\{ \begin{array}{cc} 
\;[m_3] + C' \exp\bigg(-2 \pi N^{1/2} \frac{-\mu-[m_4] k t}{[m_4]-[m_3]} \bigg) & t_{--} < t < t_{-} \\
-\frac{\mu ([m_1] [m_2] -[m_3] [m_4])+ k t \sum_{i<j<k} [m_i] [m_j] [m_k]}{([m_3] [m_4] - [m_1] [m_2])k t + \mu} & t_{-} < t < t_{+} \\
-[m_1] + C \exp \bigg(-2 \pi N^{1/2} \frac{-\mu + [m_2] k t}{[m_2] -[m_1]} \bigg) & t_{+} < t < t_{++} \end{array} \right. \ee
where (here we take $\mu>0$, so these are in ascending order):
\be t_{--}  = -\frac{\mu}{k [m_3]}, \;\;\; t_- = -\frac{\mu}{k [m_4]}, \;\;\; t_+ = \frac{\mu}{k [m_2]}, \;\;\; t_{++} = \frac{\mu}{k [m_1]} \;. \ee
Then one computes the extremal value of $\cW$ as:

\be \label{wextabjm} \cW^{ABJM}_{ext} = - \frac{2 i N^{3/2} }{3} \sqrt{2 k [m_1] [m_2] [m_3] [m_4]} \;.\ee
There is a similar solution when $\sum_i [m_i] =3$, related by $m_i \rightarrow 1-m_i$ and $\cW \rightarrow -\cW$.  However, for $\sum_i [m_i] =2$, we see the quadratic term in $\delta v$ vanishes, and we do not find a solution.

\subsubsection{$\cM_{g,p}$ partition function at large $N$}
\label{sec:mgpfunc}

Next we consider the functional computing the $\cM_{g,p}$ partition function.  Specifically, we compute the contribution to $\log Z_{\cM_{g,p}}$ from a Bethe vacuum which is, approximately at large $N$, given by a distribution of eigenvalues $u_a$ corresponding to the functions $\rho$ and $v^\alpha$, as in \eqref{lnansatz}.  Once we have found the dominant such eigenvalue distribution, as above, we may plug this in to this functional to compute the leading behavior of the partition function.  

Here we list the various ingredients, which are derived in Appendix \ref{sec:pfdetails}.  The Chern-Simons terms contribute:

\be \log Z_{\cM_{g,p}}^{CS} = -2 \pi N^{3/2} \int dt \rho(t) \sum_\alpha - p  k_\alpha t v^\alpha(t) \;. \ee
A bifundamental chiral multiplet with mass $m$, flavor flux $s$, and R-charge $r$ contributes:

\be  \log Z_{\cM_{g,p}}^{bif} = - 2 \pi N^{3/2} \int dt \rho(t)^2 G_\ell(\delta v + m) \;,\ee
where:
\be G_\ell(u) = 2 p g([u]) - (p [u] +n p - \ell) g'([u]) = p ( \frac{[u]^3}{6} - \frac{[u]}{12}) + (\ell-n p) ( -\frac{[u]^2}{2} + \frac{[u]}{2} - \frac{1}{12}) \;. \ee
Here $n=u -[u]$ is the integer part of $u$, and $\ell=s+(g-1)(r-1)$, as in \eqref{elldef}.  A vector multiplet (which, recall, does not contribute to $\cW$) contributes as above with $\delta v + m \rightarrow 0$ and $\ell \rightarrow(g-1)$, giving:

\be  \log Z_{\cM_{g,p}}^{vec} = - 2 \pi N^{3/2} \int dt \rho(t)^2  \frac{1-g}{12}  \;. \ee
Here we have imposed the constraints in \eqref{bfconst}.  Once again, since we impose the number of incoming and outgoing edges at each node are equal, the cubic terms in $\delta v$ will cancel, and this gives an expression quadratic in the $v^\alpha$. 

We also have contributions from the tail regions, where $\delta v_I + m_I \approx \hat{n}_I \in \Z$ for some $I$.  Here we find:
\be \label{tails} \log Z_{\cM_{g,p}}^{tails} =  -2 \pi N^{3/2}  \sum_{I | tail} \int_{\delta v(t)_I + m_I \approx \hat{n}_I} dt  (p \hat{n}_I - {\ell}_I - (g-1)) \rho(t) Y_I(t) \;,  \ee
where $\ell_I$ is as above, and the sum is over all such tail regions.  

Finally, for an (anti-)fundamental chiral multiplet, we have:
\be \log Z^{fun}_{\cM_{g,p}} =  -2 \pi N^{3/2} \int dt \rho(t) |t| \big( -\frac{1}{2} p (\pm v^\alpha + m) + \frac{1}{2} \ell \big) \;. \ee
We note that the expressions above are invariant under:
\be \label{mshift} (m,\ell) \rightarrow (m+1,\ell+p)\;, \ee
where we recall shifting $m \rightarrow m+1$ entails shifting the corresponding integer part, $n \rightarrow n+1$.   This reflects the fact that the fluxes $\ell$ are defined modulo $p$, as in \eqref{fluxshift}.

Let us now return to the ABJM example.  Then the partition function is a function of the masses, $m_i=[m_i]+n_i$, flavor fluxes, $s_i$, and R-charges, $r_i$, where the latter enter in the combination $\ell_i=s_i+(g-1)(r_i-1)$.  Due to the quartic superpotential, these satisfy the constraints:

\be \label{abjmcond} \sum_{i=1}^4 m_i = \sum_{i=1}^4 s_i = 0, \;\;\;  \sum_{i=1}^4 r_i =  2 \;. \ee
Then, one finds the functional computing the $\cM_{g,p}$ partition function is given by:

\bea \label{abjmfun} \log &Z_{\cM_{g,p}}^{ABJM} = -2 \pi N^{3/2} \times \\
&\int dt \bigg( -p \rho t \delta v + \rho^2 \bigg((g-1) {\delta v}^2  - \sum_i\bigg( \pm \big({\ell}_i (m_i -n_i - \frac{1}{2}) + p ( -\frac{{m_i}^2}{2} + \frac{1}{2} n_i(n_i+1) ) \big)\bigg)\delta v   \\
& -  \sum_i \bigg( \frac{1}{2}  {\ell}_i (m_i -n_i)(m_i-n_i-1) - \frac{1}{6} p ( {m_i}^3 - 3 m_i n_i (n_i+1) + n_i (n_i+1)(2 n_i+1) ) \bigg) \bigg) \;, \eea 
plus the contribution of the tails.  Now let us plug in the extremal solution in \eqref{abjmev}, which, recall, required $\sum_i [m_i]=1$.  Then we must impose:

\be \label{abjmncond} 0 = \sum_i m_i = \sum_i [m_i] + n_i \;\;\;\; \Rightarrow \sum_i n_i = -1 \;.\ee
Plugging in the eigenvalue distribution found above and evaluating the integral, one eventually obtains the following simple result:

\be \label{lnabjm} \log Z_{\cM_{g,p}}^{ABJM} = \frac{2\pi N^{3/2}}{3} \sqrt{2 k [m_1][m_2][m_3][m_4]} \bigg( -2 p + \sum_i \frac{-p n_i +s_i + (g-1) r_i}{[m_i]} \bigg) \;. \ee

\

Let us make a few comments about this formula.  First, in the case $p=0$, where $\cM_{g,p=0} \cong \Sigma_g \times S^1$, this reproduces the results of \cite{Benini:2015eyy,Benini:2016hjo}.\footnote{To compare to their results, one makes the identifications in footnote $8$
, as well as $n_i^{them}=-(s_i+(g-1) r_i)^{us}$.}  

Next, recall this formula only applies when $\sum_i [m_i]=1$; when $\sum_i [m_i]=3$, we find another solution related by $m_i \rightarrow 1 -m_i$, explicitly:

\be \label{lnabjm2} \log Z_{\cM_{g,p}}^{ABJM}\bigg|_{\sum_i [m_i]=3} = - \frac{2\pi N^{3/2}}{3} \sqrt{2 k (1-[m_1])(1-[m_2])(1-[m_3])(1-[m_4])} \ee
$$ \times \bigg(  -2p - \sum_i \frac{-p (n_i+1) +  s_i +(g-1) r_i}{1-[m_i]} \bigg) \;.$$
For $\sum_i [m_i]=2$, we do not find a solution.  We will return to this point in Section \ref{sec:holo_checks} below.

Also, note that the result \eqref{lnabjm} has the expected form \eqref{mgpdom}: 
\be  \label{mgpdom2}  \frac{1}{2 \pi i} \log Z_{\cM_{g,p}} = p \big( \cW^{I_{dom}}  - \sum_i m_i \partial_i \cW^{I_{dom}} \big)+ \sum_i s_i \partial_i \cW^{I_{dom}} + (g-1) \Omega^{I_{dom}}\;, \ee
where here $\cW^{I_{dom}}$ is given by $\cW^{ABJM}_{ext}$ in \eqref{wextabjm}, and:

\be \Omega^{I_{dom}} =  \sum_i r_i \partial_{[m_i]} \cW^{ABJM}_{ext} \;.\ee
We will see in Section \ref{sec:gq} that this relation holds also for more general quiver gauge theories.


\section{The supergravity dual}
\label{sec:sugra}

In this section our aim is to find supergravity solutions whose boundary is the manifold $\cM_{g,p}$, a circle bundle over a closed Riemann surface $\Sigma_g$, which can be locally uplifted in M-theory\footnote{For the solution to be globally uplifted to M-theory we need to impose further constraints on the magnetic flux for some particular classes of solutions. We will spell out this condition in Section \ref{11uplift}.}.   According to the AdS/CFT dictionary, we expect the on-shell action of these solutions, suitably renormalized, to match with the $\cM_{g,p}$ partition function of the dual field theory, as computed above.  We will return to this comparison in the next section. 

\subsection{Minimal $\cN=2$ gauged supergravity}

Our starting point is minimal $\cN =2$ four-dimensional gauged supergravity, whose bosonic action reads \cite{Freedman:1976aw,Fradkin:1976xz}
\be \label{act}
S = -\frac{1}{16 \pi G_4} \int d^4x \sqrt{g} \left( R +\frac{6}{l^2} -F_{\mu \nu}F^{\mu \nu} \right)\,,
\ee
where $G_4$ is the four-dimensional Newton's constant and $l $ is the AdS radius, related to the cosmological constant via $\Lambda = -3 /l^2$ . We work in Euclidean signature.

The gravitino supersymmetry variation is
\be
\delta_{\epsilon} \psi_{\mu} = \left( \nabla_{\mu} -i l^{-1} A_{\mu} + \frac{i}{4} F_{\rho \nu} \gamma^{\rho \nu} \gamma_{\mu}  \right) \epsilon \;,
\ee
with
\be
\nabla_{\mu} \epsilon = \left( \partial_{\mu}  -\frac14 \omega^{ab}_{\mu} \gamma_{ab} +\frac12 l^{-1} \gamma_{\mu} \right) \epsilon\,,
\ee
where $\epsilon$ is a Dirac spinor and  $\gamma_{\mu} $ are the generators of $\text{Cliff}(4,0)$ and so they satisfy $\{\Gamma_a, \Gamma_b \} = 2 g_{ab}$. We follow here closely the conventions of \cite{Martelli:2012sz}. The Einstein's equations coming from  \eqref{act} read
\be \label{rmunu}
R_{\mu \nu} -\frac12g_{\mu \nu} R = \frac{3}{l^2} g_{\mu \nu} +2 \left( {F_{\mu}}^{\sigma} F_{\sigma \nu} -\frac14 g_{\mu \nu} F_{\rho \sigma} F^{\rho \sigma} \right) \,,
\ee
and Maxwell's ones are
\be \label{max}
d \star F =0\,.
\ee
We will restrict our analysis to a set of solutions where $\Sigma_g$ has constant curvature, and to configurations with a real metric. Solutions to the system of equations of motion \eqref{rmunu}-\eqref{max} have been obtained in  \cite{Chamblin:1998pz} and they have the following form\footnote{In Appendix A of  \cite{Martelli:2012sz} it was shown that imposing $SU(2) \times U(1)$ symmetry (and a real metric) the configurations \eqref{met} -\eqref{afi} with $\kappa=1$ are actually the unique solutions, obtained by directly integrating the equations of motion.}
\be \label{met}
ds^2= \lambda(r)(d\tau +2s f(\theta, \phi))^2 + \frac{dr^2}{\lambda(r)}+ (r^2-s^2) \, d \Omega_{\kappa}^2 \,,
\ee
with
\be
\lambda(r) =\frac{ (r^2-s^2)^2 +(\kappa-4s^2) (r^2+s^2) -2Mr +P^2-Q^2}{r^2-s^2}\,.
\ee
and
\be \label{fff}
f(\theta, \phi) = \left\{ \begin{array}{ccc} \cos \theta d\phi \,\,\, & \text{for} \,\,\,& \kappa=1 \\ -  \theta d\phi \,\,\, &\text{for} \,\,\,&  \kappa=0  \\  - \cosh \theta d\phi \,\,\, &\text{for} \,\,\, & \kappa=-1 \end{array} \right.
\ee
In this case, $\kappa$ denotes the curvature of $\Sigma_g$: $\kappa=1$ for $S^2$, $\kappa=0$ for $\mathbb{R}^2$ and $\kappa = -1$ for $\mathbb{H}^2$. The 2d area element $d\Omega_{\kappa}^2$ reads
\be
d\Omega_{\kappa}^2 = \left\{ \begin{array}{ccc} d\theta^2 + \sin^2\theta d\phi^2 \,\,\, & \text{for} \,\,\,& \kappa=1 \\ d\theta^2 + d\phi^2 \,\,\, &\text{for} \,\,\,&  \kappa=0  \\  d\theta^2 + \sinh^2 \theta d\phi^2 \,\,\, &\text{for} \,\,\, & \kappa=-1 \end{array} \right.
\ee
The gauge field has this form
\be
A_t = \frac{-2 s Q r +P(r^2+s^2)}{r^2-s^2}\,,
\ee
\be
A_{\phi} = \left\{ \begin{array}{ccc}\cos \theta \frac{P(r^2+s^2)-2 s \, Q \, r}{r^2-s^2} \,\,\, & \text{for} \,\,\,& \kappa=1 \\ - \theta \frac{P(r^2+s^2)-2 s \, Q \, r}{r^2-s^2} \,\,\, &\text{for} \,\,\,&  \kappa=0  \\  -\cosh \theta \frac{P(r^2+s^2)-2 s \, Q \, r}{r^2-s^2} \,\,\, &\text{for} \,\,\, & \kappa=-1 \end{array} \right. \label{afi}
\ee
In these solutions, $M$ is the mass parameter and $r$ is the radial coordinate. $\tau$ parameterizes a circle fibered over a 2-dimensional constant curvature surface $\Sigma_g$ spanned by the coordinates $\theta$ and $\phi$. The fibration is due to the presence of the NUT parameter, which we denote by $s$ because of its relation with the squashing of the $U(1)$ fiber relative to the base\footnote{In previous literature (e.g. \cite{Hawking:1998ct,Chamblin:1998pz}) the NUT parameter has most often been  denoted by $n$.}.

Solutions of this kind for $\kappa=1$ were first discovered by Taub \cite{Taub:1950ez} and Newman, Unti and Tamburino \cite{Newman:1963yy}, hence the name Taub-NUT. Their structure and thermodynamics properties were later studied in \cite{Hawking:1998ct} and \cite{Chamblin:1998pz}. In the latter, (Lorentzian) solutions with NUT charge and planar and hyperbolic horizon were analyzed as well. 

In the asymptotic limit $r\rightarrow \infty$ the metric approaches
\be \label{boundary}
ds^2 = \frac{l^2}{r^2} dr^2 +r^2 \left(\frac{4s^2}{l^2} (d \psi + f(\theta, \phi))^2 + d\Omega_{\kappa}^2 \right)\,,
\ee
where we have defined $\psi = \tau/(2s)$.
 In other words, the boundary is a circle bundle over $\Sigma_g$, and in the particular case for which $f(\theta,\phi) = \cos\theta d\phi$ and $\psi $ is periodic with period $\Delta\psi = 4\pi$, the boundary is a squashed 3-sphere with squashing parameterized by $4s^2/l^2$.  We will see in the next subsection that, once one takes into account the appropriate compactifications, the boundary metric  in eq. \eqref{boundary} coincides with the one considered for the field theory computation, eq. \eqref{sgds}, up to a rescaling of coordinates.
 
The bulk solutions we have described in this section can be of the type AdS-Taub-NUT  and AdS-Taub Bolt, depending on the value of the parameters appearing in the warp factor. These solutions are characterized by different topologies, since one of the Killing vectors has a zero-dimensional fixed point set (``nut'') or a two-dimensional one (``bolt''). We discuss the requirements for the regularity for NUTs and Bolts separately below, along with the conditions of periodicity of the coordinates. We focus first on the spherical case, $\Sigma_g = S^2$, so $\kappa=1$. 

\vspace{2mm}
 
\subsection{NUTs and Bolts \label{nutsbolts}} 

\vspace{1mm}

 \subsection*{NUTs and Bolts with $S^2$ base}

\vspace{1mm}
\noindent \textit{NUTs}

\vspace{1mm}

\noindent For the NUT solution the Killing vector $\partial_{\tau}$ has a fixed point where the $S^2$ has zero radius:
\be \lambda(r=s) =0\,.
\ee This ensures that the Killing vector has a zero-dimensional fixed point. Moreover, absence of Dirac-Misner \cite{Misner:1963fr} strings constrain the period of $\tau$ to be
\be
\Delta \tau = 4 s \Delta \phi\,,
\ee
and since $\Delta \phi = 2\pi$, this yields $\Delta \tau = 8s \pi $, or equivalently $\Delta \psi = 4 \pi$ (see formula \eqref{boundary}). The coordinate $\theta$ goes from 0 to $\pi$. The last condition concerns the absence of conical singularities at the location of the nut, $r=s$ and is in the following
\be
\Delta \tau \lambda'(r=s) =4 \pi \qquad \rightarrow \qquad \lambda'(r=s) = \frac{1}{2s}\,.
\ee
In particular, for the NUT solutions the warp factor $\lambda(r)$ has a double root at $ r=s$ and the metric is defined for $r \geq s$.
The point $r=s$ is a NUT-type coordinate singularity and the metric is a smooth metric on $\mathbb{R}^4$ with the origin identified with $r=s$. For $s=1/2$ the squashing vanishes, and the boundary is that of a round $S^3$, and one recovers AdS$_4$ space. The AdS Taub-NUT space is a self-dual Einstein space: the Weyl tensor is self-dual and the gauge field has a self-dual field strength, with $Q^2-P^2=0$.

Notice that in principle, if we allow for conical singularity, we could also take into consideration quotients of the Taub-NUT space, obtaining the AdS-Taub-NUT$/\mathbb{Z}_p$ geometry. Quotients of $\mathbb{R}^4$ suffer from conical singularities at the origin, however such backgrounds were studied in \cite{Alday:2012au} and were shown to be sensible backgrounds for holography.

\vspace{4mm}

\noindent \textit{Bolts}

\noindent For the Bolt solution the Killing vector $\partial_{\tau}$ has a two-dimensional fixed point, so the only condition is that, at a radius $r_b >s$,
\be \lambda(r=r_b) =0\,,
\ee with $r_b$  a single zero of $\lambda (r)$. The absence of conical singularities at the location of the bolt requires 
\be
 \frac{r_b^2-s^2}{s \lambda'(r_b)} = \frac2p\,.
\ee
This condition ensures that the metric near the Bolt takes the form \cite{Martelli:2012sz}
\be\label{zoomed}
ds^2 = d \tilde{r}^2 +\tilde{r}^2 \left( d \left( \frac{p \psi}{2} \right) + \frac{p}{2} \cos \theta d \phi \right)^2 + (r_b^2-s^2) (d\theta^2 + \sin^2\theta d\phi^2)\,,
\ee
where we have defined a new radial coordinate $\tilde{r}$ which parameterizes the distance to the bolt in this way: $ \tilde{r} = x \sqrt{r-r_b}$ where $x=2 \sqrt{2s/p}$. The metric \eqref{zoomed} is regular at $\tilde{r}=0$ if $\frac{p}{2} \Delta \psi = 2 \pi $, hence
\be
\Delta \psi = \frac{4 \pi}{p}\,.
\ee
The metric of the Bolt is defined for $r \geq r_b$ and the topology is $\mathcal{M}_p = \mathcal{O}(-p) \rightarrow S^2$. The boundary is a biaxially squashed lens space $S^3 / \mathbb{Z}_p$. At the specific point in parameter space
\be
s = \frac{p}{4 \sqrt{p-1}}
\ee 
we obtain the quaternionic Eguchi-Hanson \cite{Eguchi:1978gw} solution. Among the class of Bolt solutions, the Eguchi-Hanson solution provides an example of regular self-dual Einstein space.
 
 \subsection*{NUTs and Bolts with $\Sigma_g$ base}
 
 The procedure to determine regularity of NUTs and Bolts carries over similarly in the planar and hyperbolic case, along the lines of \cite{Chamblin:1998pz}. We first discuss the case in which the coordinates are not compact.
 
 For the planar case, we look for self-dual solutions and we impose that the warp factor $\lambda(r)$ admits two coincident roots at $r=s$. The warp factor then assumes the form
\be
\lambda = (r-s)^3 (r+3s)\,.
\ee
In the hyperbolic case, as already noticed in \cite{Chamblin:1998pz}, NUT solutions do not exist: once imposing self-duality, the largest roots of the warp factor always lie at a radius $r_s>s$.
Bolt solutions, in contrast, are present in both planar and hyperbolic case. 

In order to find solutions with $\mathcal{M}_{g,p}$ boundary we need to compactify the coordinates by imposing suitable boundary conditions. It is instructive to consider first the metric at the boundary. Consider a 3d metric of the form 
 \be \label{boundary metric grav}
 ds_3^2 = \beta^2 (d\psi - a(\theta,\phi))^2 + d\Omega_{\kappa}^2 \,,
 \ee
where $\psi$ is compactified with period $\Delta \psi$: $\psi \sim \psi + \Delta \psi$.  Here $\theta$ and $\phi$ are coordinates on the compact Riemann surface $\Sigma_g$, which we obtain by a suitable quotient of $\R^2$ (in the case $g=0$) or $H^2$ (in the case $g>1$), and $D_g$ is the fundamental domain of this group action, as in Section \ref{sec:mgpbg}.  Then in order to have a well-defined fiber bundle we must impose
 \be \label{quan} \int_{D_g} da = p \Delta \psi \,.
 \ee
for $p \in \Z$.  To see this is necessary, let $v \in [0,1]$ be a coordinate parameterizing the boundary of $D_g$.  Then if we define:

\be h(v) = \int_0^{v} a\ee
we see that:
\be \label{h1} h(1) =\int_{\partial D_g} a = \int_{D_g} da = p \Delta \psi  \,. \ee
Then in a neighborhood of $\partial D_g$ we may shift $\psi \rightarrow \psi + h(v)$, and \eqref{h1} implies this is a single-valued coordinate transformation.  This implies $d \psi \rightarrow d \psi + a$, and so eliminates $a$ in \eqref{boundary metric grav},  and so after this coordinate transformation we may consistently identify the boundaries of $\partial D_g$ by the group action and obtain the compact space $\Sigma_g$.

For our solution, the fundamental domain for the compactification, $D_g$, is chosen as in \cite{Benini:2016hjo}, so that 
 \be
\text{vol}(\Sigma_g) = 4\pi (g-1) \,\,\, \text{for}\,\,\, g > 1
 \ee 
 and
$
 \int_{D} d \theta d \phi = 4\pi $ for $g = 1$\footnote{Notice here the slightly different conventions with respect to Section \ref{sec:mgpbg}, in order to match the conventions of \cite{Benini:2016hjo}. Here $\text{vol}(\Sigma_g) = 4\pi (g-1)$ and the Chern number $p$ appears in the periodicity of $\psi$.}. Therefore, given that $a =  \theta d\phi$ for the torus, and $a =  \cosh \theta d\phi$ in the higher genus case, \eqref{quan} yields the following conditions
 \be\label{period}
 \Delta \psi = \frac{4 \pi (g-1)}{p} \qquad \text{for } \,\, g>1, \qquad \qquad  \Delta \psi = \frac{4 \pi}{p} \qquad \text{for } \,\, g=1\,.
 \ee

Let's now turn to the Bolt solutions mentioned before. We can have Bolt solutions with topology $\mathcal{O}(-p) \rightarrow \Sigma_g$ imposing $r_b >s$, $ \lambda(r=r_b) =0$, with $r_b$  a single zero of $\lambda (r)$ and using the compactification above to get a genus $g>0$ base manifold $\Sigma_g$.

The absence of conical singularities at the location of the bolt requires 
\be
 \frac{r_b^2-s^2}{s \lambda'(r_b)} = \frac{2|g-1|}{p}\,.
\ee
This condition ensures that the metric near the Bolt takes the form (we take $\kappa = -1$ for simplicity)
\be\label{zoomed2}
ds^2 = d \tilde{r}^2 +\tilde{r}^2 \left( d \left( \frac{p \psi}{2|g-1|} \right) - \frac{p}{2|g-1|} \cosh \theta d \phi \right)^2 + (r_b^2-s^2) (d\theta^2 + \sinh^2\theta d\phi^2)\,,
\ee
where again we have defined a new radial coordinate $\tilde{r}$ which parameterizes the distance to the bolt in this way: $ \tilde{r} \propto \sqrt{r-r_b}$. The metric \eqref{zoomed2} is regular at $\tilde{r}=0$ if $\frac{p}{2|g-1|} \Delta \psi = 2 \pi $, hence we have retrieved exactly eq. \eqref{period}.

Therefore, a $\mathcal{M}_{g,p}$  boundary with $g>0$ can be filled by the Bolt\footnote{We could as well consider toroidal NUT solutions. They however present conical singularities like the $S^3/\mathbb{Z}_p$
NUTs, and we do not consider them in our holographic checks.} solution, with topology $\mathcal{O} (-p) \rightarrow \Sigma_g$ and Euclidean time period \eqref{period}.

\subsection{Supersymmetry properties \label{susyNUTsBolts}}

The supersymmetry properties of the spherical NUTs and Bolts were analyzed in detail in \cite{Martelli:2012sz}, where it was shown that configurations satisfying 
\be
P= \mp \sqrt{4s^2-1}\,, \qquad M= \pm Q \sqrt{4s^2-1}\,, \qquad Q \,\, \text{unconstrained}
\ee
preserve 1/2 of the supersymmetry, while those with
\be \label{cond_spheric}
P= \mp \frac12 (4s^2-1)\,, \qquad M= \pm 2sQ\,, \qquad Q \,\, \text{unconstrained}
\ee
are 1/4 BPS. In \cite{Martelli:2012sz} the explicit form of the Killing spinor was provided as well. We will be interested in the 1/4 BPS ones. For our purposes, we wish to generalize the BPS conditions to solutions whose boundary is $\mathcal{M}_{g,p}$. 

AdS$_4$ solutions with NUT charge and $\Sigma_g$ horizon in minimal $\mathcal{N}=2$ gauged supergravity in Lorentzian signature were presented in \cite{AlonsoAlberca:2000cs}. In the same paper, the integrability conditions for supersymmetry were analyzed, and necessary conditions for supersymmetry to be preserved were given. For the 1/4 BPS case they read
\be\label{eq_ortin}
P = \pm \frac12 (\kappa +4N^2)\,, \qquad M = \pm 2 N Q \,.
\ee
Lorentzian solutions of this form were studied also in \cite{Klemm:2013eca}: in the latter, it was shown that \eqref{eq_ortin} is a sufficient condition for preserving 1/4 of supersymmetry\footnote{See also \cite{Nozawa:2015qea} and \cite{Nozawa:2017yfl}, where the supersymmetry enhancement in the self-dual case was analyzed.}, following the procedure of \cite{Caldarelli:2003pb}. 

We work here in Euclidean signature and, taking a pragmatic approach, we explicitly construct the Killing spinor for the Euclidean solutions under consideration. 
Therefore, we map the integrability condition of \cite{AlonsoAlberca:2000cs} via
\be \label{mapAA}
 t \rightarrow  i \tau \qquad Q \rightarrow - i Q  \qquad N \rightarrow is\,.
 \ee
and we find that a necessary condition for the solutions to be 1/4 BPS is
\be \label{cond_kappa}
P= \mp \frac12 (4s^2-\kappa) \qquad M= \pm 2sQ \qquad Q \,\, \text{unconstrained}
\ee
Notice that \eqref{cond_kappa} reduces to \eqref{cond_spheric} for $\kappa=1$.  In Appendix \ref{Killing_sp} we construct the Killing spinor for the solutions satisfying the conditions \eqref{cond_kappa}, generalizing the procedure of \cite{Martelli:2012sz}, whose notation we follow closely, to the higher genus case. As one can see in Appendix, the Killing spinor has only radial dependence, so that the compactification necessary to obtain a compact Riemann surface is allowed without breaking supersymmetry. 

We notice that the signs in this equation \eqref{cond_kappa} can be reabsorbed with a change of coordinate $ \{ r \rightarrow -r, \psi -\rightarrow -\psi, \phi \rightarrow -\phi \}$ hence  without loss of generality we focus on the set of parameters
\be \label{param2}
P=-\frac12 (4s^2-\kappa) \,, \qquad M= 2sQ \,.
\ee
This condition denotes a four-dimensional subspace of BPS solutions, parameterized by $p$ (Chern class of the bundle), $g$ (genus of the base Riemann surface), $Q$ (electric charge) and $s$ (squashing parameter). As we will see in the next section, regularity imposes additional constraints on these quantities, and in particular, we will see that there are different bulk fillings for the same boundary data. 

These 1/4 BPS configurations with $\mathcal{M}_{g,p}$ asymptotics will be the gravitational configurations of our interest. When solving the BPS equations, once the $r$ component is satisfied, the remaining components of the Killing spinor equations reduce to the new minimal rigid supersymmetry equation 
\be \label{eq_bound}
\left({\nabla_{(3)}}_i - i {A_{(3)}}_i + \frac{i s}{2}\gamma_i \right)\upsilon =0\,.
\ee
for a 3d spinor $\upsilon$. Here $\gamma_i$ are the Pauli matrices $i=1,2,3$, $\nabla_{(3)}$ is the covariant derivative associated to the 3d boundary metric \eqref{boundary metric grav}.  The (global part of the) asymptotic gauge field is 
\be \label{as_bound_gf}
A_{(3)} =  \lim_{r\rightarrow\infty} A   =  P \, \tau_3  =  - \frac{1}{2}(4s^2-\kappa)\, (d\psi + f(\theta) d \phi).
\ee 
which is mapped to the background R-symmetry gauge field \eqref{rgf} by taking into account that $\beta = 2s/p $, and the fact that, as already mentioned, the boundary metric in this section is defined with the following conventions
\be \psi  \rightarrow p \psi\,,  \qquad \text{vol}(\Sigma_g) = \pi \rightarrow \text{vol}(\Sigma_g) = 4 \pi |g-1|\,.
\ee 
with respect to the field theory one \eqref{sgds}. Equations of the form \eqref{eq_bound} were shown to naturally arise on the boundary of supersymmetric AdS configurations in theories of gauged supergravity, by \cite{Klare:2012gn,Closset:2012ru}. 

As noticed in \cite{Martelli:2012sz}, one needs to supplement the expression for the gauge field in \eqref{as_bound_gf} by a flat connection. The Killing spinor in Appendix \ref{Killing_sp} is computed in a rotating frame: if one had to compute its expression in a static frame, a $e^{i \psi}$ dependence would appear. The flat connection introduces an opposite phase which cancels with the latter, so the global form of the Killing spinor to be independent of $\psi$ (see Appendix \ref{Killing_sp}).

\subsection{Moduli space of solutions}

This subsection is devoted to the study of the moduli space of the solutions. The space of the configurations with $g=0$ ($\kappa=1$) was discussed already in \cite{Martelli:2012sz}, we extend it here to the cases with $ g>0$ as well.
We focus on the cases
\be
P= - \frac12 (4s^2-\kappa) \qquad M=  2sQ \qquad Q \,\, \text{unconstrained}\,.
\ee
This class of 1/4 BPS configurations include NUTs and Bolts, depending on the value of parameter $Q$.  As already anticipated, for $\kappa=1$ the former solutions (NUTs) have as conformal boundary a squashed $S^3$, and for the latter (Bolts) the boundary is a squashed lens space $S^3/ \mathbb{Z}_p$. 

We start by analyzing the roots of $\lambda(r)$. These are
\begin{eqnarray} \label{erre}
r_1 &  = & - s -\sqrt{ \frac{-\kappa - 2Q +4s^2}{2}}\,,  \qquad r_2  = - s +\sqrt{ \frac{-\kappa -2Q +4s^2}{2}}\,, \nonumber \\
r_3 & = & s - \sqrt{ \frac{-\kappa +2Q +4s^2}{2}} \,, \qquad r_4  =  s +\sqrt{ \frac{-\kappa +2Q +4s^2}{2}}\,.
\end{eqnarray}
The largest root of $\lambda(r)$, denoted with $r_0$, can be either $r_2$ or $r_4$, depending on the specific values of $s$ and $Q$. We denote the two solutions with $r_+$ and $r_-$ respectively:
\be \label{rplusminuscases}
r_{\pm} = \pm s +\sqrt{ \frac{-\kappa\pm2Q +4s^2}{2}}\,.
\ee
We note that the requirement for a Taub-NUT solution is $r_0=r_+=r_- =s$ and gives $Q = \pm \frac12 (4s^2-\kappa)$. For this value, the warp factor $\lambda(r)$ has two coincident roots. This solution is well-defined for all value of the squashing $s>0$ and its boundary is a $p=1$ fibration over a 2d constant curvature Riemann surface $\Sigma_g$: in particular, when $g=0$, we retrieve the squashed three-sphere. 

For the Bolt, imposing regularity near $r=r_0$ amounts to \cite{Chamblin:1998pz}
\be \label{conditionBol}
 \frac{r_0^2-s^2}{ \, \lambda' (r_0)} =\frac{s\, \mathbf{p}}{2}\,.
\ee
where we have defined 
\be \mathbf{p} = \frac{p}{|g-1|} \,\,\,\,\,\,\text{for}\,\,\, \,\,\,g\neq 1\,,  \qquad \quad \mathbf{p} =p  \,\,\,\,\,\,\text{for}\,\,\, \,\, g= 1\,.
\ee
Since $r_0 >s$ for the Bolt solution, the numerator of \eqref{conditionBol} is positive. The derivative at the denominator is as well, since $r_0$ is the largest root of $\lambda(r)$ and $\lambda(r)$ goes as $r^2$ for $r \rightarrow \infty$. Therefore in our considerations we will restrict to the case $s \mathbf{p}>0$, which encompasses both $s>0, \mathbf{p}>0$ and $s<0, \mathbf{p}<0$.

We need to analyze the two cases \eqref{rplusminuscases} separately. If the largest root is $r_4$, imposing $r_0 = r_+$ yields the following condition on $Q$:
\be \label{firstpluscase}
Q= Q_{+}^{\pm} = \frac{\mathbf{p}^2\mp(16s^2- \mathbf{p})\sqrt{(16 \, s^2 + \mathbf{p})^2-128 \, \kappa \,s^2}}{128s^2}\,,
\ee
while if we impose $r_0 = r_-$ we obtain two possibilities
\be \label{second}
Q = Q_{-}^{\pm} =  -\frac{\mathbf{p}^2 \mp (16s^2+ \mathbf{p})\sqrt{(16 \, s^2 - \mathbf{p})^2-128 \, \kappa \,s^2}}{128s^2}\,.
\ee
Plugging these values of $Q$ into $r_{\pm}$ we obtain the following expressions for $r_+$ (using eq. \eqref{firstpluscase}):
\be \label{radius_bolt_sph}
r_{+}= \frac{\mathbf{p}\pm\sqrt{(16 \, s^2 + \mathbf{p})^2-128 \, \kappa \,s^2}}{16s}\,,
\ee 
and $r_-$ (using \eqref{second})
\be \label{radius_bolt_minus}
r_{-} = \frac{\mathbf{p} \mp \sqrt{(16 \, s^2 - \mathbf{p})^2-128 \, \kappa \,s^2 }}{16 s}\,.
\ee

We can see then that we obtain up to four different branches denoted by the relations \eqref{firstpluscase} and \eqref{second} \footnote{In the case $g=0$ these branches were analyzed in \cite{Martelli:2012sz}, but we appear to find another additional Bolt$_+$ solution with respect to their analysis, corresponding to the solution with $Q_{+}^-$. This branch gives a new  solution only for $p=1$, and the latter, despite being a genuinely different solution, has the same range of existence as the Bolt$_+$ with $Q^+_+$. Hence the analysis of the on-shell action for $g=0$ carries out unchanged with respect to \cite{Martelli:2012sz}. More details in the Appendix.}. We will see shortly that each of the two solutions in \eqref{firstpluscase} gives the same value $I_+$ for the on-shell action, which differs with respect to the one $I_-$ obtained for both the solutions in \eqref{second}. Hence we collectively denote both signs in \eqref{firstpluscase} with ``Bolt$_+$'' and \eqref{second} with ``Bolt$_-$.''

In order for these branches to exist, all the following conditions needs to be met:
\begin{enumerate} 
\item  the function under the square root $f_{\pm} = (16 \, s^2 \pm \mathbf{p})^2-128 \, \kappa \,s^2 $ should be positive, 
\item  respectively, $r_{\pm}$ should be indeed the largest root of the warp factor $\lambda(r)$ and \item  $r_{\pm}>s$ in order to have a Bolt. 
\end{enumerate} 
These requirements in particular set constraints on the value of the squashing parameter $s$. Indeed regular Bolt solutions might be present only in a limited interval of squashing parameter $s$: for instance, for $g=0$ the Bolt$_+$ solutions for $p=1,2$ are regular only for a finite interval $s \in [0,s_1]$, while they always exist for $p \geq 3$ \cite{Martelli:2012sz}. A detailed analysis of the moduli space for the different $g$ cases is treated in Appendix \ref{AppB}. We remind the reader that spherical NUT solutions exist for all $s>0$. Moreover, taking quotients of the latter generates the 
mildly singular NUT$/\mathbb{Z}_p$ solutions mentioned in Section \ref{nutsbolts}, which have lens space boundary and are defined for all values of $s$. 

We end this section by computing the total flux for the Bolt solutions. Indeed the field strength $F$ of the Bolt has a non-trivial magnetic flux through the Bolt surface $\Sigma_g$ at $r = r_0$, which lies inside the bulk (it is the point where the Bolt geometry caps off). The flux can be computed as   
\begin{eqnarray} \label{fluxbolt}
\int_{\Sigma_g} \frac{F}{2 \pi} & = & - \frac{2s}{r_0(s)^2-s^2} \left[ -2 Q_{\pm} (s) r_0(s) -\frac{ (4s^2-\kappa) (r_0(s)^2+s^2)}{2}  \right] \nonumber \\
&= & \left( \pm \frac{\mathbf{p}}{2} -\kappa \right) |g-1| = \pm \frac{p}{2} -(1-g)\,,
\end{eqnarray}
where the $\pm$ refers to the two classes of Bolt$_+$ and Bolt$_-$ defined above.
Notice that the form of the metric and field strength is fixed demanding regularity of the configurations, and the flux computed in \eqref{fluxbolt} follows. Its value satisfies 
\be
\int \frac{F}{2 \pi} = \frac{q}{2}
\ee
where $q = p \mod 2$, which is consistent with  $A$ being a spin$^c$ gauge field.
However, in order for the uplift to 11d to be consistent the flux at the Bolt is subject to a further condition. In particular,  depending on the geometry of the internal Sasaki-Einstein seven manifold $Y_7$, the uplift is possible only for certain values of $p$: this is discussed in  the next section.

\subsection{Uplift to 11d \label{11uplift}}

The 11d uplift ansatz for solutions of (Lorentzian) minimal gauged 4d supergravity was found in \cite{Gauntlett:2007ma}, where it was shown that any supersymmetric configuration of such 4d theory uplifts locally to a solution in eleven dimensions described by the following metric and 4-form $\mathcal{G}$:
\be \label{11dmetric}
ds_{11}^2= R_*^2 \left(\frac14 ds_4^2 +ds_{Y_7}^2 \right)\,,
\ee
\be \label{11dmetric}
ds_{Y_7}^2=  ds_6^2 + \left(\xi +\frac{A}{2} \right)^2 \,,
\ee
\be
\mathcal{G} = \frac38 R_*^3 \, \text{vol}_4 - \frac{R_*^3}4 \star_4 F \wedge d \xi\,,
\ee
where the $ds_4^2$ is the metric element in the four-dimensional spacetime, with volume form $\text{vol}_4$, $\xi$ is the contact form of $Y_7$ and K\"ahler Einstein base metric $ds_6^2$. Here $\star_4$ is defined with respect to the 4d metric and $\star_{11}$ with respect to the eleven-dimensional one. The seven-dimensional metric on $Y_7$ satisfies $R_{\alpha \beta} = 6 g_{\alpha \beta}$. The four-dimensional Newton constant $G_4$ is
\be \label{G4}
G_4 = \frac{N^{-3/2}}{16 \pi}  \sqrt{ \frac{2^5 \times 3^3 \, \text{Vol}(Y_7) } {\pi^2} }\,,
 \ee
 with
  \be
 N = \frac{1}{(2 \pi l_p)^6} \int_{Y_7} \star_{11} \mathcal{G}\,,
 \ee
and the radius $R_*$ is $R_*^6 = \frac{(2 \pi l_p)^6 N }{6 \text{Vol}(Y_7)}$.

For a solution of 4d supergravity to uplift to a properly defined M-theory configuration, some parameters should be appropriately quantized. Because of this, there are some differences among the conditions we need to impose for  NUTs and Bolts \cite{Martelli:2012sz}. 

The Taub-NUT solution has topology $\mathbb{R}^4$ and the gauge field $A$ is globally a one-form on $\mathbb{R}^4$:  the twisting $\xi + \frac12 A$ is trivial. Therefore every NUT solution uplifts without restrictions to eleven dimensions. There is no quantization condition on $\mathcal{G}$, which  is itself a globally defined four-form on $\mathbb{R}^4 \times Y_7$ \cite{Martelli:2012sz}.

For the Bolt solution, characterized by topology  $\mathcal{O}(-p) \rightarrow \Sigma_g$, the situation is different. When $Y_7$ is a regular Sasaki-Einstein manifold, $Y_7$ is a $U(1)$ bundle over the six-dimensional K\"ahler Einstein manifold $B_6$. In this case $\xi = d\eta +\sigma$ where $d\sigma$ is the Ricci form on $B_6$ divided by four. Regularity imposes that the period of $\eta$ is $2 \pi I/4k$, where $k$ is a positive integer (which is the level of the Chern-Simons theory) and $I = I(B_6) \in \{1,2,3,4 \}$ is the Fano index of $B_6$. We will restrict here to the case $k=1$ for simplicity. In order for the $U(1)$ bundle to be well-defined at the Bolt itself, the flux should satisfy \cite{Martelli:2012sz}
\be \label{quant_uplift}
\frac{4}{2I} \int_{\Sigma_g} \frac{F}{2 \pi} = m \in \mathbb{Z}\,,
\ee
since $m$ is the Chern number of the circle bundle with coordinate $\eta$ over the Bolt $\Sigma_g$. Moreover, again there is no quantization condition on $\mathcal{G}$.  The uplift in 11d for NUTs and Bolts is then slightly different. For instance, in the ABJM case, for the Bolt solutions the $U(1)$ bundle over $M_4 \times CP^3$ \eqref{11dmetric} is non-trivially fibered over 3d the boundary  space $\partial M_4$. For the NUT and NUT$/\mathbb{Z}_p$, instead, the bundle is trivial.

Eq. \eqref{quant_uplift} for Bolts is a condition involving $p$ and $g$ and  will be crucial point for the matching of the on-shell action with the partition function of the corresponding $\mathcal{N} =2$ superconformal field theory. For instance, the uplift on the round seven sphere $S^7$ ($k=1$ and $I(S^7)=4$), making use of \eqref{fluxbolt}, dictates the following constraint:
\be
\pm \frac{p}{2} +(g-1) = 2 m \qquad  m \in \mathbb{Z}
\ee
which leads to
\be \label{quant-ABJM}
p  \pm 2(g-1) =0 \MOD{4} 
\ee

Notice that there are cases for which Bolt solutions uplift for all $p$: for example the manifold $Y_7 = M^{3,2}$, since $I(M^{3,2}) =1$ \cite{Martelli:2012sz}. It is worth mentioning another specific example we will focus later on: the manifold $V^{5,2}$, whose Fano index is $I(V^{5,2}) =3$. In this case the uplift condition \eqref{quant_uplift}  using \eqref{fluxbolt} dictates the following constraint:
\be
\pm \frac{p}{2} +(g-1 ) = \frac32m \qquad  m \in \mathbb{Z}
\ee
which leads to
\be \label{quant-V52}
p  \pm 2(g-1) =0 \MOD{3}
\ee
We will see in Section \ref{sec:holo_checks} the consequences of the condition \eqref{quant-ABJM} 
and \eqref{quant-V52} 
for the matching with the dual field theory computation.

\subsection{On-shell action via holographic renormalization}

We are now ready to compute the renormalized on-shell action for the branches of solutions that we have found in the previous sections, satisfying \eqref{param2}. 
In order to compute such quantity, we need to plug in the solution into the action \eqref{act}, and perform the integral with extrema of integration $r_0$ (radial coordinate where the spacetime caps off) to $r \rightarrow \infty $ (asymptotic boundary). However, this leads to a divergent quantity, therefore we need to adopt an appropriate renormalization scheme. 

Techniques of holographic renormalization \cite{deHaro:2000vlm,Emparan:1999pm,Skenderis:2002wp,Papadimitriou:2005ii} have been extensively developed in the past years. Following these, one first regulates the action via the introduction of a radial cutoff $r_{inf} $, which is sent to infinity after adding appropriate counterterms.
The counterterms are function of the boundary 3-metric $h_{ij}$ and they read \cite{Skenderis:2002wp}
\be \label{contr}
I_{ct} = \frac{1}{8 \pi G_4} \int_{\partial M} d^3x \sqrt{h} \, \left( 2 +\frac12 R_3(h) -\Theta \right)\,,
\ee
 where the term in $\Theta$ is the Gibbons -Hawking boundary term, defined as
 \be
 \Theta_{\mu \nu} = - (\nabla_{\mu} n_{\nu} + \nabla_{\nu} n_{\mu}) \,, \qquad \Theta= {\Theta_{\mu}}^{\mu} \,,
 \ee
where $n_{\mu}$ is a unit vector normal to the boundary and $R_3(h)$ is the Ricci scalar of the induced boundary metric.
In what follows, we divide the action  in two parts, in particular we single out the contribution of the vector fields,
 \be
 S = S_{grav} + S_{F^2}\,,
 \ee
 \be
 S_{grav} = -\frac{1}{16 \pi G_4} \int d^4x \sqrt{g} \left( R +\frac{6}{l^2}  \right) \,,  \qquad  S_{F^2} = +\frac{1}{16 \pi G_4} \int d^4x \sqrt{g} F_{\mu \nu}F^{\mu \nu} \,,
 \ee
 Plugging in the solution, we obtain the on-shell action which we denote with $I$. We compute
 \be \label{grav}
 I_{grav} = \frac{1}{8\pi G_4} \frac{16 \pi^2}{\mathbf{p}} [2s r_{inf}^3 -6s^3 r_{inf}-2sr_0 +6s^3 r_0   ]\,,
 \ee
 \be \label{ctsphere}
 I_{ct} = \frac{1}{8 \pi G_4} \frac{16 \pi^2}{\mathbf{p}} [4 Q s^2-2s r_{inf}^3 +6 s^3 r_{inf} + O(r_{inf}^{-1})] \,,
 \ee
 where we have made use of
 \be
R= -12\,, \qquad  \sqrt{g} = 2s (r^2-s^2)  (-f'(\theta))\,,
\ee
\be
\sqrt{h} = s \sqrt{(s^2-r^2) \lambda(r)} (-f'(\theta))\,,
\ee
\be
\Theta=\frac{2 s Q \left(s^2-3 r^2\right)+r \left(3 s^4-14 s^2 r^2+\frac{1}{4} \left(\kappa-4 s^2\right)^2-Q^2+3 r^4+2 \kappa r^2\right)}{(s^2-r^2)^{3/2} \sqrt{\lambda(r)}}\,,
\ee
\be
R_3 = \frac{4 s^2 \left(s^4-6 s^2 r^2-4 s Q r-Q^2+r^4\right)-\kappa \left(8 s^4+s^2 \left(1-12 r^2\right)+4 r^4\right)}{2 \left(s^2-r^2\right)^3} \,. \ee
 The extremum of integration $r_0$ and the value of the gauge field contribution differ between NUTs and Bolts, whose on-shell actions are computed separately below.
 
 \subsubsection*{NUT solution}
 
In the NUT case, $g=0$, $r_0=s$ and $Q= \pm \frac12 (4s^2-1)$ hence
 \be \label{Fnut}
 I_{F^2,NUT}^{bulk} = \frac{16 \pi^2 Q^2}{8 \pi G_4} = \frac{2 \pi}{G_4} \frac{(1-4s^2)^2}{4} \,,
 \ee
 so the total free energy, obtained by summing the contributions \eqref{grav} \eqref{ctsphere} \eqref{Fnut} is
 \be
 I_{NUT} = \frac{\pi}{2 G_4} \,,
 \ee
 which was already found in \cite{Martelli:2012sz}.
 
 \subsubsection*{Bolt branches}
 
 For the Bolt solution, instead, the expression for $I_{F^2}$ is
   \be \label{FBolt}
 I_{F^2,Bolt}^{bulk} = \frac{\pi  s r_0 \left(r_0^2 \left(\left(\kappa-4 s^2\right)^2+4 Q^2\right)+s^2 \left(\left(\kappa -4 s^2\right)^2+4 Q^2\right)+8 \left(4 s^2-\kappa\right) s Q r_0 \right)}{2 G_4 \left(s^2-r_0^2\right)^2 \, \mathbf{p}}\,.
 \ee
For the Bolt$_+$ solutions $r_0 = r_+$ in \eqref{radius_bolt_sph} and $Q= Q_+$ in \eqref{firstpluscase}. These expressions are somewhat complicated but if we sum all contributions  \eqref{grav}, \eqref{ctsphere} and \eqref{FBolt}, at the end of the day the renormalized on-shell action assumes the remarkably simple form:
 \be \label{FEboltplus}
 I_{ren, Bolt+} =\frac{\pi (4(1-g)-p)}{8 G_4}\,.
 \ee
For the Bolt$_-$ solutions, conversely, we need to use $r_0 = r_-$ in \eqref{radius_bolt_minus} and $Q= Q_{\pm}$ in \eqref{second}. Again, after summing Bolt$_-$\eqref{grav}, \eqref{ctsphere} and \eqref{FBolt}, the final formula for the renormalized on-shell action is simple:
  \be \label{FEboltminus}
 I_{ren, Bolt -} = \frac{\pi  (4(1-g)+p)}{8 G_4} \,.
 \ee
 
We see from these results that the parameter $s$, which quantifies the squashing of the sphere, dropped out of the final result for both NUT and Bolts. Hence for this class of 1/4 BPS solutions the free energy does not depend on the squashing parameter, as was found in \cite{Hama:2011ea,Martelli:2012sz}. For $g=0$ the on-shell action reduces to that found in \cite{Martelli:2012sz}. For $p=0$, instead, expressions \eqref{FEboltplus}-\eqref{FEboltminus} reduce to the entropy of supersymmetric 1/4 BPS black holes in gauged supergravity, found by \cite{Caldarelli:1998hg,AlonsoAlberca:2000cs}.

Before proceeding further, we'd like to take into consideration another possible background with lens space $S^3/\mathbb{Z}_p$ boundary, which is the mildly singular AdS-Taub-NUT$/ \mathbb{Z}_p$, also named ``orbifold NUT.'' The free energy of this configuration is $1/p$ times that of the NUT solution \cite{Alday:2012au}. However, if we want to have the same boundary data as the $g=0$ Bolts, we need to consider orbifold NUTs where with the addition of $\pm \frac{p}{2} -1$ units of magnetic flux, which is the same as that of the Bolt$_{\pm}$. Therefore the on-shell action will be supplemented by  a term of the form $F^2$ \cite{Martelli:2012sz} in the following way
\begin{eqnarray}
I_{NUT/\mathbb{Z}_p + flux \pm} &=& \frac{I_{NUT}}{p} + I_{F^2 \pm} = 
 \frac{\pi}{ 2G_4 p} \left(1 + \left(\pm\frac{p}{2} -1 \right)^2  \right)\,.
\end{eqnarray}

These orbifold  configurations are acceptable solutions. However, as shown in the plot \ref{Fig3} of Appendix \ref{AppB}, their on-shell action is always higher or equal to the one of the Bolt$_+$ solutions, except in the case $p=2$ where their flux vanishes and their free energy coincide. Hence when Bolts and NUT/$\mathbb{Z}_p$ both exist, the Bolts are the favored configuration due to their lower free energy and we expect their on shell action to match the field theory computation.

We end this section by reminding the reader that the uplift of the solutions in 11d imposed the quantization condition \eqref{quant_uplift} which constrains the values of allowed $p$, depending on the geometry of the internal Sasaki-Einstein manifold $Y_7$. Reinstating the factors of $N^{3/2}$ in the above formulas, according to formula \eqref{G4}, we obtain
  \be \label{NUTY7}
 I_{NUT} = \frac{ \sqrt2 \pi N^{3/2}}3 \sqrt{ \frac{\pi^4}{3\, \text{Vol}(Y_7) }  } \,,
 \ee
 \be \label{BoltY7}
 I_{Bolt \pm} =  \frac{\sqrt2 \pi N^{3/2} (4(1-g) \mp p)}{12} \sqrt{ \frac{\pi^4}{3 \, \text{Vol}(Y_7) }  } \,. 
 \ee
Notice that the Bolt formula \eqref{BoltY7} reproduces for $p=0$ the on-shell action of supersymmetric black holes of minimal gauged supergravity \cite{Caldarelli:1998hg}, which was computed in \cite{Azzurli:2017kxo}. In the latter work (and more recently in \cite{Halmagyi:2017hmw,Cabo-Bizet:2017xdr} for the non-minimal case) it was moreover shown that the free energy of such configurations coincides with (minus) their entropy.

The on-shell action computed here corresponds to an ensemble where the magnetic fluxes and the electric chemical potentials are kept fixed \cite{Hawking:1995ap}. In other words, it refers to a canonical ensemble for the magnetic charges and a grand-canonical one for the electric ones. Since the partition function $Z_{\mathcal{M}_{g,p}}$ computed in the field theory describes supersymmetric states with the same, fixed magnetic charges and it contains a sum over all electric charge sectors with fixed chemical potentials $m_i$ \cite{Benini:2016rke}, it is natural to compare the expressions obtained in this sections \eqref{NUTY7}-\eqref{BoltY7} with those achieved in the field theory in Section \ref{sec:field theory}. The next section is devoted to this comparison.


\section{Holographic comparison for ABJM}
\label{sec:holo_checks}

\subsection{Truncating to minimal supergravity\label{trunc_minimal}}

In general, the partition function we computed in Section \ref{sec:field theory} is dual to a configuration in $\cN=2$ gauged supergravity with non-trivial magnetic fluxes and profiles for the scalar fields.  These profiles are governed by the asymptotic behavior (\ie, scalar modes) of these fields, which is determined by the background vector multiplets coupled to flavor symmetries in the dual field theory on $\cM_{g,p}$, or equivalently, by the masses, $m_i$, fluxes, $s_i$, and R-charges, $r_i$, that we choose.  In order to compare to the supergravity computation of the previous section, we must suitably restrict these parameters so that the bulk vector multiplets are trivial, and we obtain a configuration in minimal supergravity.

As above, we focus on the ABJM model for concreteness; we will return to more general models in Section \ref{sec:gq}.  Then the bulk vector multiplets are trivial when we set all of the background vector multiplets on the boundary to be equal.  Naively, this implies setting the parameters $m_i,s_i$, and $r_i$ all to be independent of $i$.  However, since $\sum_i m_i=0$, this is only possible if $m_i=0$, which leads to a trivial solution.  To get further, we recall that, due to the shift symmetry \eqref{fluxshift}, which allows us to shift the $m_i$ by integers, only the quantities $[m_i]$ and $-p m_i + s_i + (g-1) r_i$ are gauge invariant.  Thus, we should impose that:

\be \label{mincond} [m_i] \;\;\; \text{and} \;\;\; {\frak n}_i \equiv-p n_i +s_i +(g-1) r_i  \;\;\; \text{ are independent of}\; i \;. \ee
Following \cite{Benini:2015eyy}, we may interpret the fractional part of the masses, $[m_i]$, as fixing the asymptotic behavior of the scalar fields in the bulk\footnote{Notice that minimal gauged supergravity amounts to setting all scalars constant, in particular equal to their value in the vacuum (i.e. no radial flow). The value attained at the vacuum is independent of $i$.}.  Then ${\frak n}_i$ can be attributed to the net magnetic charge felt by the $i$th chiral field.  Then we conjecture that \eqref{mincond} gives the truncation to minimal supergravity for the ABJM model.

Let us see what the condition \eqref{mincond} implies for large $N$ behavior of the partition function.  For the ABJM theory, recall that, for the solution with $\sum_i [m_i]=1$, the expression for the large $N$ partition function we derived in \eqref{lnabjm} can be written as:

\be \label{abjmms1} \log Z_{\cM_{g,p}}^{ABJM} = \frac{2\pi N^{3/2}}{3} \sqrt{2 k [m_1][m_2][m_3][m_4]} \bigg( -2 p + \sum_i \frac{{\frak n}_i}{[m_i]} \bigg)  \;. \ee
Imposing \eqref{mincond}, the condition $\sum_i [m_i]=1$ forces us to set $[m_i]=\frac{1}{4}$.  The conditions \eqref{abjmcond} and \eqref{abjmncond} imply:

\be \sum_i {\frak n}_i = p+2(g-1) \;.\ee
If we set all the ${\frak n}_i$ equal, then this can only be satisfied for ${\frak n}_i \in \Z$ if we impose:

\be \label{constraint1} p+2 (g-1) = 0 \MOD{4} \;. \ee
Then we take ${\frak n}_i=\frac{1}{4}(p+2(g-1))$.  Inserting these values into \eqref{abjmms1}, we find:

\be \label{fe1} \log Z_{\cM_{g,p}}^{ABJM} = \frac{\pi N^{3/2} \sqrt{2k} }{12} \big(p +4 (g-1 ) \big) \;.\ee

Next, recall there is another solution with $\sum_i [m_i]=3$, obtained from the previous one by mapping $[m_i] \rightarrow 1-[m_i]$.  In this case the partition function is given by: 

\be \label{abjm2s} \log Z_{\cM_{g,p}}^{ABJM} = - \frac{2\pi N^{3/2}}{3} \sqrt{2 k (1-[m_1])(1-[m_2])(1-[m_3])(1-[m_4])}  \bigg(  -2p - \sum_i \frac{\hat{\frak n}_i}{1-[m_i]} \bigg) \;,\ee
where we define:
\be \label{hatelldef} \hat{\frak n}_i \equiv -p (n_i+1) +  s_i +(g-1) r_i \;,\ee
which satisfies:
\be \sum_i \hat{\frak n}_i = -p + 2 (g-1) \;.\ee
Then to set $[m_i]$ and $\hat{\frak n}_i$ independent of $i$, we take:

\be [m_i]= \frac{3}{4}, \;\;\; \hat{\frak n}_i = \frac{1}{4} (-p+2(g-1))\;, \ee
and we must impose:

\be \label{constraint2} - p+2 (g-1) = 0 \MOD{4} \;.\ee
Plugging this in to \eqref{abjm2s}, we find:
\be \label{fe2} \log Z_{\cM_{g,p}}^{ABJM} = \frac{\pi N^{3/2}\sqrt{2k}}{12}  \big( -p + 4 (g-1)\big) \;. \ee

\subsection{Holographic comparison}

At this point it is easy to compare the result obtained for the ABJM partition function on $\mathcal{M}_{g,p}$ and the on-shell action of Bolt$_{\pm}$ solutions, with uplift to M-theory on $S^7$, corresponding to the case $k=1$ in the field theory.  The field theory result in \eqref{fe1} and \eqref{fe2}  is summarized in
\be \log Z_{\cM_{g,p}}^{ABJM} = \frac{\pi N^{3/2} \sqrt{2} }{12} \big( 4 (g-1) \pm p \big) \;,\ee
subject respectively to the constraints \eqref{constraint1} and \eqref{constraint2} which read 
\be \label{quant_repeated} \pm p+2 (g-1) = 0 \MOD{4}. \ee
These results exactly match the on-shell action computed on the Bolt$_+$ (upper sign) and on the Bolt$_-$ (lower sign) solutions respectively. Indeed, plugging into formula \eqref{BoltY7}  the expression of the volume of the round seven-sphere
\be
\text{Vol}(S^7) = \frac{\pi^4}{3} \;,
\ee
one gets exactly
\be
I_{Bolt\pm}^{S^7}= - \log Z_{\cM_{g,p}}^{ABJM} \,.
\ee
Moreover, the condition \eqref{quant_repeated} maps exactly in the quantization condition \eqref{quant-ABJM} obtained in minimal supergravity. We recall that the latter condition is necessary for the uplift on $S^7$ to be globally defined for Bolt solutions.

Notice that due to the condition \eqref{constraint2} for ABJM, the Bolt$_+$ and Bolt$_-$ uplift for the same values of $p$.  The dominant configuration, namely the one which has lowest free energy, is the Bolt$_+$ for $p>0$, the Bolt$_-$ for $p<0$. 

\subsection{NUTs and Bolts for $S^3$}
\label{sec:NB}

Let us consider the special case of $\cM_{g=0,p=1} = S^3$.  One can see that the quantization condition \eqref{quant_repeated} is not satisfied here, and indeed, the $p=1$ Bolt$_\pm$ solutions are not states in the ABJM theory, so our formalism cannot reproduce their free energy. 

However,  for the NUT-type solutions there are no further constraints for the uplift, therefore the AdS-Taub-NUT is a regular configuration which uplifts to 11d on $S^7$. The NUT ($g=0$) then provides a regular filling for the squashed $S^3$ and its on-shell action is \eqref{NUTY7}
  \be
 I_{NUT}^{S^7} = \frac{ \sqrt2 \pi N^{3/2} }3   \,.
 \ee
This coincides with the ABJM free energy on the $S^3$ at the conformal point, as computed in   \cite{Jafferis:2011zi}.  

This lead us to a puzzle: why do we not recover the result of \cite{Jafferis:2011zi} for the $S^3$ partition function by our method?  The precise mapping of parameters between our partition function and theirs is:\footnote{More precisely, as described in more detail in Sections \ref{sec:gq} and \ref{sec:s3comp}, this is true in the ``physical gauge,'' in which case the R-charges $r_i$ can be tuned continuously, and appear in a combination $m_i \rightarrow m_i +r_i \equiv \Delta_i$.}

\be Z_{\cM_{g=0,p=1}}(m_i) = Z_{S^3}(\Delta_i) \;\;\;\;\;\; \text{for}\;\; \Delta_i = m_i \;. \ee
Then the superconformal R-charges, $\Delta_i = \frac{1}{2}$, correspond to the case:

\be \sum_{i=1}^4 [m_i] = 2 \;, \ee
for which we did not find a solution to the large $N$ $\cM_{g,p}$ partition function for the ABJM model in Section \ref{sec:largensol}.

To see what went wrong, recall the starting point of the computation of \cite{Jafferis:2011zi} was the integral formula, \eqref{mgpcint}, in the special case $g=0,p=1$, and they solved this at large $N$ by looking for a saddle point of this integrand.  While their starting point, \eqref{mgpcint}, and ours, \eqref{mgpsv}, are equivalent at finite $N$, these two methods of taking the large $N$ limit are evidently not equivalent.  They can be summarized as follows:

\begin{itemize}
\item {\bf Method $1$} (the method we used above for general $g$ and $p$, and used for $p=0$ in \cite{Benini:2015eyy}) - Extremize the functional, $\cW[\rho,v_\alpha]$, computing the twisted superpotential, obtaining eigenvalue distributions $\rho^\cW,v_\alpha^{\cW}$.  Then the $S^3$ partition function is obtained by plugging this into the functional $\log Z_{S^3}[\rho,v_\alpha]$, \ie:

\be \log Z_{S^3} = \log Z_{S^3}[\rho^{\cW},v_\alpha^\cW] \;. \ee
\item {\bf Method $2$} (the method used in \cite{Jafferis:2011zi})  - Directly extremize the functional $\log Z_{S^3}[\rho,v_\alpha]$, computing the integrand of the $S^3$ partition function in \eqref{mgpcint}, obtaining eigenvalue distributions $\rho^Z,v_\alpha^{Z}$.  Then the $S^3$ partition function is obtained by:

\be \log Z_{S^3} = \log Z_{S^3}[\rho^{Z},v_\alpha^Z] \;. \ee
\end{itemize}

In general, these two methods give different answers, and one must check which gives the dominant contribution to the partition function.  We see that the second method gives a solution for ABJM at the superconformal point, while the first does not.  However, as we will discuss in Section \ref{sec:gq} below, for certain other models, such as the $V^{5,2}$ model, both methods do lead to solutions.  In general, we suggest the following interpretation of these methods:

\begin{itemize}
\item Method $1$, for general $g$ and $p$, reproduces the on-shell action of the minimal SUGRA ``Bolt'' type solutions, whose boundary is  $\mathcal{M}_{g,p}$, subject to the constraint \eqref{quant_repeated}. 
\item For $\cM_{0,1}=S^3$, Method $2$ gives the on-shell action of the NUT- type solutions.
\end{itemize}
In other words, when one solution gives the dominant contribution to the free energy on the gravity side, the corresponding method gives the dominant contribution to the $\cM_{g,p}$ partition function.  We note that the configurations with the lowest free energy (Bolt$_+$ for \eqref{quant_repeated} and NUT for $g=0,p=1$) are defined for all values of the squashing.

For more general $g$ and $p$, one may ask if we can generalize Method $2$ above by starting from the integral formula, \eqref{mgpcint}, and finding a saddle, as done by \cite{Jafferis:2011zi} for $S^3$.  We comment on the case of $S^3/\Z_2$ in Section \ref{sec:s3comp} below.  For $g>0$, the integral formula \eqref{mgpint} and \eqref{mgpcint} are more subtle, as one must carefully regulate the contribution the vector multiplet at points of enhanced Weyl symmetry, as discussed in \cite{Benini:2016hjo,Closset:2017zgf}.  Thus, we leave the investigation of the large $N$ limit of this integral formula for future work.  However, given the matching of the previous subsection, and the fact that a NUT-type solution does not exist for general $g$ and $p$, we believe this gives strong evidence that Method $1$ gives the correct large $N$ behavior of the $\cM_{g,p}$ partition function for generic $g$ and $p$.   

Finally, let us mention that, for Method $1$, we can naturally take $p \rightarrow 0$: in this limit we retrieve the topologically twisted Witten index on $\Sigma_g \times S^1$ studied in \cite{Benini:2015noa,Benini:2015eyy,Benini:2016hjo}, which was used for the entropy matching of AdS$_4$ black holes. The Bolt solutions, indeed, present a non-trivial 2-cycle $\Sigma_g$ in the bulk at the location of the Bolt, which is threaded by magnetic flux. For $p=0$ (which corresponds to the black hole case) this cycle extends all the way to the boundary. The NUT solutions, whose free energy is retrieved by Method $2$, does not have this feature.


\section{General quivers}
\label{sec:gq}

In this section we discuss the more general quiver gauge theories of Section \ref{sec:qd}, and some new issues that arise in taking the large $N$ limit of their $\cM_{g,p}$ partition functions.  We derive a general result relating the partition function of these gauge theories to the extremized values of the twisted superpotential, generalizing the ``index theorem'' of \cite{Hosseini:2016tor}.  We point out a subtlety in the definition of the $\cM_{g,p}$ partition function in the presence of fractional R-charges, which is necessary for studying general quivers of the above type.  We consider the explicitly the case of the case of M-theory compactified on the $7$-manifold $V^{5,2}$, and the truncation to minimal supergravity in this case.  Finally, we discuss the truncation to minimal supergravity for general quivers of Section \ref{sec:qd}, and describe a generalization of the ``universal twist'' of \cite{Azzurli:2017kxo}.

\subsection{The $\cM_{g,p}$ partition function for general quivers}
\label{sec:mgm}

Using the ingredients provided in Section \ref{sec:largensol}, one can compute the $\cM_{g,p}$ partition function for more general quiver theories of the type discussed in Section \ref{sec:qd}.  It turns out that the final answer can be expressed in terms of the extremal twisted superpotential for the chosen theory a very simple form, given in \eqref{itnew} below.  Let us describe this computation in more detail.

Let us first review the extremization of the twisted superpotential, which was described in \cite{Hosseini:2016tor} and computed in several examples in \cite{Hosseini:2016ume}.  Let us suppose the theory has a superpotential:

\be \label{superpot} W = \sum_\alpha \prod_i {\Phi_i}^{q_i^\alpha} \;,\ee
where $\Phi_i$ is the $i$th chiral multiplet, and we suppress color indices.  Then this imposes the following constraints on the parameters appearing in the $\cM_{g,p}$ partition function:

\be \label{mgpcond} \sum_i q_i^\alpha m_i = \sum_i q_i^\alpha s_i = 0, \;\;\; \sum_i q_i^\alpha r_i =2 \;,\ee
where $m_i, s_i$, and $r_i$ are the mass, flavor symmetry flux, and R-charge, respectively, for the $i$th chiral multiplet.  In \cite{Hosseini:2016tor} it was found that a solution to the extremization of the twisted superpotential exists when we impose (translating to our notation):

\be \label{bmcond} \sum_i q^\alpha_i [m_i] = 1  \;, \ee
where $[m_i]$ is the fractional part of the mass, $m_i$.  For such a choice of $[m_i]$, we may construct an eigenvalue distribution:

\be \label{eev} \rho_*(t),\;\;\;{v^\alpha}_*(t) \;,\ee
which extremizes the functional $\cW[\rho,v_\alpha,[m_i]]$ computing the twisted superpotential.  Then we define the extremized twisted superpotential as:

\be \cW_{ext}([m_i]) \equiv \cW[\rho_*,{v^\alpha}_*,[m_i]] \;.\ee
The extremization of $\cW$ was worked out in several quivers of the above form in \cite{Hosseini:2016ume}.

Suppose we can find a set of parameters satisfying \eqref{mgpcond} and \eqref{bmcond}.  Then we may follow the procedure in Section \ref{sec:largensol} to compute the large $N$ behavior of the $\cM_{g,p}$ partition function.  Namely, we plug the extremal eigenvalue distribution \eqref{eev} into the $Z_{\cM_{g,p}}$ functional computed in Section \ref{sec:mgpfunc}.  We have worked this out in several of the examples considered in \cite{Hosseini:2016ume}, and in all cases we have found the $\cM_{g,p}$ partition function has the following simple relation to the extremal twisted superpotential, $\cW_{ext}([m_i])$:

\bea \label{itnew}  \frac{1}{2 \pi i} & \log Z_{\cM_{g,p}}(m_i,s_i,r_i) \\ &= \big(p +4 (g-1)\big) \cW_{ext}([m_i])  + \sum_i \bigg( -\big(p + 2(g-1)\big) [m_i] + {\frak n}_i  \bigg) \partial_i \cW_{ext}([m_i]) \;, \eea
where ${\frak n}_i$ is as in \eqref{mincond}, \ie:
\be {\frak n}_i =-p n_i + s_i + (g-1) r_i \;, \ee
and $n_i = m_i - [m_i]$ are the integer parts of the masses.  Using \eqref{mgpcond} and \eqref{bmcond}, one can check that these satisfy:

\be \label{fncond} \sum_i q_i^\alpha {\frak n}_i =p + 2(g-1) \;. \ee
For example, plugging in the extremal twisted superpotential for the ABJM theory in \eqref{wextabjm}, one can check that this correctly reproduces the result \eqref{lnabjm}, where in this case the extremal twisted superpotential is given by \eqref{wextabjm}.

Let us make several comments about this formula.  First, in the case $p=0$, it reduces to:

\be \label{itnewsg}  \frac{1}{2 \pi i} \log Z_{\Sigma_g \times S^1} = 4 (g-1) \cW_{ext}([m_i])  + \sum_i \big( -2(g-1) [m_i] + s_i + (g-1) r_i \big) \partial_i \cW_{ext}([m_i]) \;.\ee
This agrees with the ``index theorem'' of \cite{Hosseini:2016tor}.  

Next, recall that, due to the constraint \eqref{bmcond}, the parameters appearing in $\cW_{ext}$ are not independent, and so the derivatives are potentially ambiguous.  Specifically, we can add to $\cW_{ext}([m_i])$ a function $f([m_i])$ which vanishes at the solutions to \eqref{bmcond}, which does not change its value at allowed values of the $[m_i]$, but does change its derivatives.  Then a linear combination of derivatives:

\be \sum_i c_i \partial_i \cW_{ext}\;, \ee
is invariant under such a shift if and only if:

\be \sum_i c_i q^\alpha_i = 0, \;\;\; \text{for all superpotential constraints}\; \alpha\;. \ee
Fortunately, for the combination of derivatives appearing in the expression \eqref{itnew}, one can check that this condition is indeed satisfied, as a consequence of \eqref{bmcond} and \eqref{fncond}.

We also note that for quivers of this type, the extremal twisted superpotential is typically a homogeneous function of the $m_i$ of degree $2$ \cite{Hosseini:2016tor}.\footnote{More precisely, it is possible to fix the ambiguity mentioned above so that this condition holds.}  Indeed, this was the case for the extremal twisted superpotential for the ABJM theory, in \eqref{wextabjm}.  In this case, the expression \eqref{itnew} simplifies to:

\be \label{itnewos}  \frac{1}{2 \pi i} \log Z_{\cM_{g,p}} = p \cW_{ext}([m_i])  + \sum_i \big( -p  m_i + s_i + (g-1) r_i \big) \partial_i \cW_{ext}([m_i]) \;. \ee
In this form, the relation can be understood in terms the on-shell twisted superpotential, discussed in Section \ref{sec:mgpcomp}.  Namely, recall from \eqref{ossp} that we may write the various operators appearing in the sum over Bethe vacua in terms of the on-shell twisted superpotential and effective dilaton.  In the large $N$ limit, where the contribution from one vacuum is dominant, we may restrict our attention to the on-shell twisted superpotential and dilaton in that vacuum.  Then we claim:

\be \cW^{I_{dom}} ([m_i]) = \cW_{ext}([m_i]), \;\;\;\; \Omega^{I_{dom}}([m_i]) = \sum_i r_i \partial_i \cW_{ext}([m_i]) \;. \ee
Then inserting this into \eqref{ossp}, one derives (\ref{itnewos}).

We also note there is a symmetry of the partition function under $m_i \rightarrow -m_i$, $p \rightarrow -p$, $\cW \rightarrow -\cW$, and $k \rightarrow -k$, where $k$ denotes the various Chern-Simons levels \cite{Closset:2017zgf}.  As a result there is an additional solution of the extremization of the twisted superpotential when:
\be \label{bmcond2} \sum_i q^\alpha_i (1-[m_i]) = 1 \;. \ee
This implies:
\be \label{lcond2} \sum_i q^\alpha_i  \hat{\frak n}_i= -p + 2(g-1)\;,\ee
where $\hat{\frak n}_i$ is as in \eqref{hatelldef}.

When these conditions are satisfied, one finds:

\be \label{itnewbm}  \frac{1}{2 \pi i} \log Z_{\cM_{g,p}}= (p - 4 (g-1)) \cW_{ext}([m_i])  + \sum_i \big( -(-p + 2(g-1)) (1-[m_i]) + \hat{\frak n}_i  \big) \partial_i \cW_{ext}([m_i]) \;. \ee

Finally, we must mention an important caveat in the above computation.  We assumed that it was possible to find a set of parameters satisfying \eqref{mgpcond} and \eqref{bmcond}, however, for certain theories one cannot find such a solution in integer R-charges $r_i$.  This forces us to consider the possibility of allowing fractional R-charges, which we turn to next.

\subsection{Fractional R-charges and the R-symmetry background}
\label{sec:frac}

So far we have always assumed that the R-symmetry used to place a theory supersymmetrically on $\cM_{g,p}$ assigned all chiral multiplets integer charges.  This was necessary because our background includes an R-symmetry gauge field with non-zero flux, and so for the chiral multiplets to take values in well-defined line bundles, in general their R-charges must be integers.
However, in some of the quiver gauge theories of Section \ref{sec:qd}, some of the chiral multiplets necessarily have fractional R-charges, due to superpotential constraints, and so such theories naively seem beyond the scope of this partition function.  

Fortunately, while integer R-charges are necessary to define the $\cM_{g,p}$ partition function for all $g$ and $p$, for special choices of $g$ and $p$ we may allow certain fractional R-charges.  However, there are certain subtleties in the definition of the $\cM_{g,p}$ partition function for fractional R-charges which will be important to understand for the large $N$ computation.  The following discussion is somewhat technical, and the main points are summarized at the end of this subsection.

\

To start, let us consider the case $p=0$, corresponding to the partition function on $\Sigma_g \times S^1$.  Then the flux felt by a chiral multiplet of R-charge $r$ includes a contribution $r (g-1)$.  For this to give a well-defined vector bundle, we must then impose:

\be \label{rrest} r (g-1) \in \Z \;. \ee
For non-integer $r$, this restricts the possible values of $g$.  However, when $r$ is rational there are still infinitely many cases we may consider.

Next take $p \neq 0$.  Recall the R-symmetry gauge field, $A_R$, is given by \eqref{rgf}.  This gauge field is part of a background gauge multiplet, $\cV_R$ with the vector component given by a complex gauge field \cite{Closset:2014uda}:\footnote{We thank C. Closset for discussions about this background vector multiplet.} 
\be a_\mu^R = A_\mu^R + V_\mu + i H \eta_\mu \;. \ee
For the background discussed in Section \ref{sec:mgpbg}, we can write this as:
\be \label{aatwist} a^R = - \frac{g-1}{p} \eta + (g-1) \gamma\;, \ee
where we recall $\eta$ is the $1$-form \eqref{etadef}, and $n \gamma$ is the flat $U(1)$ gauge field on $\cM_{g,p}$ with first Chern class $n \in \Z_p$.  As in \eqref{pba}, we may construct this gauge field as the pull-back of the gauge field with flux $g-1$ on $\Sigma_g$, \ie:

\be \label{aatwist2} a^R = \pi^*((g-1) a) \;, \ee
where $a$, defined in \eqref{adef}, satisfies $\frac{1}{2\pi}\int_{\Sigma_g} da = 1$.  Now consider a chiral multiplet $\Phi$ with fractional R-charge, $r$.  Then the gauge field on $\cM_{g,p}$ coupling to this chiral multiplet includes a contribution from the R-symmetry gauge field, which can be obtained by pulling back from $\Sigma_g$, \ie:

\be a_{\Phi} = \pi^*(r (g-1) a)  =  - r \frac{g-1}{p} \eta + r (g-1) \gamma \;. \ee
Then in order for this to be well-defined, we must again impose \eqref{rrest}. 

However, recall from \eqref{flxshift} that we may equivalently write the gauge field \eqref{aatwist} as:
\be \label{aatwist3} a^R = \nu_R \eta + \pi^*(n_R a) \;, \ee
for $n_R \in \Z$ and $\nu_R \in \R$ satisfying:
\be \label{nurnrc} -p \nu_R + n_R = g-1 , \;\;\; n_R = g-1 \; \MOD{p} \;. \ee
The expression \eqref{aatwist2} corresponds to a particular choice, called the ``$A$-twist gauge'' in \cite{Closset:2017zgf}, where:
\be \label{atg} \nu_R=0, \;\;\; n_R=g-1 \;. \ee
For a general choice of $(\nu_R,n_R)$ satisfying \eqref{nurnrc}, the R-symmetry gauge field coupling to a chiral multiplet is
\be a_{\Phi} =  r \nu_R \eta + \pi^*(r n_R a)  = - r \frac{g-1}{p} \eta + r n_R \gamma \;, \ee
and the condition for this to be well-defined is:
\be \label{rq} r \; n_R \in \Z \;. \ee
Thus, by changing the parameters $(\nu_R,n_R)$, we may find backgrounds which allow more general choices of fractional R-charge.\footnote{See Appendix C of \cite{Closset:2017bse} for a related discussion of the partition function for various choices of R-symmetry gauge.}   For example, when $g-1=0\MOD{p}$, we can take the ``physical gauge'' of \cite{Closset:2017zgf}:
\be \label{phga} \nu_R=\frac{1-g}{p},\;\; n_R= 0\;,\ee
and then we may take any $r \in \R$.

For a general choice of $(\nu_R,n_R)$, one finds that the contribution of a chiral multiplet is given by:

\be \label{zchigg} \cF_\chi(m + r \nu_R)^p \Pi_\chi(m+ r \nu_R)^{r n_R + s +1-g} \;. \ee
This reproduces \eqref{zchi} in the $A$-twist gauge, \eqref{atg}.  Moreover, using \eqref{fde1}, one can check that the partition function \eqref{zchigg} does not depend on the parameters $(\nu_R,n_R)$ satisfying \eqref{nurnrc} when the R-charges are integers.  However, for fractional R-charges, the partition function will in general depend on these parameters.

To be more explicit, let us write the superpotential as in \eqref{superpot}, which imposes the constraints in \eqref{mgpcond}, which we reproduce here:

\be \label{mgpcond2} \sum_i q_i^\alpha m_i = \sum_i q_i^\alpha s_i = 0, \;\;\; \sum_i q_i^\alpha r_i =2 \;.\ee
If we work in a general R-symmetry background, as in \eqref{zchigg}, it is natural to define the parameters:
\be \tilde{m}_i = m_i + \nu_R r_i \in \C, \;\;\; \tilde{\frak n}_i = r_i n_R + s_i \in \Z \;,\ee
and these satisfy:
\be \label{mscond} \sum_i q_i^\alpha \tilde{m}_i = 2{\nu}_R , \;\;\;\; \sum_i q_i^\alpha \tilde{\frak n}_i = 2 n_R \;. \ee

The conditions \eqref{mscond} carve out different subsets of the space of allowed masses and fluxes, depending on the choice of $\nu_R$ and $n_R$.  However, many of these spaces can be related using the difference equation \eqref{fde1}.  For example, when it is possible to take all of the R-charges to be integer, that is, when it is possible to find a solution in integers $\gamma_i$ to:
\be \label{nsol} \sum_i q_i^\alpha \gamma_i = 2 , \;\;\;\;\; \gamma_i \in \Z \;, \ee
then, using the difference equation \eqref{fde1}, we may shift the mass parameters and fluxes by:
\be \tilde{m}_i \rightarrow \tilde{m}_i + \gamma_i, \;\;\;\; \tilde{\frak n}_i \rightarrow \tilde{\frak n}_i +  p \gamma_i \;,\ee
without changing the partition function.  Then this modifies the condition \eqref{mscond} by shifting $(\nu_R,n_R) \rightarrow (\nu_R+1,n_R+p)$.   Since this is just a redefinition of parameters, these correspond to equivalent backgrounds on $\cM_{g,p}$.  Note that this argument relies only on the {\it existence} of some choice of integer R-charges, but the actual R-charge we pick need only satisfy \eqref{rq}.

However, when \eqref{nsol} cannot be solved, the set of solutions to \eqref{mscond} for different $\nu_R$ are not related by a redefinition of parameters.  Then the $\cM_{g,p}$ partition function for different choices of $\nu_R \in \Z$ will give rise to inequivalent partition functions. 

\subsection*{Changing the spin-structure}

In the case of $p$ even, we may consider yet another choice of R-symmetry background, which is:
\be \label{nur12} \nu_R = \frac{1}{2}, \;\;\; n_R = g-1 + \frac{p}{2} \;. \ee
Then the R-symmetry gauge field is:
\be \label{aatwist3} a^R =  - \frac{g-1}{p} \eta +(g-1) \gamma + \frac{p}{2} \gamma \;.\ee
As above, this does not affect the field strength of the R-symmetry gauge field.  However, unlike the redefinitions above, this does change the R-symmetry gauge field itself by adding an additional flat connection, $\frac{p}{2}\gamma$, with holonomy $-1$ around the $S^1$ fiber.  In general, the presence of this additional holonomy means the Killing spinor equation, \eqref{ks2}, no longer has a solution.  

However, when $p$ is even, $H_1(\cM_{g,p},\Z_2)\supset\Z_2$, and there is an additional choice for the spin structure on $\cM_{g,p}$, differing in the sign the fermions incur as we wind around the $S^1$ fiber.  Thus, if we accompany this shift of the R-symmetry gauge field by a change of spin structure, the Killing spinor equation is unchanged, and so this defines a valid background for $p$ even.  This ambiguity in the choice of spin structure and R-symmetry gauge field was discussed in the context of holography in \cite{Martelli:2012sz}.

Note that if we take the R-charges of all chiral multiplets to be even integers, then we find the contribution of a chiral multiplet for this R-symmetry background, \eqref{zchigg}, is equal to that in the $A$-twist gauge:

\be  \cF_\chi(m + r \frac{1}{2})^p \Pi_\chi(m+ r \frac{1}{2})^{r (\frac{p}{2} + g-1) + s_i +1-g} = \cF_\chi(m)^p \Pi_\chi(m)^{(r-1)(g-1) + s_i }  \;, \ee
where we used \eqref{fde1}.  Thus, for $r \in 2\Z$, these backgrounds are equivalent.  This follows because, for such a choice of R-symmetry, all the bosons have even R-charge, and the fermions have odd R-charge, and so the effect of the change of spin structure and R-symmetry connection cancel for all fields, just as they did for the Killing spinor.   More generally, if we can find a solution to:
\be \label{nsol2} \sum_i q_i^\alpha \gamma_i = n , \;\;\;\;\; \gamma_i \in \Z \;. \ee
Then, if we shift:
\be \label{gsg} \tilde{m}_i \rightarrow \tilde{m}_i + \gamma_i, \;\;\; \tilde{\frak n}_i \rightarrow \tilde{\frak n}_i + p\gamma_i \;, \ee
this takes $(\nu_R,n_R) \rightarrow (\nu_R+\frac{n}{2},n_R+\frac{n p}{2})$.  In particular, if we can solve this for $n=1$, all choices of $\nu_R \in \frac{1}{2} \Z$ are equivalent.  However, when this is not possible, these different choices of $\nu_R \in \frac{1}{2} \Z$ in general lead to inequivalent backgrounds.  We will see below that the background \eqref{nur12} will play a special role in the large $N$ solution.\footnote{Below we will consider the background \eqref{nur12} even in the case where $p$ is not even.  While this is somewhat formal, it can be justified in cases, as in \eqref{gsg}, where we may use the shift symmetry, \eqref{fde1}, to relate backgrounds with different values of $\nu_R$.}

\subsection*{$\cM_{g,p}$ partition function for a general R-symmetry background}

With these observations, let us return to the computation of the $\cM_{g,p}$ partition function of the quiver gauge theories of Section \ref{sec:qd}.  First, we note that, for finite $N$, the computation of the $\cM_{g,p}$ partition function of a gauge theory in a general background $(\nu_R,n_R)$ is a straightforward modification of the procedure in the $A$-twist gauge, outlined in Section \ref{sec:mgpcomp}.  Namely, we simply make the following replacements in \eqref{mig}:
\be m_i \rightarrow \tilde{m}_i, \;\;\; s_i + r_i (g-1) \rightarrow \tilde{\frak n}_i \;. \ee
Then the $\cM_{g,p}$ partition function will depend on the parameters $\tilde{m}_i$ and $\tilde{\frak n}_i$.  In a general gauge, these parameters satisfy the constraints:
\be \label{mscond2} \sum_i q_i^\alpha \tilde{m}_i = 2{\nu}_R , \;\;\;\; \sum_i q_i^\alpha \tilde{\frak n}_i = 2 n_R \;. \ee
Then the only difference between the $\cM_{g,p}$ partition function for different backgrounds, $(\nu_R,n_R)$, is the set of constraints, \eqref{mscond2}, which we take the parameters to satisfy.

\

Now let us return to the large $N$ solution described in Section \ref{sec:mgm} above.  Recall that, for this solution to apply, we must find a set of parameters, $m_i,s_i,r_i$, satisfying \eqref{mgpcond}, \ie:
\be \label{mgpcond2} \sum_i q_i^\alpha m_i = \sum_i q_i^\alpha s_i = 0, \;\;\; \sum_i q_i^\alpha r_i =2 \;, \ee
as well as:
\be \label{bmcond2} \sum_i q^\alpha_i [m_i] = 1 , \;\;\; \sum_i q_i^\alpha {\frak n}_i = p + 2(g-1) \;, \ee
where:
\be \label{mdef} m_i = [m_i]+n_i, \;\;\; {\frak n}_i = -p n_i + s_i + r_i(g-1) \;. \ee
In general, it is not possible to solve all of these conditions at once.  For example, if the theory contains a superpotential term, $W=\Phi^3$, then \eqref{mgpcond2} and \eqref{bmcond2} imply the mass $m$ of $\Phi$ satisfies, respectively:
\be 3 m =0 \;\;\; \text{and}\;\;\;\; 3[m] = 1 \;, \ee
which are clearly incompatible.  However, given the discussion above, we see that the condition \eqref{mgpcond2} is the condition appropriate for the $A$-twist gauge, while for a more general background, we should instead impose the constraint \eqref{mscond2}.  Comparing to \eqref{bmcond2}, we see that if we work in the background \eqref{nur12}:
\be \label{nur12g} \nu_R = \frac{1}{2}, \;\;\; n_R =\frac{p}{2} + g-1 \;,\ee
where we must impose:

\be \label{nur12cond} \sum_i q^\alpha_i \tilde{m}_i = 1 , \;\;\; \sum_i q_i^\alpha \tilde{\frak n}_i = p + 2(g-1)\;, \ee
then we may identify:
\be  \tilde{m}_i \rightarrow [m_i], \;\;\;\; \tilde{\frak n}_i \rightarrow {\frak n}_i \;. \ee
In this background, we may always solve the constraints and find a large $N$ solution.  When a solution also exists in the $A$-twist gauge background, this is related to the one above by a shift of mass parameters as in \eqref{gsg}.  This is the case, for example, for the ABJM theory.   But even when no such solution exists in the $A$-twist gauge, we can still find a solution in the background \eqref{nur12g}.   We will consider an example of such a theory in the next subsection.  For notational simplicity, we will drop the tildes on $\tilde{m}_i$ and $\tilde{\frak n}_i$ below when the meaning is clear from context.

\

To briefly summarize the above discussion, we have seen that in the presence of chiral multiplets with fractional R-charges, there are subtly different choices of supersymmetric backgrounds on $\cM_{g,p}$, which can be parameterized by the parameters $\nu_R$ and $n_R$, and which impose the constraints \eqref{mscond2} on the parameters entering the partition function.  The restrictions \eqref{bmcond} and \eqref{fncond} which allowed us to find a large $N$ solution for the $\cM_{g,p}$ partition in Section \ref{sec:mgm} naturally force us to work with an R-symmetry background, \eqref{nur12g}, which is different than the $A$-twist gauge used in Section \ref{sec:mgpbg}.  In some cases, such as for the ABJM theory, these two backgrounds are equivalent, but in general they are distinct.

\

\subsection{Example: $V^{5,2}$ theory \label{sec:V52}}

To illustrate some of these features more concretely, let us now look at another example of a quiver gauge theory.  We take the case where the bulk gravity theory comes from the compactification of M-theory on the manifold $V^{5,2} = SO(5)/SO(3)$, whose Fano index is $I(V^{5,2}) =3$.  The dual field theory, which we will call the $V^{5,2}$ theory, has a description \cite{Martelli:2009ga} with gauge group $U(N)_{k=1} \times U(N)_{k=-1}$ with two pairs of bifundamental chiral multiplets, $A_i$ and $B_i$, $i=1,2$, in the $(N,\bar{N})$ and $(\bar{N},N)$ representations, respectively, as well as an adjoint field, $\Phi_\alpha$, for each of the two $U(N)$ factors, and superpotential:

\be W = \text{Tr} \big( \sum_\alpha {\Phi_\alpha}^3 + \sum_{i=1}^2 (B_i \Phi_1 A_i + A_i \Phi_2 B_i) \big) \;. \ee
Note the presence of the $\Phi^3$ superpotential term, which we saw in the previous section implies we may not work in the $A$-twist gauge background.  Instead we work in the background $(\nu_R=\frac{1}{2},n_R=\frac{p}{2}+g-1)$, as discussed above.  We denote the mass parameters by $m_{A_i}, m_{B_i}$, and $m_{\Phi_\alpha}$, and similarly for the fluxes ${\frak n}$.  Then to find a solution at large $N$ we impose:

\be \label{v52mcond} m_{A_i}+ m_{B_i} = \frac{2}{3} , \;\;\; m_{\Phi_\alpha} = \frac{1}{3} \;,\ee
$$ {\frak n}_{A_i} + {\frak n}_{B_i} = \frac{2}{3}(p+2(g-1)), \;\;\; {\frak n}_{\Phi_\alpha} = \frac{1}{3}(p+2(g-1)) \;. $$
Note that to find a solution with ${\frak n}_{\Phi_\alpha} \in \Z$, we must impose:
\be \label{v52gpconst} p+2 (g-1)  = 0 \MOD{3} \;. \ee
This is an example of the general feature, discussed above, that in the presence of fractional R-charges, we may only define the $\cM_{g,p}$ partition function for certain choices of $g$ and $p$.  The extremal twisted superpotential for this theory was computed in \cite{Hosseini:2016ume}, with the result:\footnote{We thank Hyojoong Kim for pointing out a factor of $\sqrt{2}$ missing in the original formulas, \eqref{cWv52} and below. The conclusions of this section, in particular eq. \eqref{InutV52}, remain unchanged.}
\be \label{cWv52} \cW = -\frac{2 i N^{3/2}}{3} \sqrt{m_{A_1} m_{A_2} m_{B_1} m_{B_2}} \;. \ee
Inserting the extremal eigenvalue distribution into the $\cM_{g,p}$ partition functional of Section \ref{sec:mgpfunc}, we obtain the result:

\be \log Z_{\cM_{g,p}}^{V^{5,2}} = \frac{2\pi N^{3/2}}{3} \sqrt{m_{A_1} m_{A_2} m_{B_1} m_{B_2}} \bigg( -2 p + \sum_i \big( \frac{{\frak n}_{A_i}}{m_{A_i}}+\frac{{\frak n}_{B_i}}{m_{B_i}} \big)\bigg)\;. \ee
We have computed this explicitly using the prescription of Section \ref{sec:largensol}, but we note it is consistent with the general result \eqref{itnew}, giving additional evidence for this conjecture.

Now let us discuss the truncation to minimal supergravity.  Similarly to the ABJM case, we look for solutions with:

\be m_{A_i} = m_{B_i} \;\;\;\text{and} \;\;\; {\frak n}_{A_i} = {\frak n}_{B_i}  \;\;\; \text{independent of} \; i \;. \ee
Then from \eqref{v52mcond}, we see we must impose:

\be m_{A_i} = m_{B_i} = \frac{1}{3}, \;\;\; {\frak n}_{A_i} = {\frak n}_{B_i} = \frac{1}{3}(p+2(g-1)) 
\;. \ee
In this case, the $\cM_{g,p}$ partition function simplifies to:

\be \label{mgpv52min} \log Z_{\cM_{g,p}}^{V^{5,2}} = \frac{4\pi N^{3/2}}{27}  \big( p+4(g-1) \big) \;.\ee
As in the ABJM case, there is an additional solution with $m \rightarrow 1-m$, where:
\be \label{mgpv52min2} \log Z_{\cM_{g,p}}^{V^{5,2}} = \frac{4\pi N^{3/2}}{27}  \big(- p+4(g-1) \big), \;\;\;\;\; -p+2 (g-1)  = 0 \MOD{3} \;. \ee

Let us now discuss the dual supergravity solutions.   In the $g=0, p=1$ case, the NUT free energy, when this configuration is embedded in a 11d theory compactified on the $V^{5,2}$ manifold, is given by
\be \label{InutV52}
I_{NUT} =  \frac{ \sqrt2 \pi N^{3/2}}3 \sqrt{ \frac{\pi^4}{3\, \text{Vol}(Y_7) }  }= \frac{16\pi  \, N^{3/2}}{27} \;,
\ee
where we have used the fact that \cite{Bergman:2001qi}
\be
\text{Vol}(V^{5,2}) = \frac{27 \pi^4}{128} \;.
\ee
The result \eqref{InutV52}  agrees with the $S^3$ partition function of the dual quiver theory computed in \cite{Martelli:2011qj}. 

Bolt solutions, instead, have free energy
\be \label{IboltV52}
I_{Bolt,\pm} =  \frac{4 \pi N^{3/2}}{27}( 4 (1-g) \mp p) \;,
\ee
and, in order to have a regular uplift to eleven dimensions on $V^{5,2}$  In this case, they are subject to the following constraint (see Section \ref{11uplift}): 
\be \label{quant-V52again}
\pm p  + 2(g-1) =0 \MOD{3} \;.
\ee
We see that the result \eqref{IboltV52} for the Bolt$+$ is reproduced by the $\mathcal{M}_{g,p}$ partition function of the $V^{5,2}$ theory computed above in \eqref{mgpv52min}, $I_{Bolt} = -  \log Z_{\cM_{g,p}}^{V^{5,2}} $, and we also find the same constraint on $p$ and $g$ as in \eqref{v52gpconst}, with a similar relation for the Bolt$-$ and \eqref{mgpv52min2}.   

However, something unexpected happens for $p=1$. From the lower sign of \eqref{quant-V52again} we see that $p=1$ Bolt$_-$  are a solution of the theory, namely they lift up to M-theory compactified on $V^{5,2}$. Their free energy is 
\be \label{54bolt-}
I_{Bolt_-}^{p=1} =  \frac{20 \pi N^{3/2}}{27} \;,
\ee
which is higher than that of the corresponding NUT solution \eqref{InutV52}
\be
I_{Bolt_-}^{p=1} = \frac54 I_{NUT} \;.
\ee

We are left then with the following puzzle: the NUT and $p=1$ Bolt$_-$ have the same boundary data, in particular they are both regular\footnote{We remind that the Bolt$_-$ solution exist only for the range of squashing $s \in [0, \frac{\sqrt{5-2\sqrt6}}{4}]$.} fillings for the squashed $S^3$. However, our field theory result for these configurations exactly matches with \eqref{54bolt-}. The field theory computation carried out previously in \cite{Martelli:2011qj}, which utilized a similar method to that of \cite{Jafferis:2011zi}, matches \eqref{InutV52}.  The two methods used here are the same as those discussed in Section \ref{sec:NB}.  As we observed there, it seems that the two methods reproduce different saddle points, which are realized  in the dual supergravity picture in form of two solutions with different topology, namely the NUT and $p=1$ Bolt configurations.  

In this regard, it would be interesting to retrieve the $\mathcal{M}_{g,p}$ partition function for the quiver dual to the M-theory reduction on the SE7 manifold $M^{32}$ \cite{Martelli:2008si,Benini:2011cma}. In that case, there is no restriction for the uplift: Bolt solutions are present for every $p$.  In particular, the Bolt$_+$ uplifts in the case with boundary $S^3$, and has a lower free energy than the NUT solution. The $S^3$ free energy for this quiver was computed in \cite{Amariti:2011uw,Gang:2011jj} and once again it coincides with the on-shell action of the corresponding NUT solution. Given that this theory is chiral, the matrix model is not under good computational control with current methods and therefore computing the $\mathcal{M}_{g,p}$ partition function is a difficult task. We leave this for future investigation.

\subsection{Minimal supergravity for general quivers and the universal twist}

Now consider a general quiver of the type studied in Section \ref{sec:qd}.  We would like to find a general analogue of the truncation to minimal supergravity discussed for the ABJM and $V^{5,2}$ models above.  To do this, we make the following assumption.  For general mass parameters, $m_i$, we expect the partition function we have computed to be dual to a configuration in gauged supergravity.  More precisely, in the case $p=0$ it was argued in \cite{Benini:2015eyy} that one must first extremize the partition function as a function of the mass parameters, $m_i$, in order to compare to the supergravity solution.  This reflects the attractor mechanism in the gravity theory, which fixes the behavior of scalar fields in the bulk as they approach the horizon \cite{Benini:2015eyy,Benini:2016rke,Cabo-Bizet:2017xdr}  However, in the minimal case, there are no scalars, and so no such extremization is necessary.  

We conjecture that a similar story holds for general $p$.  Namely, we expect that for the case of generic $m_i$, an extremization principle is necessary to compare the result to a solution in gauged supergravity.  We leave the study of this general, non-minimal case for future work.  However, we argue that the truncation to minimal supergravity corresponds to the case where we have chosen the $m_i$ such that no extremization is necessary.  

Let us see how this works in more detail.  We have argued in Section \ref{sec:mgm} that the large $N$ solution is given by:

\be \label{itnew4}  \frac{1}{2 \pi i} \log Z_{\cM_{g,p}}(m_i) = (p +4 (g-1)) \cW_{ext}(m_i)  + \sum_i \big( -(p + 2(g-1)) m_i + {\frak n}_i \big) \partial_i \cW_{ext}(m_i) \;. \ee
In examples above we specialized to the minimal supergravity case by choosing judicious values of the $m_i$ and ${\frak n}_i$.  

For a general theory, we may take an approach similar to that in \cite{Azzurli:2017kxo}.   As observed in \cite{Hosseini:2016tor}, and as we  will discuss in more detail in Section \ref{sec:s3comp}, for a general quiver of the type above, one finds the relation:

\be \label{ws3r4} \frac{1}{2\pi i} \log Z_{S^3}(\Delta_i) = -4 \cW_{ext}(m_i) , \;\;\;\; \text{for}\;\;\Delta_i =2 m_i \;, \ee
where $\log Z_{S^3}(\Delta_i)$ is the partition function on the round sphere as a function of continuous trial R-charges, $\Delta_i \in (0,2)$.  Then $F$-maximization \cite{Jafferis:2010un} implies that the twisted superpotential is extremized, as a function of the $m_i$, when we set:

\be m_i = \frac{1}{2} \Delta^{SC}_i \;, \ee
where $\Delta_i^{SC}$ are the superconformal R-charges.  Now suppose that we also set:
\be {\frak n}_i  = \frac{1}{2} \Delta^{SC}_i (p +2(g-1) ) \;. \ee
This implies that the second term in \eqref{itnew4} vanishes.  Moreover, the first term is already extremized as a function of the $m_i$, by the argument above, and so applying the extremization procedure does not affect $Z_{\cM_{g,p}}(m_i)$.  Thus, we conjecture that this gives a general solution to the truncation to minimal supergravity.  In this case, we find the general relation:

\be  \frac{1}{2 \pi i} \log Z_{\cM_{g,p}}(m_i) = (p +4 (g-1)) \; \cW_{ext}(m_i = \frac{1}{2}  \Delta^{SC}_i) \;. \ee
Using \eqref{ws3r4} and the fact that the $S^3$ free energy agrees with the free energy of the NUT solution, we may rewrite this as:

\be \label{mtfe} \log Z_{\cM_{g,p}}(m_i) = - \frac{1}{4}(p +4 (g-1)) \; \cI_{NUT} \;. \ee
Note the dependence on $p$ and $g$ is precisely the one we found in supergravity above.  This can be thought of as a generalization of the ``universal twist'' of \cite{Azzurli:2017kxo} to the case $p \neq 0$.  

Moreover, in order for the fluxes to be well-defined, we must impose:

\be \label{msgtc} \frac{1}{2} \Delta^{SC}_i (p +2(g-1) ) \in \Z \;. \ee
When $\Delta^{SC}_i$ is rational, as in the cases above, this allows for families of solutions.  Note the free energy, \eqref{mtfe}, and uplift condition, \eqref{msgtc} agree with those found in the special cases of the ABJM and $V^{5,2}$ models above.  We conjecture these map to the M-theory uplift conditions for the corresponding supergravity backgrounds.

Finally, we note that there is a similar truncation corresponding to the Bolt$_-$ solution.  Here we work in the background $(\nu_R = -\frac{1}{2},n_R =-\frac{p}{2} + g-1)$, and find:

\be \label{msbm}  \frac{1}{2 \pi i} \log Z_{\cM_{g,p}}(m_i) = (-p +4 (g-1)) \; \cW_{ext}(m_i =- \frac{1}{2}  \Delta^{SC}_i) , \;\;\;\;  \frac{1}{2} \Delta^{SC}_i (-p +2(g-1) ) \in \Z \;. \ee
In cases where both truncations exist for the same $g$ and $p$, such as for the ABJM theory discussed in Section \ref{sec:holo_checks}, these typically correspond to different choices of masses and fluxes in the field theory, leading to two different truncations to minimal supergravity.  In this case, one of the solutions (namely, the Bolt$_+$ for $p>0$ or the Bolt$_-$ for $p<0$) is dominant over the other, and this will be picked out as the preferred solution when we perform the extremization of masses outlined at the beginning of this subsection.

\section{Comparison with $S^3$ partition function}
\label{sec:s3comp}

In this final section, we briefly discuss a curious relation between the extremal value of the twisted superpotential, $\cW_{ext}([m_i])$, and the large $N$ limit of the $S^3$ partition function, as a function of trial R-charges $\Delta_i$, which was first observed by \cite{Hosseini:2016tor}, and which was used at the end of the previous section.  Namely, they observed that:\footnote{Here and below we rewrite their observations in our notation.}

\be \label{obs} \frac{1}{2 \pi i} \log Z_{S^3}(\Delta_i) = -4 \cW_{ext}([m_i]) \;,  \ee
where $Z_{S^3}$ is the $S^3$ partition function, which was computed in the large $N$ limit by \cite{Jafferis:2011zi}, provided that we identify the trial R-charges $\Delta_i$ and the $[m_i]$ via:

\be \label{d2m} \Delta_i = 2 [m_i] \;. \ee
In particular, recall one can find a solution to the extremization of $\cW$ when one imposes the following conditions, for the superpotential constraints \eqref{superpot}:

\be \sum_i  q_i^\alpha [m_i] = 1  \;\;\; \Leftrightarrow \;\;\; \sum_i q_i^\alpha \Delta_i = 2  \;, \ee
which reproduces the constraint on the R-charges appearing in the $S^3$ partition function.  Since the $S^3$ partition function appears as a special case of the $\cM_{g,p}$ partition function, namely $S^3=\cM_{g=0,p=1}$, it is natural to ask if this relation can be understood in our framework.  

More precisely, it was argued in \cite{Closset:2017zgf} that the $\cM_{g=0,p=1}$ partition function in the ``physical gauge'', \eqref{phga}, reproduces the standard partition function on the round $S^3$ \cite{Kapustin:2009kz,Jafferis:2010un,Hama:2010av}, namely:

\be Z_{\cM_{g=0,p=1}}(m_i) = Z_{S^3}(\Delta_i),\;\;\;\; \text{where}\;\; \Delta_i = m_i \;. \ee
However, note that the identification between the mass parameters and trial R-charges here is not the one, \eqref{d2m}, appearing in the relation \eqref{obs}.  Moreover, we saw in Section \ref{sec:holo_checks} that, in the case of the superconformal R-charges for the ABJM theory, $\Delta_i=\frac{1}{2}$, we did not find a solution for the $\cM_{g=0,p=1}$ partition function using our method, while the relation \eqref{obs} above still holds in this case.  Thus, we do not see a direct relation between our computation of the $\cM_{g=0,p=1}$ partition function and the observation \eqref{obs}.  

However, it turns out we can approach this relation from another point of view, through the lens space $S^3/\Z_2$.  The partition function on this space was first computed by Benini, Nishioka, and Yamazaki  \cite{Benini:2011nc} (see also \cite{Alday:2012au}), who considered more generally the case $S^3/\Z_p$.  In general, their background on $S^3/\Z_p$ is different from the one considered above and in \cite{Closset:2017zgf}, as the latter includes a flat R-symmetry gauge field, while the former does not.  However, we claim that one actually recovers the partition function of BNY in the case $p=2$ from the $\cM_{g,p}$ partition function if one takes the background of \eqref{nur12}, \ie:

\be \label{nur122} \nu_R = \frac{1}{2}, \;\;\; n_R = 0  \;. \ee
Then \eqref{rq} implies we may take the $r_i \in \R$.  To see this, we observe that the partition function of a chiral multiplet on the $S^3/\Z_2$ background of \cite{Benini:2011nc,Alday:2012au} can be written as:

\be Z^{BNY}_{\chi,S^3/\Z_2}(m,s;r) = \cF_\chi( \frac{m+r}{2})^2 \Pi_\chi(\frac{m+r}{2})^{1+s} \;, \ee
where $s \in \Z_2$ is the torsion flux and $r \in \R$ is the R-charge, as one can straightforwardly show from the infinite product formula for the chiral multiplet given in \cite{Benini:2011nc,Alday:2012au}.  This agrees with \eqref{zchigg} up to a redefinition $m \rightarrow m/2$.  One may check this relation also holds for the Chern-Simons and gauge contributions, and so we find, for a general gauge theory:

\be Z^{BNY}_{S^3/\Z_2}(m_i,s_i,r_i) = Z_{\cM_{g=0,p=2}}(\frac{m_i}{2},s_i,r_i) \;\;\;\; \text{for} \;\; (\nu_R,n_R)=(\frac{1}{2},0) \;. \ee 

Recall the background \eqref{nur122} is precisely the one which allows us to take a large $N$ limit of the $\cM_{g,p}$ partition function for a general quiver.  Then, from \eqref{itnew} specialized to the case $g=0,p=2$, we find:

\be \frac{1}{2 \pi i} \log Z_{\cM_{g=0,p=2}}(m_i,{\frak n}_i)= -2 \cW_{ext}([m_i])  + \sum_i {\frak n}_i   \partial_i \cW_{ext}([m_i]) \;. \ee
Here we impose:
\be \sum_i q_i^\alpha [m_i] =1, \;\;\; \sum_i q_i^\alpha {\frak n}_i = 0 \;. \ee
Then a simple choice for the ${\frak n}_i$ is to take them all to be zero, in which case we find the result:
\be \label{lens2} \frac{1}{2 \pi i} \log Z_{\cM_{g=0,p=2}}(m_i,{\frak n}_i=0)= -2 \cW_{ext}([m_i])  \;. \ee

In fact, we can make a stronger statement.  We claim that in the case ${\frak n}_i=0$, and for arbitrary $m_i$, the functionals computing $-2\cW$ and $\frac{1}{2 \pi i} \log Z_{\cM_{g=0,p=2}}$ for arbitrary eigenvalue distributions are identical.  For example, for the contribution of a bifundamental chiral multiplet, we have, from Section \ref{sec:largensol}:
\be \cW_{bif}  = i N^{3/2} \int dt \rho(t)^2 g([ \delta v + m]) \;, \ee
\be  \frac{1}{2 \pi i} \log Z_{\cM_{g=0,p=2}}^{bif} = i N^{3/2} \int dt \rho(t)^2 G_\ell(\delta v + m) \;, \ee
where recall $g(u)=-\frac{1}{12} u(u-1)(2u-1)$, and in the present case:
\be G_\ell(u) = 4 g([u]) - (2 [u] -1 ) g'([u]) \;. \ee
Then one finds the difference of the two functional gives:
\be -2 \cW_{bif} -  \frac{1}{2 \pi i} \log Z_{\cM_{g=0,p=2}}^{bif}  =  i N^{3/2} \int dt \rho(t)^2 \frac{1}{12} (2 [\delta v + m] - 1)  \;. \ee
However, when we impose the constraint \eqref{bfconst} and sum over all bifundamental chirals, we find that this vanishes.  One can similarly check the other ingredients in the two functionals agree, and so we have:

\be -2\cW_{bif}[\rho,v^\alpha,m_i] = \frac{1}{2 \pi i} \log Z_{\cM_{g=0,p=2}}[\rho,v^\alpha ,m_i,{\frak n}_i=0 ] \;. \ee
This immediately implies that their extremized values, \eqref{lens2}, agree.  Moreover, it implies that the two methods discussed in Section \ref{sec:NB} will give the same result on this space, since one is extremizing the same functional in both cases.  

This is indeed what happens for the supergravity solutions. For $p=2$ the Bolt$_+$ has zero flux, and its on-shell action coincides with the NUT$/\mathbb{Z}_2$  one. The two branches  are actually continuously connected: they join at the point in phase space corresponding to $s= 1/2\sqrt2$.

For instance for ABJM we have:
\be
I_{Bolt_+\, p=2}^{S^7} = -\frac{\pi N^{3/2} \sqrt{2} }{24} \big( 2 p - 8 \big)= \frac{\pi  N^{3/2} }{3 \sqrt2 } \,,
\qquad
I_{NUT/\mathbb{Z}_2}^{S^7} =  \frac12 I_{NUT}^{S^7} = \frac{ \pi N^{3/2} } {3 \sqrt2 }\,.
\ee
Given the proposed correspondence between the two methods and the free energy of the NUT and Bolt solutions, respectively, this suggests that these two supergravity solutions will agree for the case with conformal boundary $S^3/\Z_2$, not just for minimal supergravity, but for general $\cN=2$ gauged supergravity theories.  It would be interesting to explore the special role played by $S^3/\Z_2$ in more detail. 

Finally, it was argued in \cite{Alday:2012au} that the large $N$ limit of the partition function on the BNY lens space, $S^3/\Z_p$, is related to that of $S^3$ by:

\be \label{alday} \log Z_{S^3/\Z_p} = \frac{1}{p} \log Z_{S^3}  \;. \ee
This relation arises because the bulk fillings for these boundary manifolds are simply $\Z_p$ quotients of $AdS_4$, and so their on-shell action is related by a factor of $\frac{1}{p}$ to the bulk dual of the $S^3$ partition function.  Then if we combine \eqref{alday} with \eqref{lens2}, we arrive at the same relation \eqref{obs}, noted by \cite{Hosseini:2016tor}.


\section{Discussion}
\label{sec:disc}

In this paper we accomplished the task of computing the large $N$ limit of the $\mathcal{M}_{g,p}$ partition function for three-dimensional SCFTs with holographic M-theory duals and showing that it matches with the on-shell action of the minimal supergravity Bolt solutions. A few remarks are in order here.

First of all, in solving the supergravity equations of motion we have made some assumptions on the form of the solutions. For instance, for $g=0$ we have imposed $SU(2) \times U(1)$ symmetry in the bulk, and we looked for configurations with a real metric. In principle we cannot exclude the existence of further M-theory solution with the same boundary data, which yield a different value for the on-shell action. It would be interesting to investigate further the phase space of solutions obtained when one releases one or more assumptions mentioned above.

One obvious extension of this work consists of incorporating vector multiplets in the gravity side, namely working with Bolt configurations which arise as solutions of $\mathcal{N}=2$ gauged supergravity with vector multiplets. Solutions of the $U(1)$ Fayet-Iliopoulos STU model can be lifted up to 11 dimensions \cite{deWit:1982bul,Cvetic:1999xp,Azizi:2016noi}, and the analysis of Euclidean NUT or Bolt solutions in this framework is currently work in progress. Notice that Lorentzian solutions with non-trivial scalar profiles and NUT charge in theories of FI-gauged supergravity were found in \cite{Colleoni:2012jq,Erbin:2015gha}.

With the addition of vectors and scalars, we expect a more intricate phase space to arise for the NUT and Bolt configurations. Phase transitions involving different solutions can arise\footnote{See for instance \cite{Bobev:2016sap} and \cite{Bobev:2017asb} for further studies on this topic.} and this should be reflected in the field theory computation.  It would also be interesting to study the quantum corrections to the on-shell action, and match them to the finite $N$ result in the field theory, as was done recently in the context of AdS$_4$ black holes in \cite{Liu:2017vbl}. 

As discussed in section \ref{sec:holo_checks}, it would be interesting to study the large $N$ limit of the integral formula \eqref{mgpcint} for the $\cM_{g,p}$ partition function, and compare this to the large $N$ limit we found using the sum formula \eqref{mgpsv}.  In particular, we may hope to compare these different large $N$ solutions to different (leading or sub-leading) saddles in the supergravity partition function, such as the NUT and Bolt solutions above.

Lastly, one can consider more general quiver gauge theories, including those with massive type IIA supergravity duals \cite{Romans:1985tz}, as considered in \cite{Hosseini:2016tor,Hosseini:2017fjo,Benini:2017oxt}.  It would also be desirable to obtain the partition function for the quiver theory dual to the M-theory reduction on $M^{32}$. As mentioned at the end of Section \ref{sec:V52}, for this theory all Bolt solutions are allowed, including in particular the $p=1$ Bolt$_+$, which is a regular filling for the squashed $S^3$ for a finite range of squashing parameters. It would be interesting to see if the field theory computation is able to reproduce its on-shell action $I_{Bolt_+ \, p=1} =\frac{3}{4} I_{NUT} $, which is lower than the $S^3$ free energy computed in \cite{Amariti:2011uw,Gang:2011jj}. The question of finding the correct vacuum of the theory is of course entangled with the fact that, as already mentioned, we cannot rule out the presence of (perhaps less symmetric) branches  of solutions with the same boundary data. A deeper understanding of these points would be desirable.

All these questions are left for future investigation and we hope to report back in the near future.

\section*{Acknowledgments}

We would like to thank D. Berenstein, C. Closset, M. Crichigno, G. Horowitz, S. M. Hosseini, K. Hristov, Heeyeon Kim, Hyojoong Kim, D. Klemm, J. Louko, D. Martelli, A. Passias, S. Pufu, E. Shaghoulian, J. Sparks, and A. Zaffaroni for discussions. We acknowledge support from the Simons Center for Geometry and Physics, Stony Brook University for hospitality during some steps of this paper.  BW was supported in part by the National Science Foundation under Grant No. NSF PHY11-25915.

\appendix

\section{Explicit Killing spinors}\label{Killing_sp}

This appendix is devoted to the explicit construction of the Killing spinor for the solutions described in Section \ref{susyNUTsBolts}. Let's recall the Killing spinor equation:
\be \label{KillSE}
\delta_{\epsilon} \psi_{\mu} = \left(  \partial_{\mu}  -\frac14 \omega^{ab}_{\mu} \gamma_{ab} +\frac12  \gamma_{\mu}  -i A_{\mu} + \frac{i}{4} F_{\rho \sigma} \gamma^{\rho \sigma} \gamma_{\mu}  \right) \epsilon =0
\ee
The necessary conditions for the solutions to preserve supersymmetry, as already mentioned, can be found from the results of \cite{AlonsoAlberca:2000cs} performing the following parameter map
\be
\theta \rightarrow i \theta \qquad t \rightarrow i t \qquad \phi \rightarrow i \phi \qquad Q \rightarrow i Q \qquad s \rightarrow i s\,.
\ee
The necessary conditions read
\be
MP -Qs(\kappa - 4 s^2)=0 \,, 
\ee
and
\be
\mathcal{B}_+ \mathcal{B}_- =0 \,.
\ee
with
\be
\mathcal{B}_{\pm} = (M \pm s Q)^2 -s^2 (\kappa \pm P -4 s^2)^2 -(\kappa \pm 2  P -5 s^2)(P^2-Q^2) \,,
\ee
which reduce, in the case $\kappa=1$ to those found by \cite{Martelli:2012sz,Martelli:2013aqa} and are solved, for solutions preserving only $1/4$ of supersymmetry, by \eqref{cond_kappa}. We are interested in constructing the Killing spinor for this latter class of solutions: the procedure we will use follows from that of \cite{Martelli:2012sz} with minor modifications which allow for general bases $\Sigma_g$.

\subsection*{Spinors and vierbeins}
We introduce the orthonormal frame  conveniently chosen in \cite{Martelli:2012sz}:
\be
e^1  = \tau_1 \sqrt{r^2-s^2}\,, \qquad 
e^2  =  \tau_2 \sqrt{r^2-s^2}\,,  \qquad
e^3  =  2s\, \tau_3 \sqrt{\frac{\lambda(r)}{r^2-s^2}}\,, \qquad
e^4  = dr  \sqrt{\frac{r^2-s^2}{\lambda(r)}} \,.
\label{frame}
\ee
Depending on $\Sigma_g$, we choose the following form for $\tau_{1,2,3}$:
\begin{itemize}
\item for $S^2$ ($g=0$) we chose the $SU(3)$ left-invariant vielbeins
\be
\tau_1  = \cos \psi d \theta + \sin \theta \sin \psi d \phi \qquad
 \tau_2  =  -\sin \psi d \theta+ \cos \psi \sin \theta d\phi  \qquad
 \tau_3  =  d \psi + \cos \theta d \phi
\ee
which in particular satisfies $\tau_1 + i \tau_2 = e^{-i \psi} (d \theta + i \sin \theta d \phi)$.
\item for $\mathbb{R}^2$ (or, upon compactification $g=1$) we choose
\be
\tau_1  =   \sin \psi d \theta + \cos \psi d \phi \qquad
 \tau_2  = -\cos \psi d \theta+ \sin \psi  d\phi  \qquad
 \tau_3  =  d \psi -  \theta d \phi
\ee
\item  for $\mathbb{H}^2$ (or, upon compactification $g>1$) we have
\be
\tau_1  = \sin \psi d \theta + \sinh \theta \cos \psi d \phi \qquad
 \tau_2  =  -\cos \psi d \theta+ \sin \psi \sinh \theta d\phi  \qquad
 \tau_3  =  d \psi - \cosh \theta d \phi
\ee
\end{itemize}
We take the following basis of four-dimensional gamma matrices, which is the same used in \cite{Martelli:2012sz}:
\bea\label{gamma}
\gamma_i &=& \left(\begin{array}{cc} 0 & \sigma_i \\ \sigma_i  & 0\end{array}\right)\,, \qquad \gamma_4 \ = \ \left(\begin{array}{cc}0 &  i\mathbf{I}_2 \\ 
- i \mathbf{I}_2 & 0\end{array}\right), \qquad \gamma_5 = \gamma_1\gamma_2\gamma_3\gamma_4\  =\  \left(\begin{array}{cc} \mathbf{I}_2 & 0 \\ 0 &  -\mathbf{I}_2 \end{array}\right)\,
\eea
where $\sigma_i$, $i =1,2,3$ are the Pauli matrices. The supersymmetry parameter $\epsilon$ is a Dirac spinor. It is convenient to write it in terms of its positive and negative chirality parts $\epsilon  =   \left(\epsilon_+ , \epsilon_-  \right) \,,$ where
\be
 \qquad \left( \frac{1+\gamma_5}{2} \right)\epsilon =\left(\begin{array}{c}\epsilon_+ \\ 0 \end{array} \right)\,,\qquad \left( \frac{1-\gamma_5}{2} \right)\epsilon =\left(\begin{array}{c}0 \\ \epsilon_- \end{array} \right)\,.
\ee
In what follows we moreover further distinguish the single parts of  $\epsilon_+$ and $\epsilon_-$ as $
\epsilon_\pm  =  \left(\epsilon^{(+)}_\pm , \epsilon^{(-)}_\pm \right) $,
so that the 4d Dirac spinor $\epsilon$ has this form
\be
\epsilon =  \left(\begin{array}{c} \epsilon^{(+)}_+ \\  \epsilon^{(+)}_- \\ \epsilon^{(-)}_+ \\   \epsilon^{(-)}_-  \end{array}  \right) .
\ee

\subsection*{Radial component of the Killing spinor equation}

We write the warp factor  $\lambda(r)$ as
\be
\lambda(r) = (r-r_1)(r-r_2)(r-r_3)(r-r_4)\,,
\ee
where $r_{\alpha}$, $\alpha=1,2,3,4$, are the four roots of the warp factor, and their explicit form is in \eqref{erre}. If we decompose the $r$ component of the Killing spinor equation \eqref{KillSE}  into chiral parts, we obtain
\be \label{eqnepsplus}
\partial_r \, \epsilon_+ = - \frac{i}{2} \sqrt{\frac{r^2-s^2}{\lambda(r)}}\ \left[  \epsilon_- + \frac{\kappa- 2Q-4s^2}{2(r- s)^2}\, \sigma_3\,  \epsilon_- \right] \,,
\ee
and
\be \label{eqnepsminus}
\partial_r \, \epsilon_- = +\frac{i}{2}\sqrt{\frac{r^2-s^2}{\lambda(r)}}\, \left[\epsilon_+ +  \frac{\kappa+ 2Q-4s^2}{2(r+ s)^2}\,\sigma_3 \, \epsilon_+ \right] \,.
\ee
In analogy to \cite{Martelli:2012sz}, we impose the following projections on the Killing spinor:
\be \label{projections}
\epsilon_-^{(+)} = \epsilon_+^{(+)}  = 0 \qquad \quad
\epsilon_-^{(-)} = i \sqrt{\frac{(r-s)}{(r+s)}\frac{(r-r_1)(r-r_2)}{(r-r_3)(r-r_4)}}\epsilon_+^{(-)}\,.
\ee
Using the projection relation \eqref{projections} we find a solution to equations \eqref{eqnepsplus} and \eqref{eqnepsminus} of the form
\be\label{divided_spinor}
\epsilon =  \left(\begin{array}{c} F(r) \\  G(r) \end{array} \right) \otimes \upsilon,
\ee
where $\upsilon$ is a two-component spinor which does not depend on $r$, and the two radial functions $F(r)$ and $G(r)$ read:
\be
F(r)= \sqrt{\frac{(r-r_3)(r-r_4)}{r-s}} \,, \qquad G(r) =  i \sqrt{\frac{(r-r_1)(r-r_2)}{r+s}} \,.
\ee

\subsubsection*{3d Killing spinor equation}

Once we have inserted \eqref{divided_spinor} the remaining components $t,\theta, \phi$ of the Killing spinor equation \eqref{KillSE} reduce to the following three-dimensional equation the spinor $\upsilon$:
\be\label{KillingSpE3}
\big({\nabla_{(3)}}_i - i {A_{(3)}}_i + \frac{i s}{2}\gamma_i \big)\, \upsilon =0\,.
\ee
Here ${\nabla_{(3)}}_i$ and $\gamma_i$ are respectively the  covariant derivative and gamma matrices 
for the 3d boundary metric $
ds_3^2 = \tau_1^2 +\tau_2^2 + 4s^2 \tau_3^2
$
with spin connection
\be \nonumber
 \omega_{\theta}^{02} = -s \sin \psi  \qquad \omega_{\theta}^{12} = -s \cos \psi \qquad \omega_{\phi}^{02}= s \cos \psi f(\theta) 
 \ee
 \be
 \omega_{\phi}^{01} = - (\kappa-2s^2) f( \theta) \qquad \omega_{\phi}^{12} = - s f(\theta) \sin \psi \qquad \omega_t^{01} = 2s^2 -\kappa 
\ee
where $f(\theta)$ is defined in \eqref{fff}. The 3d gauge field appearing in \eqref{KillingSpE3} is the asymptotic value of the gauge field $A$, hence it has the form 
\be
A_{(3)}   =  P \, \tau_3  =  - \frac{1}{2}\, (4s^2-\kappa)\, \tau_3.
\ee
Choosing the vielbeins $
\bar{e}^1 = \tau^1$, $\bar{e}^2 = \tau^2$, $\bar{e}^3 = 2s \tau^3 $
 it is easy to see that the components of equation \eqref{KillingSpE3} are solved by
\be\label{1/4BPSupsilon}
\upsilon =  \left(\begin{array}{c}0 \\ \upsilon^{(0)} \end{array} \right) \,,
\ee
where $\upsilon^{(0)}$ is a constant. The full Killing spinor then reads
\be
\epsilon  =   \left(\begin{array}{c} 0 \\  \sqrt{\frac{(r-r_3)(r-r_4)}{r-s}} \upsilon^{(0)} \\ 0 \\ i \sqrt{\frac{(r-r_1)(r-r_2)}{r+s}}  \upsilon^{(0)}  \end{array}  \right).
\ee
The solution therefore preserves two supersymmetries out of eight, it is then 1/4 BPS. Notice that the Killing spinor only has radial dependence, hence no further supersymmetries are broken while taking the quotient to obtain a higher genus surface. Let us mention that if we had chosen the static vielbein, which for instance for $\kappa=1$  reads
\be
\tau_1  =  d \theta  \qquad
 \tau_2  = \sin \theta d\phi  \qquad
 \tau_3  =  d \psi + \cos \theta d \phi
\ee
the Killing spinor would have had no radial dependence, provided the asymptotic gauge field $A^{(3)}$ is supplemented by flat connection term $A^{(3)} = P \, \tau_3 + A_{flat} $ where $A_{flat} = \kappa \, d\psi$.

\section{Moduli space of gravity solutions \label{AppB}}

We analyze here the range of existence of the Bolt$_+$ and Bolt$_-$ solutions, depending on the value of parameters $p$ and $g$. Let's remind the reader that we have formally found four branches of solutions \eqref{firstpluscase} and \eqref{second}. However, as we will see below, not all of them are regular and in our formalism one needs to check this explicitly case by case. What happens in general is that some branches are defined in a specific $s$ interval. A given solution branch, if it's not defined everywhere, either joins other another branch (i.e. NUT solution), or ends ``annihilating'' another branch (as happens for the $p=1$ Bolt$_-$ solutions, which both end at the same point). This is consistent with an analysis of these solutions using Morse theory\footnote{We thank D. Berenstein for discussions on this topic.}.

In order to obtain regular solutions one needs to ensure that the three following conditions are satisfied: \begin{enumerate}\item $r_b >s$ \item $f_{\pm} >0$ \item$r_+(Q_+) > r_-(Q_+)$ for Bolt$_+$, $r_+(Q_-) < r_-(Q_-)$ for Bolt$_-$ \end{enumerate}

We checked this explicitly for the cases below. The procedure used to analytically obtain the range of parameters for the $g=0$ case is spelled out in section 5.2.1 of \cite{Martelli:2012sz}. Generalizing it to higher genus case corresponds to incorporating the following modifications in their procedure
\be
p \rightarrow \mathbf{p} = \frac{p}{|g-1|} \,\,\,\,\text{for}\,\,\,\,g \neq 1 \qquad \quad p\rightarrow p \,\,\,\, \text{for}\,\,\,\,g=1
\ee
and
\be
f_{\pm} \rightarrow f_{\pm} = (16 \, s^2 \pm \mathbf{p})^2-128 \, \kappa \,s^2 \,.
\ee
We summarize here the main results of our analysis. 

\subsection*{$p=1$}
\begin{itemize}

\item $g=0$ Two regular Bolt$_+$ branches exists for $s \in [0, \frac{-1 + \sqrt2}{4}]$ and two regular Bolt$_-$ branches exist for $s \in [0, \frac{\sqrt{5-2\sqrt6}}{4}]$. Notice that this interval does not include the round $S^3$ which is obtained by $s =1/2$. 

\item $g=1$ for this case only one regular Bolt$_+$ solution (with $Q_+^+$) exists for all $s>0$. The Bolt$_-$ branch with $Q_{-}^+$ exist for $s \in [0, \frac14]$. At this point $s =1/4$, this joins the NUT solution, which has $Q = -2s^2$, but presents conical singularities.

\item $g>1$ the Bolt$_+$ with $Q_+^+$ exists for $s \in [0,\mathbf{f}(g)]$, while the other Bolt$_+$ with $Q_{+}^-$ exists for $s \geq \mathbf{f}(g)$. Therefore the two Bolt$_+$ branches cover together the entire $s$ axis. For example, $\mathbf{f}(2) \sim 1/4$, $\mathbf{f}(3) = 1/4\sqrt2 $ etc. Moreover, one out of the two Bolt$_-$ solutions (the one corresponding to $Q_-^-$) exists for all $s>0$. 

\end{itemize}

\subsection*{$p=2$}

\begin{itemize}

\item $g=0$ We have one Bolt$_+$ solution in the interval $s \in [0,\frac{1}{2 \sqrt2}]$. At the point $s=1/2\sqrt2$ the solution joins the NUT solution since $r_+(Q_+) =s$. The latter extends for all $s$. The two Bolt$_-$ branches exist in the interval $s \in [0,\frac{2-\sqrt2 }{4}]$.

\item $g=1$ the Bolt$_+$ solution with $Q_+^+$ exists for $s>0$, and the $Q_{-}^+$ Bolt$_-$ solution exist in the interval $s \in [0, \frac{1}{2\sqrt2}]$, merging at the point $s = 1/2 \sqrt2 $ with the NUT branch.

\item $g>1$ The Bolt$_-$ solution corresponding to $Q_-^-$ exists regular for all $s>0$. The Bolt$_+$ with $Q_+^+$ exists for $s \in [0,\mathbf{h}(g)]$, while the other Bolt$_+$ with $Q_{+}^-$ exists for $s \geq \mathbf{h}(g)$. Therefore the two Bolt$_+$ branches cover together the entire $s$ axis. For example, $\mathbf{h}(2) \sim 1/2 \sqrt2$, $\mathbf{h}(3) = 1/4 $,  $\mathbf{h}(4) = 1/2\sqrt6 $ etc.

\end{itemize}

\subsection*{$p \geq 3$}
\begin{itemize}

\item $g=0$ Bolt$_+$ solution exists for all $s>0$. The two Bolt$_-$ solutions instead are regular only for $s \in [0, \frac{1}{4} (\sqrt{p+2} -\sqrt2)\, ]$.

\item $g=1$ Bolt$_+$ solution with $Q_+^+$ exists for all $s>0$. The $Q_-^+$ Bolt$_-$ solution exist for $s \in [0, \frac{\sqrt{p}}{4}]$ and at the end point it merges with the planar NUT.

\item $g>1$ One Bolt$_-$ branch is once again present for all $s>0$. In analogy to the previous cases, two Bolt$_+$ solutions exist, with domains $s \in [0, \mathbf{l}_p(g)]$ and $s \geq \mathbf{l}_p(g)$. For instance, for $g=3,p=6$ we have $\mathbf{l}(3, p=6) = \sqrt3/4$.

\end{itemize}

\subsection*{Plots of the on-shell action}

\begin{figure}[ht]
\begin{center}
\includegraphics[width=7.5cm]{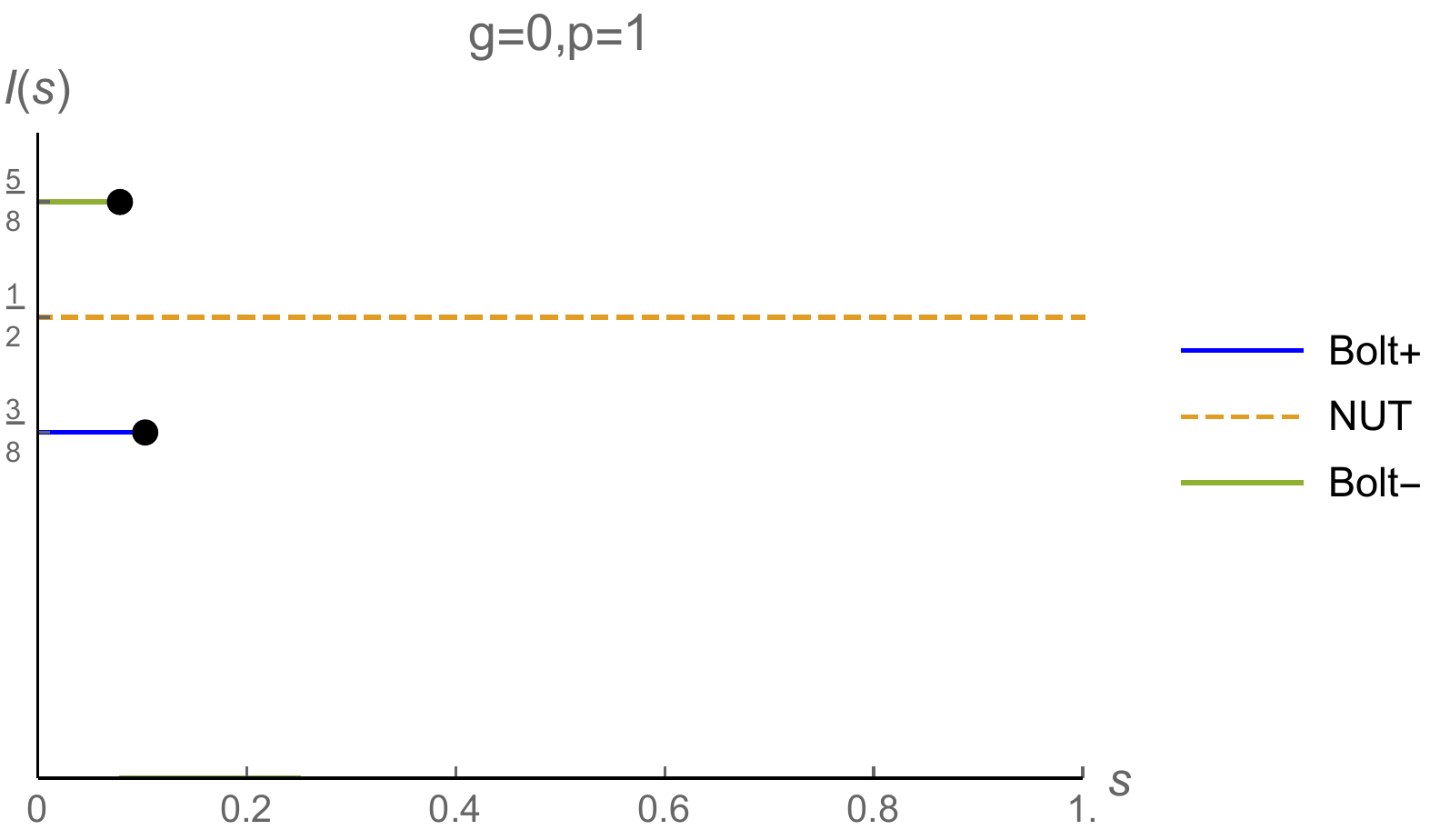}
\includegraphics[width=7.5cm]{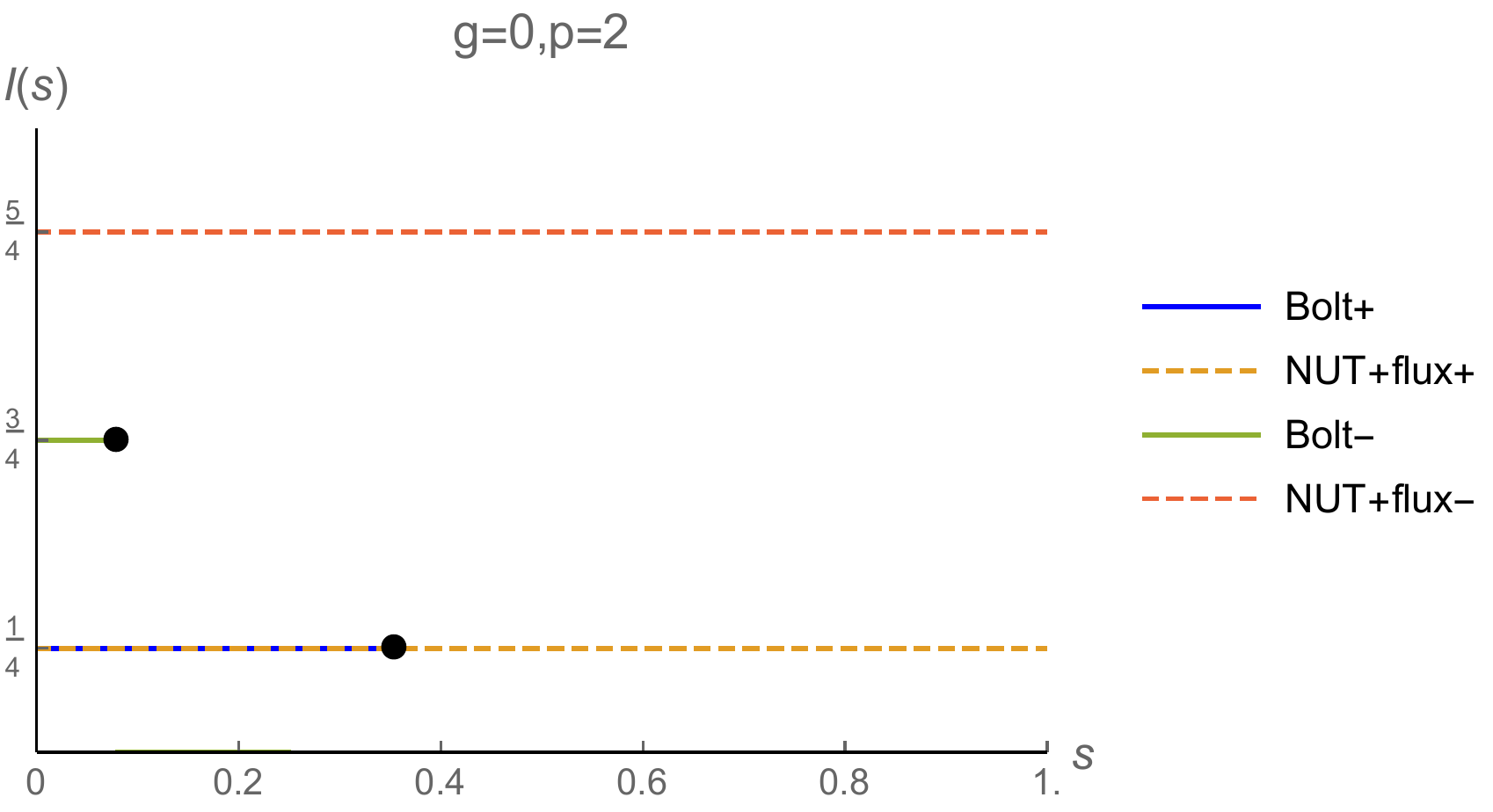}
\includegraphics[width=7.5cm]{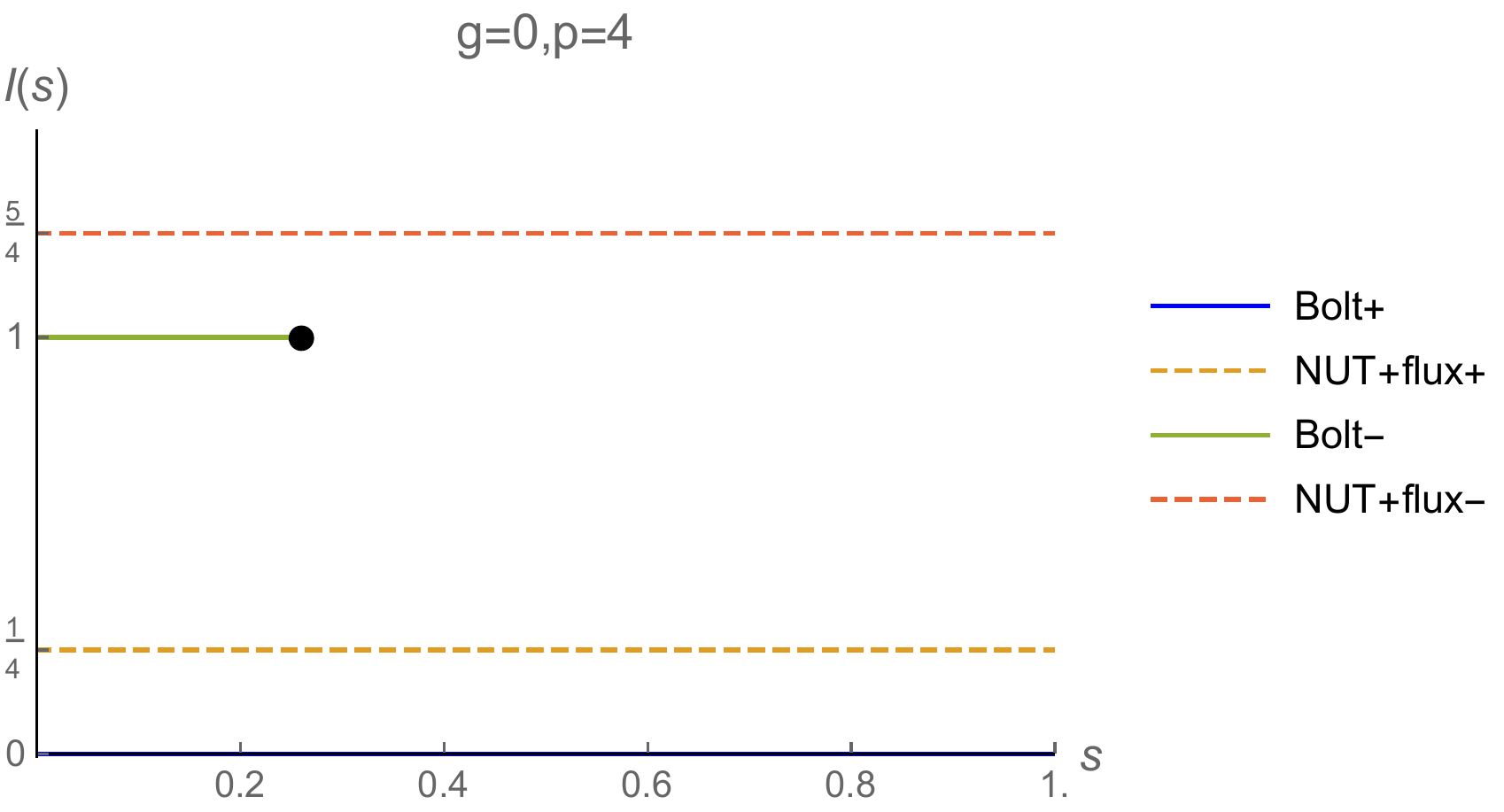}
\includegraphics[width=7.5cm]{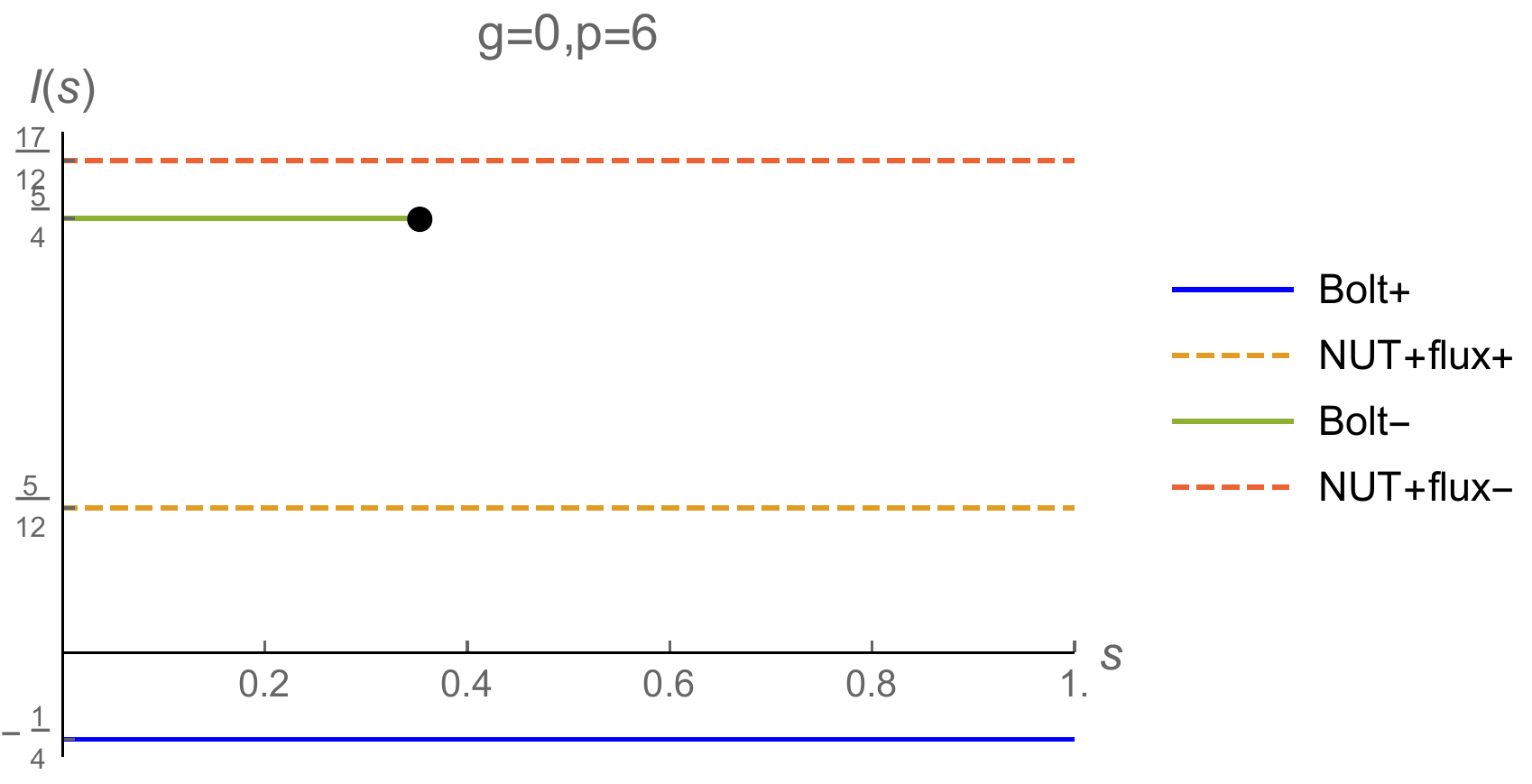}
\end{center}
\caption{\label{Fig1}Value of the on-shell action in units of $\pi/ G_4$ for the branches of solutions with $g=0$ for the values $p=1,2,4,6$ respectively. In the picture the Bolt$_+$, Bolt$_-$ solutions and the mildly singular NUT$/\mathbb{Z}_p$ plus $\pm \frac{p}2 -1$ unit of magnetic flux are depicted. }
\end{figure}

\begin{figure}[!ht]
\begin{center}
\includegraphics[width=7.5cm]{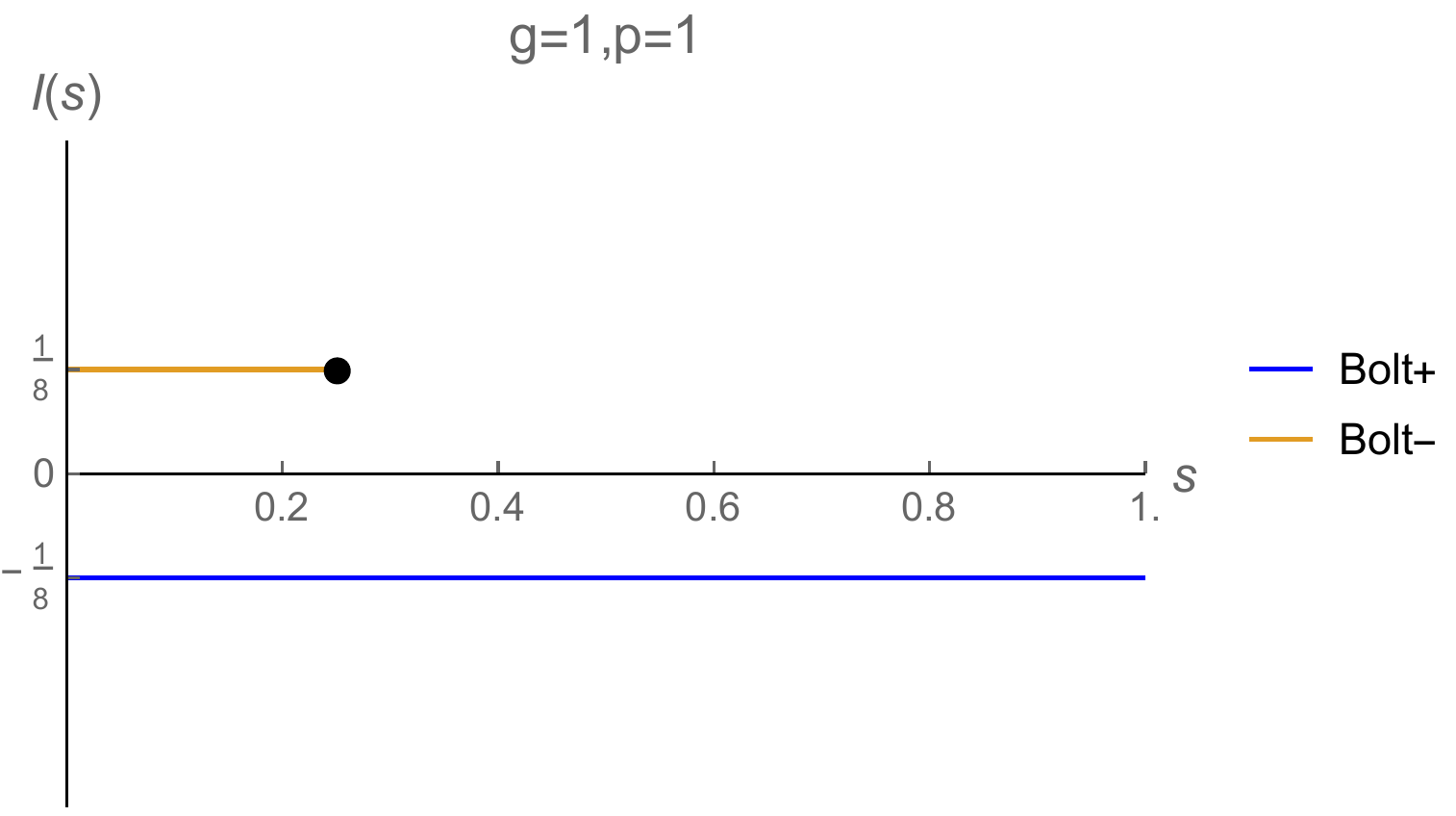}
\includegraphics[width=7.5cm]{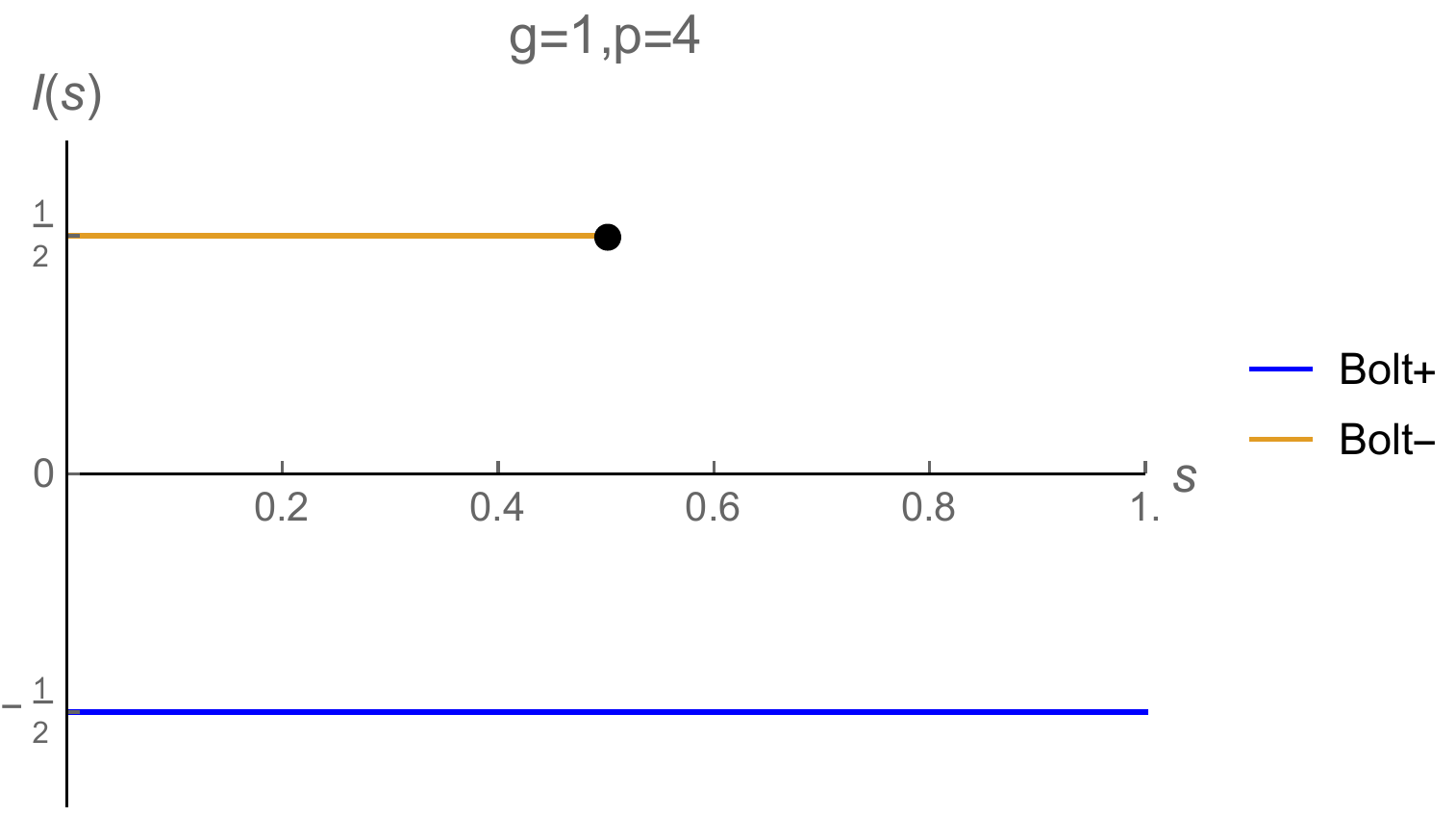}
\includegraphics[width=7.5cm]{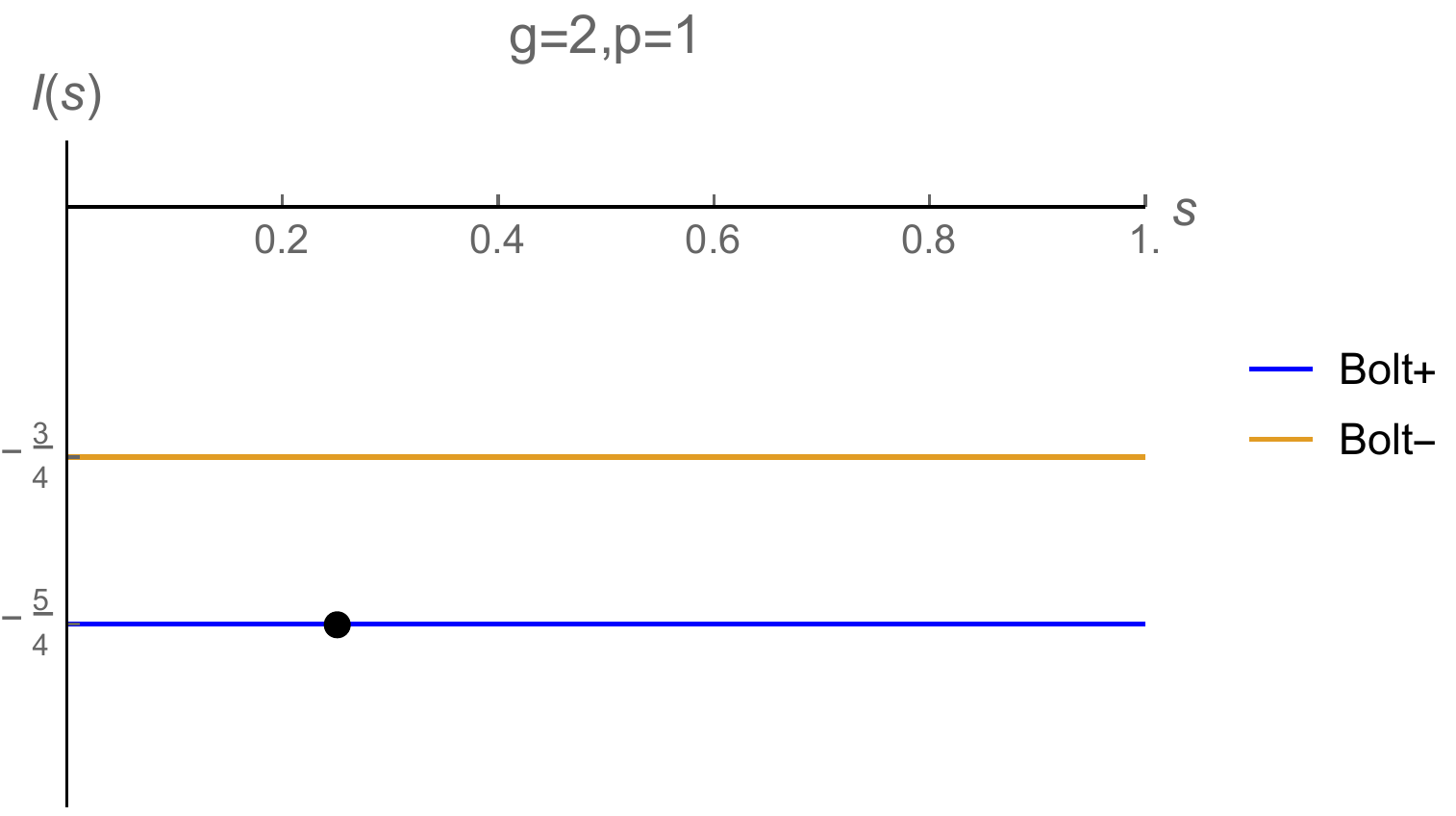}
\includegraphics[width=7.5cm]{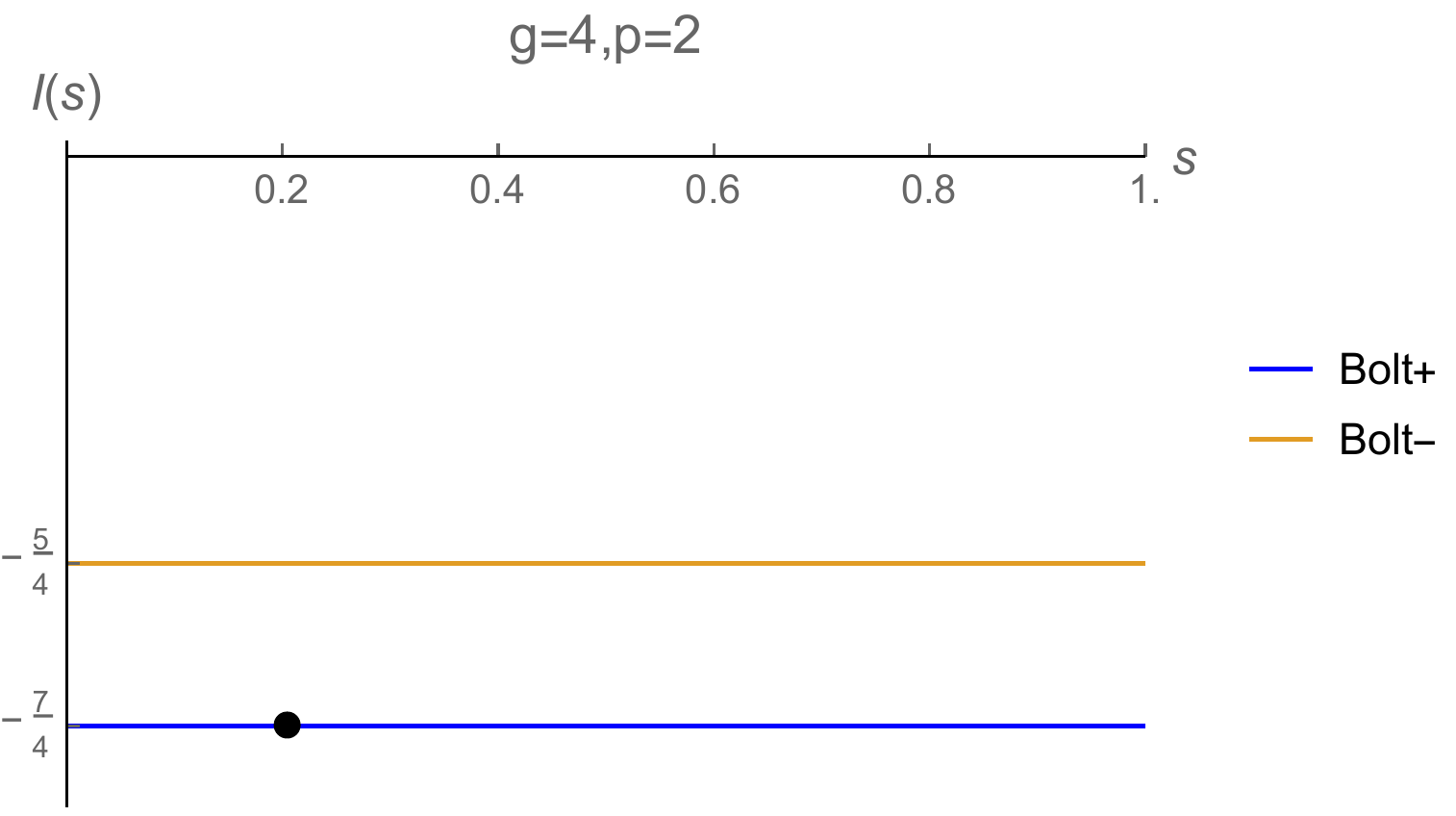}
\end{center}
\caption{\label{Fig2}Examples of moduli space for Bolts with $g>0$. The first row refers to the cases $g=1,p=1,4$, and the last two plots refer to $g=2, p=1$ and $g=3, p=4$ respectively. }
\end{figure}

\begin{figure}[!ht]
\begin{center}
\includegraphics[width=10.5cm]{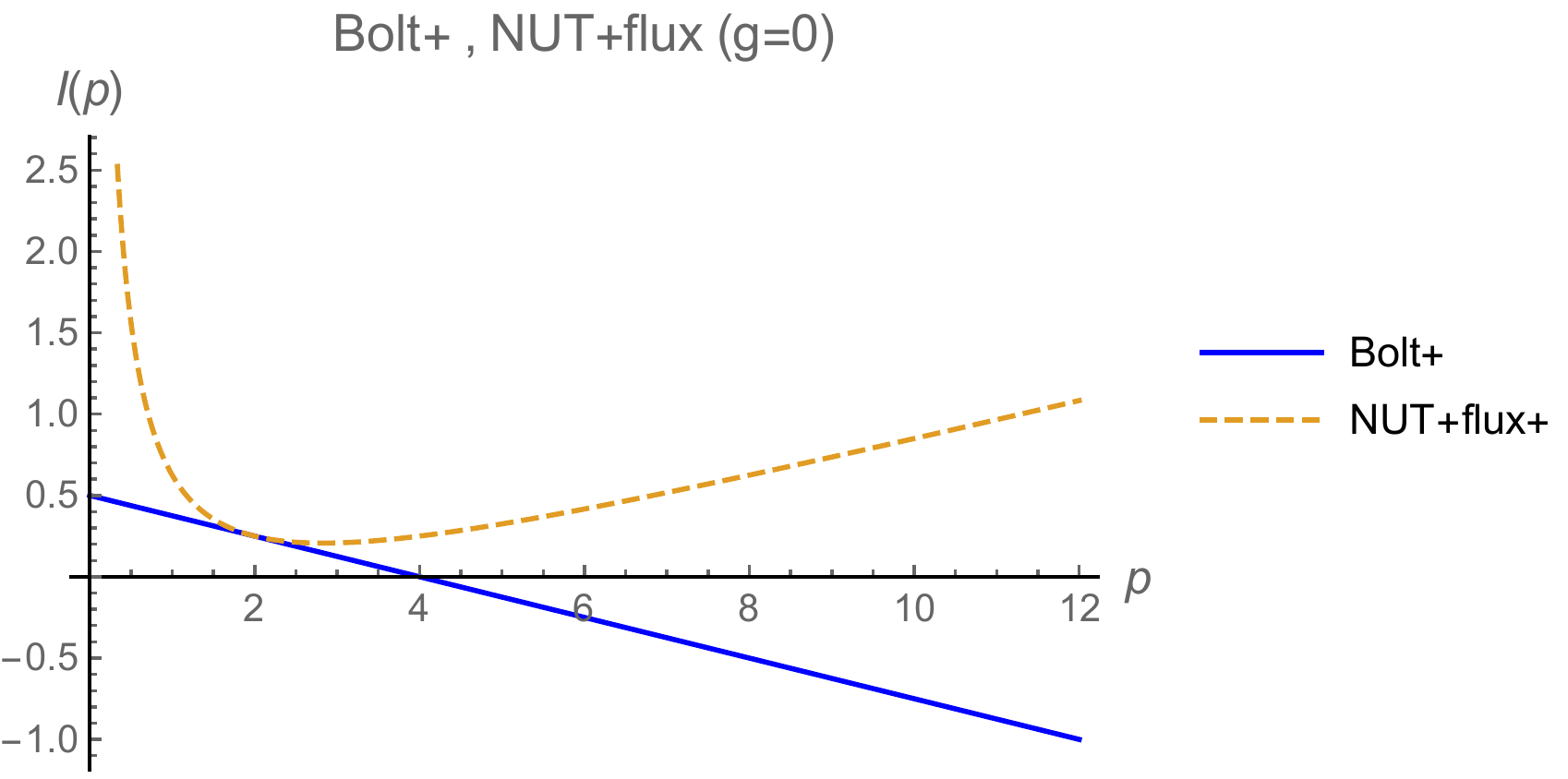}
\end{center}
\caption{\label{Fig3}This plot is showing the behavior of the on-shell action (in units of $\pi/G_4$) for the Bolt$_+$ solution and the NUT$/\mathbb{Z}_p$ with the same units of flux for $1<p<12$, as a function of $p$. One can see that the Bolt$_+$ has always lower free energy, so dominates the ensemble.}\end{figure}

Plots of the free energy of the solutions ($y$ axis) in function of the squashing parameter $s$ ($x$ axis), for the sample values of $g=0,1,2$ and for various values of $p$ are shown in Fig. \ref{Fig1} and Fig. \ref{Fig2}. The horizontal line represent the range of existence of the given solution branch.

The Bolt$_+$ solution is depicted in blue and it has the same flux as the NUT$/\mathbb{Z}_p$ with $\frac{p}{2}-1$ units of flux (dashed, orange). The Bolt$_-$ branches are depicted in green, and have the same flux as  the NUT$/\mathbb{Z}_p$ with $-\frac{p}{2}-1$ flux (dashed, red). For the ABJM case Bolt$_+$ and Bolt$_-$ satisfy the quantization condition for the same values of $g,p$\footnote{This is not true for theories like $V^{5,2}$, see discussion in section \ref{sec:V52}.}.
The last set of plots in Fig. \ref{Fig3} finally show that the Bolt$_+$ solution is always favored with respect to the other branches. 

\section{Details of partition function calculations}
\label{sec:pfdetails}

In this appendix we present more details on the computation of the twisted superpotential and $\cM_{g,p}$ partition function for large $N$ quiver gauge theories of the type discussed in Section \ref{sec:qd}.  These theories may have bifundamental chiral multiplets, fundamental chiral multiplets, and Chern-Simons terms, and we refer to the main text for a description of the constraints on the matter content.  Here we recall that we take the large $N$ ansatz:

\be \label{ansatz} u_a^\alpha = v^\alpha_a + i N^{1/2} t_a \;. \ee
In the large $N$ limit, $t_a$ becomes a continuous variable $t$, and we define:

\be \rho(t) = \frac{1}{N} \sum_a \delta(t-t_a) \;, \ee
normalized by $\int dt \rho(t) = 1$, and $v^\alpha_a \rightarrow v^\alpha(t)$.

\subsection{Contributions to twisted superpotential}

We start with the contributions to the twisted superpotential.  These were derived already in \cite{Benini:2016rke,Hosseini:2016tor}, however we review them here for completeness, and to establish the relation of our notations.  We also employ a slightly different argument in some places than used in these works.  

Our starting point is the expression for the twisted superpotential at finite $N$ in \eqref{wfn}.  Let us review the various ingredients in the $U(N)$ quiver gauge theories.  For the CS terms, we have:
\be \cW_{CS} = \sum_{\alpha,a} \frac{1}{2} k_\alpha {u_a}^\alpha({u_a}^\alpha -1) \;. \ee
If we take the ansatz (\ref{ansatz}), this becomes:

\be \cW_{CS} = \sum_{\alpha,a}  k_\alpha \bigg( -\frac{N}{2} {t_a}^2 + i N^{1/2} t_a (v_a^\alpha - \frac{1}{2}) + \frac{1}{2} ({v^\alpha_a}^2-v^\alpha_a) \bigg) \;. \ee
Imposing the constraint $\sum_\alpha k_\alpha =0$, we may write this to leading order in $N$ as:

\be \cW_{CS} = \sum_{\alpha,a}  i N^{1/2} k_\alpha t_a v_a^\alpha   \;\;\; \rightarrow \;\;\; i N^{3/2} \int dt \rho(t) \sum_\alpha k_\alpha t v^\alpha(t)  \;. \ee

Next we compute the contribution of bifundamental and adjoint fields.  Here we will take a slightly different approach from \cite{Benini:2015eyy}, and instead use the following argument, modified from that of \cite{Jafferis:2011zi}.  We will consider the large $N$ limit of functions of the form:

\be \label{biftrick} \sum_{a,b=1}^N f(u_a - \hat{u}_b) \;, \ee
where $u_a$ and $\hat{u}_b$ are the eigenvalues for two of the $U(N)$ gauge groups (or possibly the same one, in the case of adjoint fields), and $f$ is some function, which may depend on other parameters which we suppress.  Let us rewrite this as:

\be \label{lsl}  \sum_{a,b=1}^N f(u_a - \hat{u}_b) =  \sum_{a \neq b}^N f(u_a - \hat{u}_b)  + \sum_{a=1}^N f(u_a - \hat{u}_a)\;. \ee
We expect the second term to be subleading in $N$, but we will return to its contribution below.  Focusing on the first term for now, and plugging in the ansatz (\ref{ansatz}), this can be written as:

\be\sum_{a \neq b}^N f(v_a - \hat{v}_b + i N^{1/2} (t_a - t_b) )  \rightarrow \;\;\; N^2 \int dt dt' \rho(t) \rho(t')  f( v(t) - \hat{v}(t') + i N^{1/2} (t-t') ) \;. \ee
Now, expecting most of the contribution to come from the region near $t=t'$, we make a change of variables $t' \rightarrow \xi = N^{1/2} (t-t')$.  Then this becomes, to leading order:

\be \label{bfint} N^{3/2} \int dt  \rho(t)^2  PV \int_{-\infty}^\infty d \xi  f( \delta v(t)  + i \xi ) \;,\ee
where we defined $\delta v(t) = v(t) - \hat{v}(t)$, and we take the principle value to exclude the contribution from $t=t'$.

In the present case, a single bifundamental chiral multiplet contributes:
\be \cW_{bif} = \sum_{a,b=1}^N \cW_\pi(u_a - \hat{u}_b + m) \;, \ee
where we have defined the contribution of a chiral multiplet in the parity-preserving regularization 
(\ie, without the level $-\frac{1}{2}$ CS term) as:\footnote{Here and below, we work on the principle branch of the polylogarithms, defined so that $\text{Li}_s(e^{2\pi i u}) \rightarrow 0$ as $u \rightarrow i \infty$, which fixes the branch in the upper half $u$-plane, and extending continuously along vertical lines to the lower half plane.  Then the resulting functions are periodic under $u \rightarrow u+1$, with branch cuts along $\text{Re}(u) \in \Z$ for $\text{Im}(u)<0$.}

\be \cW_\pi(u) =  \frac{1}{(2 \pi i)^2} \dilog(e^{2 \pi i u})  + \frac{1}{4}[u]([u]-1) + \frac{1}{24} \;, \ee
where, as in the main text, we define, for complex $u$:

\be [u] = u-n , \;\;\; n \in \Z, \;\;\; \text{such that} \;\; 0 < \text{Re}([u]) \leq 1 \;. \ee
Here we have used the freedom to change branch, taking $\cW \rightarrow \cW + n u + m$, $n,m \in \Z$, to make the twisted superpotential periodic under $u \rightarrow u+1$, which will be convenient below.  

Inserting this in (\ref{bfint}), we find:

\be \cW_{bif}  = N^{3/2} \int dt \rho(t)^2 PV \int_{-\infty}^\infty d \xi \cW_\pi(\delta v + m +  i \xi ) \;. \ee
Using:

\be \int du \cW_\pi(u) = \frac{1}{(2 \pi i )^3} \text{Li}_3(e^{2 \pi i u}) + \frac{1}{24} [u]([u]-1)(2[u]-1) \;, \ee
and cutting off the integral over $\xi$ at some $\xi_{max}$, we find:
\be \cW_{bif}  = -i N^{3/2} \int dt \rho(t)^2 \bigg( \frac{1}{(2 \pi i )^3} \text{Li}_3(e^{2 \pi i u}) + \frac{1}{24} [u]([u]-1)(2[u]-1) \bigg) \bigg|_{u=-i \xi_{max} + \delta v + m}^{u=i \xi_{max} + \delta v + m} \;. \ee
Next we use:

\be \label{li3} \frac{1}{(2 \pi i)^3} \text{Li}_3(e^{2 \pi i u})  -  \frac{1}{(2 \pi i)^3}  \text{Li}_3(e^{-2 \pi i u})  = -\frac{1}{12} [u] ([u]-1) (2 [u]-1) \;, \ee
as well as $\text{Li}_3(e^{2 \pi i u} ) \rightarrow 0$ as $\text{Im}\; u \rightarrow \infty$, to find:

\be \label{wbifd} \cW_{bif}  = i N^{3/2} \int dt \rho(t)^2 \bigg( \frac{1}{2} g([\delta v+ m]+ i \xi_{max}) + \frac{1}{2} g([\delta v +m] - i \xi_{max}) \bigg) \ee
$$ =  i N^{3/2} \int dt \rho(t)^2  \bigg( \frac{1}{2} ([\delta v + m]-\frac{1}{2}) \xi_{max}^2+g([\delta v+m]) \bigg) \;, $$
where we defined:
\be \label{gdef} g(u) = - \frac{1}{12}u(u-1)(2u-1) \;. \ee
However, we see this diverges as we take the cutoff $\xi_{max} \rightarrow \infty$.  

To deal with this divergence, we make two observations.  First, we only expect that certain choices of matter content and superpotential terms lead to theories with appropriate supergravity duals, with $N^{3/2}$ scaling of the degrees of freedom.  Specifically, as in Section \ref{sec:qd}, we impose:

\begin{itemize}
\item The number of incoming and outgoing edges (bifundamental chirals) at any node ($U(N)$ gauge factor) in the quiver are equal.
\item The sum of all masses of bifundamentals charged under a node is zero.\footnote{More precisely, it need only be integer for the argument below to work, and in the context of the twisted superpotential the integer part of the masses is irrelevant, so imposing the sum is precisely zero is unnecessary.  However, when computing the partition function below, the integer part will be important, and we will need to impose this condition.}
\end{itemize}

Then when we sum over the contributions of all chiral multiplets, we find the coefficient of $\xi_{max}^2$ is:

\be -\frac{1}{2} \sum_{I} ( v_{\alpha_I} - v_{\beta_I} + m_I - n_I - \frac{1}{2}) = \frac{1}{2} \sum_{I} (n_I + \frac{1}{2}) \;, \ee
where $I$ runs over the bifundamentals, with the $I$th fundamental connecting the $\alpha_I$th and $\beta_I$th gauge groups, and $n_I = v_{\alpha_I} - v_{\beta_I} + m_I - [v_{\alpha_I} - v_{\beta_I} + m_I] \in \Z$, and we have used the conditions above to cancel the dependence on the $v^\alpha$ and $m_I$.  Then the RHS, while not necessarily zero, is $\frac{1}{2}$ times an integer.  Then there is one more fact we must use, which is that the twisted superpotential is only defined modulo changes of branch.  For finite $N$, such changes of branch take the form:

\be \cW \rightarrow \cW + n^a u_a + n^i m_i + n , \;\;\;\; n^a,n^i,n \in \Z \;. \ee
For infinite $N$, we choose to focus on changes of branch which preserve Weyl symmetry of the gauge group, and consider only those terms which appear at leading order in $N$.  Then there are two basic choices, which lead to the following terms when we plug in the ansatz (\ref{ansatz}):\footnote{Note the condition that the twisted superpotential be finite fixes the choice of change of branch of the second type above, while the first type changes the twisted superpotential by a finite piece, and so is not specified uniquely. }

\be \sum_{a,\alpha} \hat{n}^\alpha u^\alpha_a \;\;\;  \rightarrow \;\;\; i N^{3/2} \bigg( \sum_\alpha \hat{n}^\alpha \bigg)\int dt \; t \rho(t) \ee

\bea \label{bc2} &\sum_{a<b} ( \sum_\alpha (n^\alpha u^\alpha_a - \hat{n}^\alpha u^\alpha_b) + n^i m_i + n')  \\
 & \rightarrow N^{3/2} \int dt \rho(t)^2 \int_0^{\xi_{max}} d \xi ( \bigg(\sum_\alpha n^\alpha \bigg) i \xi + \sum_\alpha (n^\alpha -\hat{n}^\alpha) v^\alpha  + n^i m_i + n') \\
& = i N^{3/2} \int dt \rho(t)^2 \bigg( \frac{1}{2} \bigg(\sum_\alpha n^\alpha\bigg) \xi_{max}^2 - i \xi_{max} ( \sum_\alpha (n^\alpha-\hat{n}^\alpha) v^\alpha + n^i m_i + n') \bigg) \;,\eea
where we must take $\sum_\alpha n^\alpha = \sum_\alpha \hat{n}^\alpha$ so that this term is of order $N^{3/2}$.  Then we note that, after summing over the contributions from all chirals, the divergent terms from (\ref{wbifd}) are precisely of the form of the second change of branch above, so we may remove them.  Then we find the contribution of a bifundamental chiral multiplet is:  

\be \cW_{bif}  = i N^{3/2} \int dt \rho(t)^2 g([\delta v + m]) \;.\ee
Note that while this is cubic in $\delta v$, the first consistency condition above implies that the cubic terms for each $v_{\alpha}$ cancels when we sum over all chirals, and so the final functional is always a quadratic function of the $v^\alpha$.

Finally, let us return to the subleading contribution in (\ref{lsl}):

\be \sum_{a=1}^N f(u_a - \hat{u}_a)  \rightarrow N \int dt \rho(t) f(\delta v(t)) \;. \ee
For the case of a bifundamental chiral multiplet, this gives:

\be \label{sl}  N \int dt \rho(t) \cW_\pi(\delta v + m) = N \int dt \rho(t)\bigg( \frac{1}{(2 \pi i)^2} \dilog(e^{2 \pi i (\delta v+m)}) + \frac{1}{4} ([\delta v+m])([\delta v+m]-1) + \frac{1}{24} \bigg) \;. \ee
Naively this is subleading in $N$, however, note that as $\delta v+m$ approaches an integer, the derivative of this expression with respect to $\delta v$ diverges.  This means the subleading term will start to compete with the leading term, and we will have to take it into account below.  Specifically, let us write, for $n \in \Z$:

\be \label{Yforma} \delta v(t) +m = n + C e^{-2 \pi N^{1/2}Y_I(t)} \;. \ee
Then one finds the following additional contribution to the derivative of $\cW$, to leading order in $N$:

\be \frac{\delta \cW}{\delta (\delta v)} = \cdots - i N^{3/2} \int \rho(t) Y_I(t) \;.\ee
We will have to take this term into account when finding the extremal value of $\cW$.

Finally, for an (anti-)fundamental chiral, we have the following contribution to the twisted superpotential:

\be \cW_{fun} = \sum_{a=1}^N \cW_\pi(\pm u_a + m) = \sum_{a=1}^N \cW_\pi(\pm (v^\alpha_a + i t_a N^{1/2}) + m) \ee
$$\rightarrow N \int dt \rho(t) \cW_\pi(\pm (v^\alpha + i t N^{1/2}) + m) \;. $$
Using the limit:

\be \cW_\pi(u) \rightarrow \pm \bigg( \frac{1}{4} [u]([u]-1) + \frac{1}{24} \bigg) \;\;\;\; \text{as} \;\;\;\; \text{Im} u \rightarrow \pm \infty \;, \ee
we find this becomes:
\be N \int dt \rho(t) \big( \mp \frac{1}{4} t |t| N + \frac{i}{2} |t| N^{1/2} ([\pm v^\alpha +  m]-\frac{1}{2}) \big)  + \cdots \;. \ee
If we impose that the total number of fundamentals and anti-fundamentals are equal, the $O(N^2)$ term cancels, and this becomes, to leading order:
\be \cW_{fun} =  i N^{3/2} \int dt \rho(t) \frac{1}{2} |t|  ([\pm v^\alpha + m] - \frac{1}{2}) \;. \ee

\subsection{Contributions to $\log Z_{\cM_{g,p}}$}

Next we consider the contributions to $Z_{\cM_{g,p}}$.  Here the starting point is the finite $N$ expression for the partition function as a sum over Bethe vacua, \eqref{opdef}.  Specifically, we consider the contribution from a given Bethe vacuum, and take the eigenvalues ${u}_a$ in this vacuum to be given by the distribution $\rho(t),v^\alpha(t)$, as above.  Then we compute the functional:
\be \log Z_{\cM_{g,p}}[\rho,v^\alpha(t)] =\bigg( p \log \cF({u}_a,m_i) + (g-1) \log \cH(u_a,m_i) + s_i \log \Pi_i(u_a,m_i) \bigg) \bigg|_{u_a \rightarrow (\rho,v^\alpha)} \;. \ee

First we have the contribution of the Chern-Simons terms.  These contribute only through the fibering operator:
\be\label{zcs} -p \pi i \sum_\alpha \sum_a k_{\alpha}  {u_a^\alpha}^2 \rightarrow - 2 \pi N^{3/2} \int dt \rho(t) \sum_\alpha - p k_\alpha t v^\alpha(t) \;.  \ee

Next consider the contribution of a bifundamental chiral.  This is given by:
\be \label{lzbif} \log Z_{\cM_{g,p},bif}= \sum_{a,b} p \log \cF_\pi(u_a - \hat{u}_b + m)  + \ell \log \Pi_\pi(u_a - \hat{u}_b + m) \;, \ee
where:
\be \log \cF_\pi(u) = \frac{1}{2 \pi i} \dilog(e^{2\pi i u})  + u \log (1- e^{2 \pi i u}) - \frac{1}{2} \pi i u^2 + \frac{\pi i}{12} \;, \ee
$$ \log \Pi_\pi(u) = - \log(1- e^{2 \pi i u}) + \pi i u + \frac{\pi i}{2} \;. $$
and we write $\ell=s+(g-1)(r-1)$, where $s$ is the net flux felt by the chiral from background vector multiplets, and $r$ is its R-charge.  We note that this can be written in terms of the twisted superpotential as:

\be \label{fpw} p \log \cF_\pi(u)  + \ell \log \Pi_\pi(u) = 2 \pi i\bigg( p \cW_\pi(u) - ( p u -\ell) \partial_u \cW_\pi(u) \bigg) \ee
$$ = 2 \pi i \bigg( p \cW_\pi([u]) - ( p [u] + p n  -\ell) \partial_{[u]} \cW_\pi([u]) \bigg) \;, $$
where $n=u-[u] \in \Z$, and we have used the fact that $\cW$ is periodic under $u \rightarrow u+1$.  Then we have:

\be \log Z_{\cM_{g,p},bif}\approx N^{3/2} \int dt \rho(t)^2 PV \int_{-\infty}^\infty d \xi  \bigg( p \log \cF_\pi(\delta v + m + i \xi)  + \ell \log \Pi_\pi(\delta v + m + i \xi) \bigg)\;. \ee
One can compute this directly, but we can also use (\ref{fpw}) to compute:

\be \log Z_{\cM_{g,p},bif}  = -2 \pi N^{3/2} \int dt \rho(t)^2 \bigg( \frac{1}{2} G_\ell(\delta v+ m+ i \xi_{max}) + \frac{1}{2} G_\ell(\delta v +m - i \xi_{max}) \bigg) \ee
$$ =  -2 \pi N^{3/2} \int dt \rho(t)^2  \bigg( - \frac{1}{2} ( p (\delta v + m) - \ell) \xi_{max}^2+G_\ell(\delta v+m) \bigg) \;, $$
where we defined (here $n=u-[u]$):

\be G_\ell(u) = 2 p g([u]) - (p [u] +n p - l) g'([u]) = p ( \frac{[u]^3}{6} - \frac{[u]}{12}) + (\ell-n p) ( -\frac{[u]^2}{2} + \frac{[u]}{2} - \frac{1}{12}) \;.\ee

Although the vector multiplet does not contribute to the twisted superpotential, it does contribute to the partition function, appearing in the same way as an adjoint chiral multiplet of R-charge $2$.  Then we can obtain its contribution from the above formula by taking $\delta v + m \rightarrow 0$ and $\ell=(g-1)$, giving:

\be  \log Z_{\cM_{g,p},vec} = - 2 \pi N^{3/2} \int dt \rho(t)^2 (g-1)( \frac{1}{2} \xi_{max}^2 - \frac{1}{12}) \;. \ee

We again find a divergence that must be dealt with.  This is resolved by imposing the condition:

\be \sum_{I \in \alpha} s_I=0 \;\; \text{ and } \;\; \sum_{I \in \alpha} (r_I-1) +2 = 0 \;, \ee
where the sum is taken over all bifundamental chirals charged under the $\alpha$th gauge group, and this must hold for all gauge groups, and adjoints are counted twice.  These two conditions can be summarized by:
\be \sum_{I \in \alpha} \ell_I=0 \;, \ee 
where in the sum we include the contribution of the vector multiplet, with $\ell_I=g-1$.  Then, using also the condition $\sum_{I \in \alpha} m_I=0$ imposed above, we find the coefficient of the term quadratic in $\xi_{max}^2$ vanishes.  Thus, we are left with the following finite contribution from a bifundamental:

\be \label{zmgpbif} \log Z_{\cM_{g,p},bif} = - 2 \pi N^{3/2} \int dt \rho(t)^2 G_\ell(\delta v + m) \;. \ee

In addition, there is the diagonal contribution from the bifundamental:

\be \sum_{a=1}^N p \log \cF_\pi(u_a - \hat{u}_a + m)  + \ell \log \Pi_\pi(u_a - \hat{u}_a + m) \ee

$$ = N \int dt \rho(t) \bigg( (p (\delta v+m) - \ell) \log(1- e^{2 \pi i (\delta v+m)}) + \cdots \bigg) \;. $$
Here, since the contribution is subleading, we keep only those terms which can diverge.  Specifically, the log gets large in the tail region, where:

\be \delta v(t) +m = \hat{n} + C e^{- 2 \pi N^{1/2} Y(t) } \;, \ee
and one finds a contribution:

\be -2 \pi N^{3/2} \sum_{tails} \int_{\delta v(t) + m \approx \hat{n}} dt  (p \hat{n} - \ell) \rho(t) Y(t) \;.   \ee

For an (anti-)fundamental chiral, we have the following contribution to the twisted superpotential:

\be \cW_{fun} \rightarrow  N \int dt \rho(t) \bigg( p \cF_\pi(\pm (v^\alpha + i t N^{1/2}) + m) +  \ell \Pi_\pi(\pm (v^\alpha + i t N^{1/2}) + m) \bigg)  \;.\ee
Using the limits:

\be \left\{ \begin{array}{cc} \cF_\pi(u) \rightarrow \pm \bigg(- \frac{\pi i}{2} u^2  + \frac{\pi i}{12} \bigg)  \\ 
\Pi_\pi(u) \rightarrow \pm \bigg( \pi i u - \frac{\pi i}{2} \bigg) \end{array} \right. \;\;\;\; \text{as} \;\;\;\; \text{Im} u \rightarrow \pm \infty \;,\ee
we find this becomes:
\be N \int dt \rho(t) \big( \pm \frac{\pi i}{2} p t |t| N + N^{1/2} ( \pi p |t| (\pm v^\alpha + m) - \pi t \ell) \big) \;. \ee
Imposing the number of fundamentals and anti-fundamentals are equal, the $O(N^2)$ term cancels and we find:

\be -2 \pi N^{3/2} \int dt \rho(t) |t| \big( -\frac{1}{2} p (\pm v^\alpha + m) + \frac{1}{2} \ell  \big) \;. \ee

Finally, we also have the contribution from the Hessian:

\be  \log {\text H} = \log  \det \left( \begin{array}{cc} \displaystyle \frac{\partial^2 \cW}{\partial u_a \partial u_b} & \displaystyle \frac{\partial^2 \cW}{\partial \hat{u}_a \partial u_b} \\ 
\displaystyle \frac{\partial^2 \cW}{\partial u_a \partial \hat{u}_b} &\displaystyle  \frac{\partial^2 \cW}{\partial \hat{u}_a \partial \hat{u}_b} \end{array} \right) \;. \ee
Then, naively this has order $N \log N$ if all the components of the matrix stay finite, but near the tail regions there are divergences which cause it to contribute at leading order.  As argued in \cite{Benini:2015eyy}, one finds:

\be \log H \approx  2 \pi N^{3/2} \sum_{tails}  \int_{t | \delta v(t) + m \approx n} dt \rho(t) Y(t)  \;. \ee

\bibliographystyle{ieeetr}
\bibliography{bib3d}{}

\end{document}